\documentclass[11pt, a4paper]{article}
\usepackage{amsmath,amsthm,amssymb}
\usepackage{enumerate}
\usepackage{graphicx,xcolor,cancel}
\usepackage[toc,page]{appendix}
\usepackage{nicefrac, xfrac}
\usepackage{comment}
\usepackage{mathtools}
\usepackage{braket,mathdots}
\usepackage{array}
\usepackage{tabularray}
\usepackage{relsize}
\usepackage{fnpct}
\newcommand\Tstrut{\rule{0pt}{3.6ex}}

\numberwithin{equation}{section}

\newtheoremstyle{bfnote}
  {}{}
  {\itshape}{}
  {\bfseries}{.}
  { }{\thmname{#1}\thmnumber{ #2}\thmnote{ (#3)}}

\newtheorem{theorem}{Theorem}[section]
\newtheorem{corollary}{Corollary}[section]

\newtheorem{definition}{Definition}[section]

\newtheorem{lemma}{Lemma}[section]

\newtheorem{proposition}{Proposition}[section]

\theoremstyle{bfnote}
\newtheorem{mainresult}{Main Result}
\theoremstyle{definition}
\newtheorem{remark}{Remark}[section] 
\usepackage{tabularray}

\usepackage[bookmarks=false,hidelinks]{hyperref} 
\usepackage[top=2.8cm,bottom=2.8cm,left=1.9cm,right=1.9cm]{geometry}
\usepackage{tikz}
\usetikzlibrary{shapes.geometric,decorations.markings,intersections,calc,positioning,arrows.meta}

\makeatletter
\g@addto@macro\bfseries{\boldmath}
\makeatother

\DeclareMathOperator{\Tr}{Tr}

\newcommand{\li}{\lambda^i}
\newcommand{\lj}{\lambda^j}
\newcommand{\bli}{\bar{\lambda}^i}
\newcommand{\blj}{\bar{\lambda}^j}
\newcommand{\eqa}{\begin{eqnarray}}
\newcommand{\eeqa}{\end{eqnarray}}
\newcommand{\beq}{\begin{equation}}
\newcommand{\eeq}{\end{equation}}

\newenvironment{proofC}[1][Proof]{\textbf{#1.} }{\ \rule{0.5em}{0.5em}}

\newcommand{\der}[2]{\frac{\partial #1}{\partial #2}}

\begin{document}

\title{Random matrix ensembles and integrable differential identities}

\author{Costanza Benassi$^{\;a)}$, Marta Dell'Atti$^{\;b)}$, Antonio Moro$^{\;a)}$
\footnote{Corresponding author: Antonio Moro; email: antonio.moro@northumbria.ac.uk}
 \\
\\
\small{$^{a)}$School of Engineering, Physics and Mathematics, Northumbria University Newcastle, UK}\\
\small{$^{b)}$Faculty of Mathematics Informatics and Mechanics, University of Warsaw, Poland} \\
}
\date{}

    \maketitle

\begin{abstract}
Integrable differential identities, together with ensemble-specific initial conditions, provide an effective approach for the characterisation of relevant observables and state functions in random matrix theory. We develop this approach for the unitary and orthogonal ensembles. In particular, we focus on a reduction where the probability measure is induced by a Hamiltonian expressed as a formal series of even interaction terms. We show that the order parameters for the unitary ensemble, that is associated with the Volterra lattice, provide a solution of the modified KP equation. The analogous reduction for the orthogonal ensemble, associated with the Pfaff lattice, leads to a new integrable chain. A key step for the calculation of order parameters for the orthogonal ensemble is the evaluation of the initial condition by using a map from orthogonal to skew-orthogonal polynomials. The thermodynamic limit leads to an integrable system (a chain for the orthogonal ensemble) of hydrodynamic type. Intriguingly, we find that the solution to the initial value problem for both the discrete system and its continuum limit are given by the very same semi-discrete dynamical chain.\\


\end{abstract}

\tableofcontents{}

\section{Introduction}\label{sec:intro}
Random matrix ensembles appear in a variety of contexts in mathematics and physics and represent a fundamental mathematical object for the description of certain universal properties of complex systems. First introduced by Wishart with the derivation of correlation matrix distributions for multivariate random variables~\cite{wishart}, in the 1950s Wigner further developed the concept of random matrix ensemble to formulate a statistical theory of energy spectra of heavy nuclei~\cite{wigner_1951,wigner_1955,wigner_1957}. This specific application in physics inspired studies on random matrices with a real spectrum, such as symmetric and Hermitian matrix ensembles. Further seminal contributions by Mehta, Gaudin and Dyson~\cite{mehta_1960,mehta_gaudin_1960,gaudin_1961,DysonI} led to the development of a mathematical theory of random matrices, intended as the abstract mathematical object underpinning -- in Dyson's own words -- ``a new kind of statistical mechanics, in which we renounce exact knowledge not of the state of a system but of the nature of the system itself"~\cite{DysonI}. Dyson's perspective elucidates the fundamental reasons behind the relevance of random matrix ensembles as a universal framework to model and understand complexity in mathematics, physics and applied sciences. Notable examples, just to mention a few, are: the use of Hermitian matrix ensembles to model metric fluctuations in $2D$ quantum gravity~\cite{Witten} and quantum topological field theory~\cite{difrancesco}; their application to  graph enumeration~\cite{Brezin:1977sv,Ercolani2002AsymptoticsOT} and problems in enumerative geometry~\cite{Witten,Kontsevich}; the discovery of the remarkable properties of the partition functions and their relationship with integrable hierarchies of nonlinear PDEs~\cite{vanM_Toda_orthogonal,vanMoerbeke2000,Adler1999ThePL,adler2002}; Riemann-Hilbert problems and nonlinear steepest descent~\cite{Bleher_Its,McL_Miller};  the intriguing connections with the moments of the distribution of the zeros of the Riemann zeta-function on the critical line~\cite{keating_snaith}. The body of work on random matrix theory, in light of its {nearly century long} history, is extremely vast and its applications have extended well beyond the classical fields from which it has originated. Hence, it is out of the scope of the present work to give a detailed and exhaustive account.

In this paper, we consider random matrix ensembles specified by the pair {$({\cal M}^{(\beta)}_n, \omega)$ where ${\cal M}^{(\beta)}_n$} is a space of $n\times n$ random matrices with real eigenvalues and $\omega(\mathbf{t})$ is a probability measure, invariant with respect to a suitable group of transformations on {{${\cal M}^{(\beta)}_n$. We shall focus on the cases $\beta=1,2$}, that is: $i$) ${\cal M}^{(2)}_n={\cal H}_{n}$, where~${\cal H}_{n}$  is the space of Hermitian matrices of order~$n$; and $ii$) ${\cal M}^{(1)}_n ={\cal S}_{n}$, where ${\cal S}_{n}$  is the space of symmetric matrices of order $n$. For both {spaces ${\cal H}_n$ and ${\cal S}_n$,} the measure $\omega(\mathbf{t})$ is {assumed to be} of the form
\begin{equation}
\label{intro:measure}
d\omega(\mathbf{t}) = \text{e}^{-H(M;\, \mathbf{t})}\; dM,
\end{equation}
where $dM$ is the Haar measure, invariant with respect to the unitary {($U(n)$)} and the orthogonal {($O(n)$)}  groups for Hermitian and symmetric matrices respectively.
The Hamiltonian {function} $H(M; \mathbf{t})$ specifies the weight of a configuration as a function of a set of parameters (coupling constants) $\mathbf{t} =(t_{1},t_{2},\dots)$, and it is formally defined as follows
\begin{equation}   
\label{intro:ham}
H(M; \mathbf{t})={\rm Tr}\bigg(\frac{M^2}{2} -\sum_{j\geq 1} t_j\, M^j\bigg).
\end{equation}
 {The ensembles $({\cal H}_n,\omega )$ and $(\mathcal{S}_n,\omega)$ {are}, respectively, {the} unitary and orthogonal ensembles, based on the {above mentioned} symmetry properties of the measure \eqref{intro:measure}.} The particular case $\mathbf{t}=\mathbf{0}$ corresponds to the celebrated Gaussian Unitary Ensemble (GUE) and the Gaussian Orthogonal Ensemble (GOE) for, respectively, ${\cal M}^{(2)}_n ={\cal H}_{n}$ and ${\cal M}^{(1)}_n ={\cal S}_{n}$. Gaussian ensembles have been studied extensively since the early developments of random matrix theory~\cite{Meh2004}.
We notice that in statistical thermodynamics, in the definition of the measure~\eqref{intro:measure}, it is customary to divide the exponent by the temperature $T$. However, given the form of the Hamiltonian \eqref{intro:ham}, one can always {`absorb'} $T$ into the definition of the coupling constants $\mathbf{t}$ or, equivalently, set $T=1$.  \\
The probability density distribution induced by the measure is defined as
\begin{equation}
\label{intro:probdensity}
P^{(\beta)}(M; \mathbf{t})\, dM = \frac{{\rm e}^{-H(M; \mathbf{t})}}{Z^{(\beta)}_n(\mathbf{t})}\,dM,
\end{equation}
where $Z_n^{{(\beta)}}(\mathbf{t})$ is the partition function
\begin{equation}
\label{intro:Zn}
Z^{(\beta)}_{n}(\mathbf{t}) = \int_{{\cal M}^{(\beta)}_{n}}{\rm e}^{-H(M; \mathbf{t})}\;  dM.
\end{equation}
The partition function $Z_{n}^{{(\beta)}}(\mathbf{t})$ is, indeed, key for {the study} of the properties of the probability density~\eqref{intro:probdensity} and therefore the calculation of expectation values
\[
\mathbb{E}^{(\beta)}_{n}[f(M)] := \int_{{\cal M}^{(\beta)}_{n}} f(M)\,P^{(\beta)}(M; \mathbf{t})\; dM,
\]
where the function $f(M)$ is an observable defined on ${\cal M}^{(\beta)}_{n}$. Beside the partition function, it is convenient to consider, especially if one is interested in the behaviour of observables for  large $n$, the Helmholtz free energy
\begin{equation}
\label{intro:free_energy}
F^{(\beta)}_{n}(\mathbf{t}) := \frac{1}{n} \log Z^{(\beta)}_{n}(\mathbf{t}).
\end{equation}
As an immediate application of the Helmholtz free energy, one can calculate by direct differentiation the expectation values of the traces 
\begin{equation}   
\mathbb{E}^{(\beta)}_n[{\rm Tr} (M^{j})] = n\, \der{F^{(\beta)}_{n}}{t_{j}} = \int_{{\cal M}^{(\beta)}_{n}} {\rm Tr} (M^{j})\, P^{(\beta)}(M; \mathbf{t}) \, dM.
\end{equation}
In the following, it will be convenient to work with an equivalent (up to a scaling factor and a shift) definition of free energy, that is
\begin{equation}
\label{intro:phi}
\Phi^{(\beta)}_n(\mathbf{t}) := n F^{(\beta)}_n(\mathbf{t}) - \log C^{(\beta)}_n,
\end{equation}
where
\begin{equation}
\label{C_volume}
C_n^{(\beta)} = n!  \,\pi^{\frac{1}{4}(\beta n(n-1))} \prod_{k=1}^{n} \frac{\Gamma \left(1+\frac{\beta}{2}\right)}{\Gamma \left(1+k\frac{\beta}{2}\right)} 
\end{equation}
is a suitable constant whose value depends on whether the chosen matrix ensemble is orthogonal ($\beta = 1$) or unitary ($\beta =2$).

Important questions in random matrix theory are concerned with the study, in the space of couplings~$\mathbf{t}$, of the partition function, the density distribution of eigenvalues, the correlation functions and observables, such as the expectation values of traces mentioned above at both finite and large $n$. Numerous classical results are available for particular choices of sets of coupling constants $\mathbf{t}$. These include, just to mention a few, the celebrated Wigner's semi-circle law for the GUE eigenvalues distribution~\cite{Meh2004},  Fredholm determinantal formulae for the correlation functions~\cite{tracy_widom}, level spacing distributions and their asymptotic evaluation~\cite{Bothner}, the use of the nonlinear steepest descent method~\cite{Bleher_Its}, Riemann-Hilbert and orthogonal polynomials  methods for large $n$ asymptotic expansion for the free energy and the density distribution of eigenvalues~\cite{McL_Miller, Borot,grava_claeys_mcl,martinez}. The general problem of evaluating partition functions and observables for an arbitrary number of coupling constants remains open. However, we note that for Hermitian {matrix models} a general large $n$ asymptotic formula for the free energy has been proved in~\cite{Charlier}.

The approach we follow below builds upon a celebrated conjecture of Witten~\cite{Witten}, proved by Kontsevich~\cite{Kontsevich}, implying that the partition function of a Hermitian random matrix model is identified with the $\tau$-function of a particular solution of the KdV hierarchy. The extensive body of results obtained by Adler and van Moerbeke starting from the mid 1990s (see e.g.~\cite{vanM_Toda_orthogonal,vanMoerbeke2000,Adler1999ThePL,adler2002}) further established that the Witten-Kontsevich theorem is a manifestation of the universal integrable structure encoded in the algebraic structure of random matrix ensembles. For instance, for a general class of {unitary} ensembles~{with partition function} of the form \eqref{intro:Zn}, Adler and van Moerbeke showed that a suitable factorisation of the moments matrix yields the Lax matrix {of the Toda hierarchy}, evaluated {for} a particular solution. Equivalently, one can say that the {sequence of partition functions $\{Z^{(2)}_{n}(\mathbf{t}) \}_{n \in \mathbb{N}}$ for the unitary ensemble~\eqref{intro:Zn} is identified (up to a constant factor) with the $\tau$-function of a particular solution to the Toda hierarchy}. {A similar analysis for the orthogonal ensemble shows that the sequence of partition functions $\{Z^{(1)}_{2n}(\mathbf{t}) \}_{n \in \mathbb{N}}$ is identified with the $\tau$-function of a novel integrable hierarchy referred to as Pfaff lattice}~\cite{Adler1999ThePL,adler2002}. 

\begin{table}[t]   
    \centering \footnotesize
    \begin{tblr}{
        hlines, vlines,
        colspec = {c c c},
        row{1} = {font=\bfseries, c},
        rowsep = 2ex,
        colsep = 0pt,
    }
    & & {~~All couplings~~} & {~~Even couplings only ($t_{2n+1} = 0$)~~}  \\
    
    & ~~\rotatebox[origin=c]{90}{\textbf{Unitary ensemble}}~~  & 
        \vspace{-1cm}
        ~~ $ L_{\text{Toda}} = \begin{pmatrix}
            a_{1} &  b_{1} & 0  & 0 & \cdots \\[.2ex]
b_{1} & a_{2} & b_{2}  & 0 & \ddots \\[.2ex]
0 & b_{2} & a_{3}  & b_{3} & \ddots \\[.2ex]
\vdots & \ddots & \ddots & \ddots & \ddots
        \end{pmatrix}$ ~~
    
     & \qquad ~~$ L_{\text{Volterra}} = \begin{pmatrix}
            0 &  b_{1} & 0  & 0 & \cdots \\[.2ex]
b_{1} & 0 & b_{2}  & 0 & \ddots \\[.2ex]
0 & b_{2} & 0  & b_{3} & \ddots \\[.2ex]
\vdots & \ddots & \ddots & \ddots & \ddots
        \end{pmatrix}$~~ \\
    
   & ~~\rotatebox[origin=c]{90}{\textbf{Orthogonal ensemble}}~~ & ~~$
   L_{\text{Pfaff}} = \begin{pmatrix}
            ~0~ & ~1~ & ~0~ & ~0~ & ~0~ & ~0~ & \cdots\\[.1ex]
w^{-1}_1& ~v^0_1~ & w^{0}_{1}\hspace*{-0.1ex}& 0 & 0 & 0 &\ddots\\[.1ex]
 v_1^{-1} & w^{1}_{1}\hspace*{-0.1ex}& -v_1^0 & 1 & 0 & 0 & \ddots\\[.1ex]
 w^{-2}_{1}\hspace*{-0.1ex}& v_1^{1} & \hspace{-0.1ex}w^{-1}_2& ~v_2^0 & ~w^{0}_{2}& 0 & \ddots \\[.1ex]
 v_1^{-2} & w^{2}_{1}\hspace*{-0.1ex}& v_2^{-1} & ~w^{1}_{2}& -v_2^0 & 1 & \ddots \\[.1ex]
\vdots& \ddots &\ddots& \ddots & \ddots& \ddots & \ddots 
        \end{pmatrix}$~~ 
    
    & ~~$L_{\substack{\text{even} \\ \text{Pfaff}}} = \begin{pmatrix}
            ~0~ & ~1~ & ~0~ & ~0~ & ~0~ & ~0~ & \cdots\\[.1ex]
w^{-1}_1& ~0~ & w^{0}_{1}\hspace*{-0.1ex}& 0 & 0 & 0 &\ddots\\[.1ex]
 0 & w^{1}_{1}\hspace*{-0.1ex}& 0 & 1 & 0 & 0 & \ddots\\[.1ex]
 w^{-2}_{1}\hspace*{-0.1ex}& 0 & w^{-1}_2& 0 & ~w^{0}_{2}\hspace*{-0.1ex}& 0 & \ddots \\[.1ex]
 0 & ~w^{2}_{1}\hspace*{-0.1ex}& 0 & w^{1}_{2}\hspace*{-0.1ex}& 0 & 1 & \ddots \\[.1ex]
\vdots& \ddots &\ddots& \ddots & \ddots& \ddots & \ddots \\
        \end{pmatrix}$~~
    \end{tblr}
  \caption{Lax matrices for the lattice hierarchies associated with the unitary ensemble (Toda, Volterra) and the orthogonal ensemble (Pfaff, even Pfaff) respectively. The entries for Toda are the variables $a_n$, $b_n$, with $n \in \mathbb{N}$; while for Volterra only the variables $b_n$ appear. {The entries for Pfaff are denoted as $v_{n}^{k}$, $w^{k}_n$, with upper index $k \in \mathbb{Z}$ and lower index $n \in \mathbb{N}$. The symbols $w^{k}_n$, for a given $k$, are located on the $(2 |k| -1)$-th lower diagonal, while the symbols $v_{n}^{k}$ are located on the $2 |k|$-th lower diagonal.  The lower index $n$ refers for $k<0$ to the $n$-th odd site and for $k>0$ to the $n$-th even site along the diagonal. The even Pfaff Lax matrix contains the symbols~$w^{k}_n$ only.}}
    \label{tab:even}
\end{table}

{Our study below will be  specifically concerned with the partition function and the Lax equations for the unitary and orthogonal ensembles}. We shall see that the integrable structure of the orthogonal ensemble in connection with the Pfaff lattice presents a higher level of complexity compared to the case of the unitary ensemble. The Lax matrix  of the Toda lattice is tridiagonal and symmetric, and provides two sets of order parameters for the unitary ensemble. The Lax matrix for the Pfaff lattice is { a Hessenberg matrix, i.e. lower triangular with one non-trivial upper diagonal,} featuring an infinite set of order parameters for the orthogonal ensemble. A comparison between the forms of the two Lax matrices is shown in Table~\ref{tab:even}.
A classical result by Weyl~\cite{Meh2004, Weyl} allows us to reduce the partition function~\eqref{intro:Zn}, for both {$(\mathcal{H}_n, \omega)$ and $(\mathcal{S}_n,\omega)$}, to {a} so-called $\beta$-integral
\begin{equation}
\label{intro:partition_n_eigenvalues}
Z^{(\beta)}_n(\mathbf{t})= \frac{C^{(\beta)}_n}{n!} \int_{\mathbb{R}^n} |\Delta_n(z)|^{\beta} \, \prod_{i=1}^n \rho(z_i; \mathbf{t})  \,dz_i \,, \qquad \rho(z; \mathbf{t})=\exp \bigg(- \frac{z^2}{2} + \sum\limits_{k\geq 1}t_k\,z^k \bigg) ,
\end{equation}
where the integration variables $z_i$ are the eigenvalues of the matrix $M$ in the integral~\eqref{intro:Zn}. Here,~$\Delta_n(z)$ denotes the Vandermonde determinant, i.e.\ 
$\Delta_n(z)=\prod_{1\leq i < j \leq n}(z_i-z_j)$, and $C^{(\beta)}_n$ is the constant obtained from the integration { in \eqref{intro:Zn} of} the non-diagonal degrees of freedom and defined in \eqref{C_volume},  { i.e.\ the volume of the relevant invariant subgroup associated to the Haar measure \cite[eq.\ 1.2.15]{EynardReview}}. Below, we shall focus on the $\tau$-function, i.e.\ the integral
\begin{equation}  
\label{intro:tau_n} 
\tau^{(\beta)}_n(\mathbf{t}):=\dfrac{Z^{(\beta)}_n(\mathbf{t})}{C^{(\beta)}_n}=\frac{1}{n!}\int_{\mathbb{R}^n} |\Delta_n(z)|^{\beta} \, \prod_{i=1}^n \rho(z_i; \mathbf{t}) \,dz_i \,.
\end{equation}
An important step in the study of the above $\tau$-function is the observation that $\tau_{n}^{{(\beta)}}(\mathbf{t})$ can be expressed in terms of the moments of a one variable probability measure induced by the Hamiltonian~\eqref{intro:ham}.
More specifically, $\tau_{n}^{{(2)}}(\mathbf{t})$ can be calculated as the determinant  of the $n\times n$ moments matrix {$m_{n}(\mathbf{t})$} of entries 
\begin{equation*} 
\left (m_n(\mathbf{t}) \right)_{ij} = \left(\,x^i,\, x^j\,\right )_\mathbf{t}, \qquad  x \in \mathbb{R}, \qquad i,j =0,\dots,n-1,
\end{equation*}
where $(\,\cdot\,\,,\,\cdot\,)_\mathbf{t}$ is the inner product
\begin{equation}
\left(\,f\,,\, g\,\right)_\mathbf{t} = \int_{\mathbb{R}}f(x)\,g(x)\,\rho(x; \mathbf{t})\,dx,
\label{intro:inner_product}
\end{equation}
with $\rho(x;\mathbf{t})$ defined in~\eqref{intro:partition_n_eigenvalues}.
A remarkable result proven in~\cite{vanM_Toda_orthogonal} is  that the factorisation of the semi-infinite {extension of the moments matrix $m_{n}(\mathbf{t})$, such that $n \to \infty$}, of the form
\[
m_{\infty}(\mathbf{t}) = A(\mathbf{t})^{-1} \left(A(\mathbf{t})^{\top} \right)^{-1},
\]  
where $A(\mathbf{t})$ is lower triangular with non-zero diagonal elements,
yields the Lax matrix of the Toda hierarchy
\[
L_{\textup{Toda}}(\mathbf{t}) = A(\mathbf{t}) \,\Lambda \,A(\mathbf{t})^{-1} \,, 
\]
evaluated on the particular solution specified by the sequence~$\{\tau_{n}^{{(2)}}(\mathbf{t}) \}_{n\in \mathbb{N}}$. Here, $\Lambda$ is the shift matrix such that $(\Lambda v)_{i} = v_{i+1}$.
An analogous result has been proven in the case $\beta = 1$~\cite{adler2002}. Specifically, the $\tau$-function $\tau_{2n}^{{(1)}}(\mathbf{t})$ -- calculated for symmetric matrices of even order --  is expressed as the Pfaffian of the skew-symmetric moments matrix {$m_{2n}(\mathbf{t})$}, of entries
\begin{equation*} 
(m_{2n}(\mathbf{t}))_{ij} = \langle\, x^i, y^j \,\rangle_\mathbf{t}, \qquad x\in \mathbb{R},~y \in \mathbb{R}, \quad i,j =0,\dots,2n-1 
\end{equation*}
where $\langle \,\cdot\,\,, \,\cdot \,\rangle_\mathbf{t}$ denotes the skew-symmetric inner product
\begin{equation}
 \langle\, f\,,\, g\,\rangle_{\mathbf{t}} = \frac{1}{2}\int_{\mathbb{R}^2} f(x)\, g(y)\, \text{sgn}(y-x)\, \rho(x; \mathbf{t})\,\rho(y; \mathbf{t})\,dx\,dy.
    \label{intro:skew}
\end{equation}
The following (unique) factorisation of the semi-infinite skew-symmetric moments matrix holds (see e.g.~\cite{vanMoerbekenotes})
\[
m_{\infty}(\mathbf{t}) = S(\mathbf{t})^{-1} J \left(S(\mathbf{t})^{\top}\right)^{-1},
\]
with $J$ such that $J^2 = -I$, where is $I$ is the semi-infinite identity matrix. This decomposition leads to the Lax matrix 
\[
L_{\textup{Pfaff}}(\mathbf{t}) = S(\mathbf{t}) \,\Lambda \,S(\mathbf{t})^{-1},
\]
evaluated on a particular solution of the Pfaff lattice hierarchy.
Unlike the Toda lattice, where $L_{\textup{Toda}}$ is tri-diagonal, $L_{\textup{Pfaff}}$ is constituted of $2 \times 2$ non-zero blocks with respect to which is lower triangular {(as shown in Table~\ref{tab:even}, first column)}.

For both unitary and orthogonal ensembles, the hierarchy is specified by {a} system of compatible equations for the Lax matrix {$L$} of the form
\begin{equation}   
\label{intro:hierarchy}
\der{L}{t_{j}} = \left [B(L^j)\,, \, L \right ], \qquad j \in \mathbb{N},  
\end{equation}
where {$B(L^j)$} is a suitable projection of the {{$j$}-th power} of the Lax matrix with respect to a certain algebra splitting (more details are provided in Sections~\ref{sec:toda} and~\ref{sec:SME_Pfaff} for {$L = L_{\text{Toda}}$ and $L = L_{\text{Pfaff}}$, respectively}). 
The entries of the Lax matrix can be interpreted as statistical observables or, in the large~$n$ limit, as state functions for the matrix ensemble. These observables can be expressed in terms of the relevant sequence of $\tau$-functions, or equivalently the free energy \eqref{intro:phi}. 
Hence, the Lax equations~{\eqref{intro:hierarchy}} constitute a set of nonlinear differential identities governing the evolution of observables in the space of coupling constants~{$(t_1,t_2,\dots,t_j, \dots)$}. These are identified with the time variables of the integrable hierarchy, {{with} the $j$-th equation in~\eqref{intro:hierarchy} describing the evolution in the time~$t_j$.}

The {particular solution of the differential identity associated with the given random matrix ensemble} is fixed by the initial value $L(\mathbf{0})$. In particular,  $\tau^{{(\beta)}}_{n}(\mathbf{0})$ yields the partition function of the Gaussian ensemble (either GUE or GOE, {for, respectively, $\beta =2$ and $\beta = 1$}). Therefore, the integrable differential identity~\eqref{intro:hierarchy} and the related observables for a given choice of the parameters~$\mathbf{t}$ are specified via the Gaussian ensemble, for which numerous exact analytical results are indeed available (e.g.~\cite{Meh2004}). Depending on the specific initial condition, integrability unlocks, in principle, the application of a range of techniques to tackle the study of order parameters, e.g.\ IST (inverse scattering transform) method, Riemann-Hilbert problem and the nonlinear steepest descent method.
We note that the Witten-Kontsevich theorem -- stating that the partition function of 2D quantum gravity and the one of a related Hermitian random matrix model is given by the $\tau$-function of a particular solution of the KdV equation with initial condition specified by a Virasoro constraint -- can be interpreted as a realisation of the method of integrable differential identities outlined above,  where the required integrable differential identities are given by the KdV hierarchy. In~\cite{HermitianPRE} integrable differential identities for the unitary ensemble have been used to derive the Volterra hierarchy and demonstrate the occurrence of phase transitions associated with the onset of a dispersive shock in the order parameters {appearing in} the Lax matrix.

We also note that the approach based on integrable differential identities has been introduced independently in the context of classical thermodynamics and statistical mechanics in~\cite{MoroAnnals,Moro:2012,Moro_Barra_2014} for the solution of the van der Waal model and its virial extensions; in~\cite{Moro:2012,MoroAnnals,Moro_Barra_2014,Moro_giglio_2016,Giglioetal_2021} for the solution of mean field spin models; in~\cite{Moro2014,Moro_Barra_Agliari_DelloSchiavo,MoroPotts} for the solution of a range of mean field liquid crystals models~\cite{Moro_DeMatteis_Giglio_2018,Moro_Giglio_DeMatteis_2024}.  These results formalised and  extended the approach to the solution of the classical Curie-Weiss model as reported independently in~\cite{brankov_zagrebnov} and~\cite{choquard_wagner} and in the two consequential papers~\cite{guerra_sum,Barra_Genovese}.

In this paper we focus on the unitary and orthogonal matrix ensembles. We study the differential identities, i.e.\ the  Lax equations of the form mentioned above, and the associated initial conditions in the case of finite~$n$ and in the thermodynamic limit as $n \to \infty$. The main objects of study are the sequence of $\tau$-functions $\{\tau^{{(\beta)}}_{n}(\mathbf{t}) \}_{n\in \mathbb{N}}$ and the derivatives of $\tau_{n}^{{(\beta)}}(\mathbf{t})$ with respect to the parameters~$\mathbf{t}$.  The explicit calculation of the initial condition in terms of the matrix size $n$ is crucial to determine the scaling properties of the observables and perform the large $n$ limit. The thermodynamic order parameters are obtained by constructing a suitable interpolation function for the entries of the Lax matrix consistently with the scaling properties of the initial condition. For the unitary ensemble we shall see that the interpolation function is expressed as a power series in the lattice spacing~$\varepsilon = 1/N$ of the Toda chain, where $N$ is a large (thermodynamic) scale such that $n/N$ is finite for large $n$. The thermodynamic limit corresponds to the continuum limit of the Toda {lattice}. At the leading order, the order parameters satisfy an integrable system of hydrodynamic type referred to as dispersionless Toda lattice (dTL). 

In the {special} case where odd couplings are absent, i.e.\ $t_{2j+1}\!=\!0$, the Toda {hierarchy associated with the remaining even couplings $t_{2j}$ yields  the Volterra lattice, which, in the continuum limit, gives, at the leading order, a scalar integrable hierarchy of hydrodynamic type, namely the Hopf hierarchy.} The initial condition is readily evaluated in terms of a Selberg's integral with Gaussian weight. This procedure is equivalent to the calculation of the recurrence coefficient of Hermite polynomials that are associated with the GUE~\cite{Meh2004}. We adopt a similar approach to tackle the analogous problem for the {orthogonal} ensemble. {However,} some major differences occur due to the higher complexity of the Pfaff lattice integrable structure. {Setting the odd couplings to zero, i.e.\ $t_{2j+1}=0$, we obtain a reduced Pfaff lattice that can be viewed as the analogue of the Volterra lattice relative to the Unitary ensemble. We refer to this reduction as {\itshape even Pfaff lattice}. We also note that unlike the Toda lattice, where the Lax matrix is symmetric and} tridiagonal, and the entries have a simple explicit expression in terms of the $\tau$-function, the Lax matrix for the Pfaff lattice is block lower triangular and no simple explicit formula for its entries in terms of the $\tau$-function is available. Table~\ref{tab:even} shows, for the sake of comparison, the forms of Toda and Pfaff Lax matrices and their respective reductions. }

{{For the orthogonal ensemble, we evaluate the initial condition explicitly. This is achieved by exploiting a suitable map from orthogonal to skew-orthogonal polynomials, as a a direct calculation via Selberg's integrals is not possible in this case}. In addition, the scaling properties of the initial condition suggest that the correct asymptotic expansion of some of the order parameters, suitably interpolating the entries of the Lax matrix, {{is required to be} singular in the large $n$ limit. This procedure leads to a novel integrable chain. Remarkably, the same integrable chain emerges from the analysis of the Lax equations with the prescribed initial condition at both finite $n$ and at the leading order in the thermodynamic limit. {We emphasise that the chain so obtained can be viewed as the analogue of the Hopf equation obtained for the unitary ensemble.}

\noindent We now briefly summarise the main results obtained in this paper. 

\begin{mainresult}[Unitary Ensemble]\label{res:unitary_ensemble}
\noindent We prove that, when the Hamiltonian~\eqref{intro:ham} contains even powers only -- and the Toda lattice sub-hierarchy, associated with even couplings $t_{2n}$ only, reduces to the Volterra lattice hierarchy (see e.g.~\textup{\cite{HermitianPRE}}) --  the  variation of the chemical potential, i.e.\ $\Delta \mu_{n}^{{(2)}} = \mu_{n+1}^{{(2)}} - \mu_{n}^{{(2)}}$, where $\mu_{n}^{{(2)}} = \Phi_{n+1}^{{(2)}} - \Phi_{n}^{{(2)}}$, gives a solution of the modified KP (mKP) equation (see Theorem~\ref{thm:mKP_volterra}). 
\end{mainresult}
We observe that the entries of the Toda lattice Lax matrix for the unitary ensemble are expressed in terms of the ensemble expectation values of the traces {$\Tr (M^j)$, i.e.\ of non-negative powers of random Hermitian matrices of order $n$ ($M\in{\cal H}_{n}$)}. 
Recalling that the Toda Lax matrix is tri-diagonal and symmetric, the diagonal entries are given in terms of the expectation values of the traces, whilst the off-diagonal entries give the variation of the chemical potential with respect to the matrix size~(see~\eqref{eq:expect_Toda}). In the case where the Hamiltonian~\eqref{intro:ham} contains both even and odd powers, i.e.\ $\mathbf{t} = (t_1, t_2, \dots)$ includes both even and odd couplings, the $\tau$-function $\tau^{{(1)}}_{n}(\mathbf{t})$ gives a solution of the KP hierarchy (see e.g.\ \cite{Adler_KP}). The above result establishes a similar relationship between the modified KP equation and the Volterra chain that emerges when the Hamiltonian~\eqref{intro:ham} contains even powers only, i.e.~$t_{2n+1}\!=\!0$.

\begin{mainresult}[Pfaff lattice]\label{res:pfaff_lattice_proof}
We give a rigorous proof of a result obtained in {\textup{\cite{benassi2021symmetric}}}, namely, that the reduced Pfaff lattice equation restricted to the even couplings, i.e.\ $t_{2n+1} = 0$, is equivalent to a double infinite chain of the following form
\begin{align*}   
\partial_{t_{2}} w^{k}_{n} = {\cal P}_{3}^{k}\big(w^{0},w^{1}, w^{k} \big) + {\cal P}_{2}^{k}\big(w^{0}, w^{k\pm 1} \big) \qquad k \in \mathbb{Z},~ n \in \mathbb{N},
\end{align*}
where {${\cal P}_{3}^{k}$} and {${\cal P}_{2}^{k}$} are homogeneous polynomial expressions of, respectively, degree $3$ and $2$ in the variables~{$w^k_n$, {i.e.\ the entries of} the 
{even Pfaff Lax matrix} {shown} in Table~\ref{tab:even}.} For the sake of simplicity, we dropped the explicit dependence on the subscripts (see equations~\eqref{pfaff_lattice} for further details). 
\end{mainresult}

\noindent
This result is proved in part~\eqref{thm:semi-discrete_syst} of Theorem~\ref{theo:discrete_chain}. We clarify that here we focus on the analysis of the first member of the Pfaff hierarchy, and provide explicitly the equations that describe the deformations with respect to the coupling constant $t_{2}$. However, higher members of the hierarchy can be obtained proceeding in the same way by considering higher members of the Pfaff-Lax hierarchy. A detailed analysis of the solutions and their singularities in the thermodynamic limit in analogy to the unitary ensemble given in~\cite{HermitianPRE} will be discussed elsewhere.

\begin{mainresult}[Orthogonal Ensemble]\label{res:orthogonal_ensemble}
We provide explicit expressions for the initial condition of the Lax matrix evaluated on the solution to the reduced Pfaff lattice associated with the GOE. This allows us to specify the order parameters as solutions of the integrable differential identity, i.e.\ the Lax equations for the reduced Pfaff lattice.
\end{mainresult}

\noindent 
We prove this result by observing  that the entries of the Lax matrix of the Pfaff lattice can be obtained as coefficients of the recursion relation of the skew-orthogonal polynomials introduced in~\cite{Adler2000_skew}. In part~\eqref{thm:initial_datum} of Theorem~\ref{theo:discrete_chain}, we give these coefficients explicitly using the map from orthogonal to skew-orthogonal polynomials discovered in~\cite{adler2002}. In analogy with the unitary case, we interpret the entries of the Pfaff lattice in terms of expectation values and chemical potential (see~\eqref{chemicpot2}-\eqref{corrvar}).

\begin{mainresult}[Reduced Pfaff lattice: continuum limit] \label{res:reduced_pfaff_continuum_limit}
The reduced Pfaff lattice admits an integrable continuum limit obtained through a singular asymptotic expansion that is consistent with the large $n$ asymptotics of the Lax matrix evaluated on the initial datum. 
\end{mainresult}

\noindent 
We show that a suitable singular limit yields a closed double infinite chain of hydrodynamic type (Theorem~\ref{thm:chain_mixed} and Corollary~\ref{cor:quasilinear_system}). The positive semi-infinite part of the hydrodynamic chain decouples from the negative part. Following~\cite{chain}, in Theorem~\ref{thm:SMEintegrability} we prove that the hydrodynamic chain is integrable in the sense of Ferapontov, namely it admits infinitely many hydrodynamic reductions. In Proposition~\ref{prop:classification_chain} we also prove that the integrability conditions, {referred to as} the Gibbons-Tsarev system, are included in the classification given by Odesskii and Sokolov {in}~\cite{Odesskii_2010,HydroOdesskiSokolov}.

\begin{mainresult}[Reduced Pfaff lattice and Orthogonal Ensemble]\label{res:reduced_pfaff_new_system}
The solution to the Lax equation for the reduced Pfaff lattice with initial condition fixed by the GOE satisfies the following one-dimensional chain, {specified by a sequence $\{W^{k}(\mathbf{t})\}$, with upper index $k \geq -1$},
\begin{gather}
\label{intro:universal_chain}
\begin{aligned}
    &\partial_{t_2} W^{-1} = 2(W^{-1})^2 \, W^1\, \\[1ex]
    &\partial_{t_2} W^k = 2\,W^{-1} \left( (k+1)W^{k+1} - W^1\,W^k - (k-1)W^{k-1} \right), \qquad k > 0 \,. 
\end{aligned}
\end{gather}
Remarkably, the chain~\eqref{intro:universal_chain} characterises the solution of the Lax equation both at finite $n$ and in the thermodynamic limit $n \to \infty$.
\end{mainresult}

\noindent 
This result combines Theorems~\ref{thm:reduction_pfaff} and~\ref{thm:reduction_cont}. To the best of our knowledge, the chain~\eqref{intro:universal_chain} has not been considered in the literature to date.

From the discussion above, there is a clear parallel between the approaches adopted for the unitary and orthogonal ensembles. However, it is important to emphasise that, for the orthogonal ensemble, the Lax equations hold for ensembles of symmetric random matrices of even order only, and all considerations on the partition functions follow from the sequence of $\tau$-functions $\{\tau^{{(1)}}_{2n}(\mathbf{t}) \}_{n \in \mathbb{N}} $ defined as the Pfaffian of the moments matrix of even order. The extension of the method of differential identities to the odd $n$ case remains open. The entries of the Lax matrix for the unitary ensemble define a sequence~$\{u_{n}\}_{n\in \mathbb{N}}$ that is a solution of the Volterra hierarchy and provides natural order parameter, as illustrated in~\cite{HermitianPRE}. The sequence $\{u_n\}_{n\in\mathbb{N}}$ discussed in \cite{HermitianPRE} corresponds to a rescaling of the sequence~$\{B_n\}_{n\in \mathbb{N}}$ introduced in Section~\ref{sec:toda}.
For $n \to \infty$ the order parameter $u_{n}$ changes incrementally in the transition from even to odd $n$. Introducing a meso-scale, say $N$, where $N$ is sufficiently large, the interpolated order parameter $u(x)$, such that $u_{n} = u(n/N)$, exhibits phase transitions within certain regions of the space of couplings, associated with the onset of a dispersive shock. These transitions correspond to increments in $x$ proportional to the scale $\varepsilon = 1/N$, such that $u_{2n} \to u_{2n+1}$. The application of the approach adopted in this work to the sequence~$\{\tau^{{(1)}}_{2n+1} (\mathbf{t})\}_{n\in \mathbb{N}}$ of $\tau$-functions for the Pfaff lattice is expected to be relevant in the study of phase transitions with respect to increments in $n$. Using the unitary ensemble, as studied in~\cite{HermitianPRE,ClaJorKel2016,ClarJordaLour2023,Clarkson25}, as a reference, in certain regions of the couplings~$\mathbf{t}$ the structure of the dispersive shock exhibits increased complexity where abrupt changes in the order parameters occur with respect to increments of suitable multiples of the scale $\varepsilon$, e.g.\ $u_{2n} \to u_{2n+2}$. This suggests that, in the case of the orthogonal ensemble, where we restrict the analysis to the even sequence of Pfaff $\tau$-functions, phase transitions may still be detected by abrupt changes in the order parameters associated with the even sequence $\{\tau^{{(1)}}_{2n}(\mathbf{t}) \}$. A further step in this direction will require, on one hand, the study of the thermodynamic limit (a chain of hydrodynamic type, whose solutions are expected to develop, generically, gradient catastrophe singularities at ``finite time"), and, on the other hand, a direct analysis of the chain~\eqref{intro:universal_chain}.

This work is organised as follows. In Section \ref{sec:HME}, after reviewing the unitary ensemble and its relationship with the Toda lattice, we focus on the case where the Hamiltonian \eqref{intro:ham} includes even interactions only, and the partition function $\tau^{{(2)}}_n(\mathbf{t})$ corresponds to a $\tau$-function of the Volterra lattice. We show that the Volterra lattice variables, constituting the relevant statistical mechanical quantities for the ensemble, provide a solution to the modified Kadomtsev-Petviashvili equation (Main Result~\ref{res:unitary_ensemble}). In Section \ref{sec:SME_gen} we turn our attention to the orthogonal ensemble and its relationship with the Pfaff lattice. As for the unitary ensemble, we focus on the case of even interactions only, where the partition function of the ensemble corresponds to a $\tau$-function of the reduced even Pfaff hierarchy. We provide explicit expressions for the even Pfaff hierarchy in terms of a double infinite chain for the entries of the Lax matrix (Main Result \ref{res:pfaff_lattice_proof}). We identify the solution of the lattice for orthogonal ensemble by deriving explicit initial conditions for the Lax matrix entries and interpret them in terms of relevant statistical observables of the matrix ensemble  (Main Result \ref{res:orthogonal_ensemble}).  In Section 4, we investigate the thermodynamic limit of the orthogonal ensemble as a suitable continuum limit of the reduced even Pfaff lattice, exploiting the asymptotic properties of the initial datum. We find a novel hydrodynamic chain and prove its integrability in the sense of Ferapontov, { a definition of integrability based on the existence of hydrodynamic reductions, introduced in \cite{chain} as recalled in Definition \ref{def:int_chain} below (see also Main Result \ref{res:reduced_pfaff_continuum_limit})}. Furthermore, we classify the hydrodynamic chain so obtained according to the framework provided in \cite{Odesskii_2010}. In Section \ref{sec:thermolimit} we show that both the even Pfaff lattice and its continuum limit admit a reduction that is compatible with the initial condition. In both cases, the reduction consists of the same system of differential equations \eqref{intro:universal_chain} (Main Result \ref{res:reduced_pfaff_new_system}).

\section{Unitary ensembles and integrable differential identities}\label{sec:HME}

We start by reviewing the connection between the unitary ensemble and Toda lattice with a particular attention to its reduction to the Volterra lattice. We focus on the sequence $\{\tau_{n}(\mathbf{t}) \}_{n\in \mathbb{N}}$, which are simultaneously $\tau$-functions of the Toda and the KP hierarchy~\cite{vanM_Toda_orthogonal,Adler_KP}, and establish the analogous result for the sequence of $\tau$-functions of the Volterra chain and the modified KP equation, which constitutes our Main Result \ref{res:unitary_ensemble}. Furthermore, we conclude the section by discussing the thermodynamic limit for both Toda (see e.g.~\cite{deift_Toda}) and Volterra lattices~\cite{HermitianPRE}. {As in this section we focus on the unitary ensemble so that $\beta = 2$, for the sake of simplicity we drop the upper index $\beta$, by setting $Z_{n} :=Z_{n}^{(2)}$, $\tau_{n} :=\tau_{n}^{(2)}$  and similarly for all the quantities specified for $\beta =2$.}

\subsection{Toda {and} Volterra hierarchies{,} and ensemble observables}
\label{sec:toda}
The unitary ensemble is specified by the probability density distribution
\begin{equation}
\label{probdist}
    P(M;\mathbf{t}) \; dM = \frac{\text{e}^{-H(M;\mathbf{t})}}{Z_{n}(\mathbf{t})} \, dM,
\end{equation}
with partition function and Hamiltonian given respectively by
\begin{equation}\label{eq:parti_hami_HME}
Z_{n}({\bf t}) = \int_{{\cal H}_{n}} \text{e}^{-H(M;\mathbf{t})} \; dM\,, \qquad   H(M;\mathbf{t}) = \Tr \left( \frac{M^{2}}{2} - \sum_{k=1}^{\infty} t_{k}\, M^{k} \right )
\,.
\end{equation}
The partition function $Z_n(\mathbf{t})$ is defined by integration with respect to the Haar measure on the space of Hermitian matrices of order $n$ ($\mathcal{H}_n$), and depends explicitly on the coupling constants of the theory, i.e.\ ${\mathbf{t}} = (t_{1},t_{2},\dots)$. The case~$\bf{t}\!=\!0$ corresponds to the GUE, where the entries of the random matrix are independent and identically distributed \cite{Meh2004}.
Based on a classical result by Weyl~\cite{Weyl}, the partition function can be expressed via the following integral over the eigenvalues (see \eqref{intro:tau_n} with~$\beta=2$):
\begin{equation}
\label{taun}
\tau_{n}({\bf t})  =  \frac{1}{n!}\int_{\mathbb{R}^n}\Delta_{n}^{2}(z) \, \prod_{i=1}^n \, \rho(z_i\,; \mathbf{t})\,dz_i \,,
\end{equation}
where
\begin{equation}
\label{rho}
\rho(z; \mathbf{t}) = \exp \Big( -\frac{z^2}{2} + \sum_{k=1}^{\infty} t_{k}\, z^{k} \Big)\,,
\end{equation}
is the weight function and $\Delta_n(z)$ is the Vandermonde determinant 
\begin{equation} \label{eq:vandermonde} 
\Delta_{n}(z) =  \prod_{1 \le k < \ell \le n} \!\! (z_{k}-z_{\ell})\,.  
\end{equation} 
In the following, with a slight abuse of terminology, we shall refer to $\tau_{n}(\mathbf{t})$ as the partition function, as it coincides with the function $Z_{n}({\bf t})$ up to an $n$-dependent multiplicative constant which does not play any role in the calculation of expectation values.
{Among} the various approaches for the study of the partition function and associated observables, we will consider the one where $\tau_{n}({\bf t})$ is viewed as a particular solution of a system of integrable nonlinear differential equations. As we shall see, this approach allows to infer some important features of the partition function and related observables, e.g.\ their scaling properties, directly from the properties of the equations they satisfy. \\
\noindent We first recall the following result by Adler and van Moerbeke~\cite{vanM_Toda_orthogonal,Adler_KP,vanMoerbekenotes} which
establishes a link between the sequence of partition functions $\{\tau_{n}({\bf t}) \}_{n\geq 0}$ and a particular solution of the open Toda lattice hierarchy.

\begin{theorem}[\cite{vanMoerbekenotes}, Theorem 3.1]
The partition function $\tau_{n}(\mathbf{t})$ of the unitary ensemble is a $\tau$-function of a particular solution to the semi-infinite Toda lattice system. Therefore, the semi-infinite tri-diagonal symmetric matrix of the form
\begin{equation}
L(\mathbf{t}) = \left(
\begin{matrix}
a_{1} &  b_{1} & 0  & 0 & \dots \\
b_{1} & a_{2} & b_{2}  & 0 & \dots \\
0 & b_{2} & a_{3}  & b_{3} & \dots \\
\vdots & \ddots & \ddots & \ddots & \ddots
\end{matrix}
\right),
\end{equation}
where 
\begin{equation}
\label{anbndef}
a_{n}(\mathbf{t}) = \partial_{t_1}\log \frac{\tau_{n+1}(\mathbf{t})}{\tau_n(\mathbf{t})} \qquad b_{n}(\mathbf{t}) =  \frac{\sqrt{\tau_{n+1}(\mathbf{t})\,\tau_{n-1}(\mathbf{t})}}{\tau_n(\mathbf{t})} \qquad n  \ge 1\,,
\end{equation}
is a Lax matrix for the Toda lattice satisfying the compatible system of equations
\begin{equation}
\label{LaxToda}
\der{L}{t_{k}} = \frac{1}{2}\left [\, (L^{k} )_{\mathfrak{s}}\,,L \,\right],
\end{equation}
where $(L^{k})_{\mathfrak{s}}$ denotes the skew-symmetric projection\footnote{Given a semi-infinite matrix $A\in \mathfrak{gl}(\infty)$, $(A)_{\mathfrak{s}} = A_+ - (A_+)^{\top}$, where $A_+$ denotes the upper triangular projection of the matrix~$A$. The skew-symmetric projection is part of the decomposition of the algebra of semi-infinite matrices $\mathfrak{gl}(\infty)=\mathfrak{s}+\mathfrak{b}$, where $\mathfrak{b}$ represents a projection onto lower triangular matrices defined as $(A)_{\mathfrak{b}}=A_-+(A_+)^{\top}$, and $A_-$ is the projection on the lower triangular part.\label{foot:splitting_hermitian} \label{foot:toda}} of the $k$-th power of the matrix $L(\mathbf{t})$. Moreover, the tri-diagonal matrix $L$ admits the eigenvector $p(z;\mathbf{t}) =\big(p_{0}(z;\mathbf{t}),p_{1}(z;\mathbf{t}),\dots,p_{n}(z;\mathbf{t}),\dots\big)^{\!\top} $ which satisfies the equation
\begin{equation}\label{eq:orthonormal_poly}
L(\mathbf{t}) \,p(z;\mathbf{t}) = z \,p(z;\mathbf{t}),
\end{equation}
where the entries $p_{n}(z;\mathbf{t})$ are $n$-th degree polynomials, orthogonal with respect to the weight $\rho (z;\mathbf{t})$ as in~\eqref{rho}.

\end{theorem}
For example, the first equation of the hierarchy 
\begin{subequations}
\label{Todat1}
   \begin{align}
       &\partial_{t_1} a_{n} = b_{n}^{2} - b_{n-1}^{2} \\
       &\partial_{t_1}b_{n} = \frac{b_{n}}{2}  \left(a_{n+1} - a_{n} \right),
   \end{align} 
\end{subequations}
describes changes of the sequence of partition functions $\tau_{n}(\mathbf{t})$, via the Flaschka variables $a_{n}(\mathbf{t})$ and~$b_{n}(\mathbf{t})$ defined in~\eqref{anbndef}, with respect to  the coupling constant $t_{1}$.
Similarly, the second equation of the hierarchy reads as follows
\begin{subequations}
\begin{align}
&\partial_{t_2}{a_{n}} = \left(a_{n} + a_{n+1} \right) b_{n}^{2} - \left(a_{n-1} + a_{n} \right) b_{n-1}^{2}  \\
& \partial_{t_2}{b_{n}} = \frac{b_{n}}{2} \left(b_{n+1}^{2} - b_{n-1}^{2} + a_{n+1}^{2} - a_{n}^{2} \right) .
\end{align}
\end{subequations}
From the result above, it follows that the sequence of order parameters $a_{n}(\mathbf{t})$ and $b_{n}(\mathbf{t})$, associated with ensembles of Hermitian random matrices of order $n$ with Hamiltonian~\eqref{eq:parti_hami_HME}, is a solution of the Toda lattice hierarchy.

Moreover,  $a_{n}(\mathbf{t})$ and $b_{n}(\mathbf{t})$ correspond to specific statistical mechanical observables for the unitary ensemble. In fact, given the free energy
\begin{equation}
\label{PHI}
\Phi_{n}(\mathbf{t}) = \log \tau_{n}(\mathbf{t}),
\end{equation}
the chemical potential and its variation are defined respectively as
\begin{equation}
\label{ChemPot}
\mu_{n}(\mathbf{t}) = \Phi_{n+1}(\mathbf{t}) - \Phi_{n}(\mathbf{t}), \qquad \Delta \mu_{n}(\mathbf{t}) = \mu_{n+1}(\mathbf{t}) - \mu_{n}(\mathbf{t}).
\end{equation}
Now, recalling that the expectation value $\mathbb{E}_{n}(\varphi)$ of an observable $\varphi$ is defined as
\begin{equation}\label{eq:expectation_value}
\mathbb{E}_{n}(\varphi) = \frac{\displaystyle\int_{\mathbb{R}^{n}} \varphi \; \Delta^{2}_n(z) \, \prod_{i=1}^{n} \rho(z_i; \mathbf{t})\,dz_{i}}{\displaystyle\int_{\mathbb{R}^{n}} \Delta^{2}_n(z)\, \prod_{i=1}^{n} \rho(z_i; \mathbf{t})\, dz_{i}}\,,
\end{equation}
the definitions~\eqref{taun} and~\eqref{anbndef} imply that
\begin{subequations}\label{eq:expect_Toda}
\begin{align}
&a_{n}(\mathbf{t})=  \partial_{t_1} \mu_n(\mathbf{t}) = \mathbb{E}_{n+1}\!\left(\sum_{i=1}^{n+1} z_{i} \right) - \mathbb{E}_{n}\!\left(\sum_{i=1}^{n} z_{i} \right)   \\[.5ex]
&\log b_{n}(\mathbf{t}) = {\frac{1}{2}} \,\Delta \mu_{n-1}(\mathbf{t})\,.
\end{align}
\end{subequations} 
The solution of interest is selected by the initial value 
\begin{equation}
\label{abinit}
a_{n}({\bf 0}) =  \partial_{t_1} \log \frac{\tau_{n+1}(\bf t)}{\tau_{n}({\bf t})} \Big|_{\mathbf{t}=\mathbf{0}} =0,\,  \qquad  b_{n}({\bf 0}) = \sqrt{\frac{\tau_{n+1}({\bf 0})\,\tau_{n-1}({\bf 0})}{\tau_n^2({\bf 0)}}} = \sqrt{n} \,, 
\end{equation}
which follows from the calculation of the Selberg type integral~\cite{Meh2004}
\[
\tau_{n}({\bf 0}) = \frac{1}{{n!}} \int_{\mathbb{R}^{n}} \Delta_{n}^{2}(z)\, \prod_{i=1}^{n} \rho(z;\mathbf{0})\, d z_{i} = \frac{(2\pi)^{n/2}}{n!} \prod_{j=1}^{n} \Gamma (1+ j)\,,
\]
with $\rho(z;\mathbf{0})$ defined in~\eqref{rho}, and the observation that
\[
\partial_{t_1}{\tau_{n}(\mathbf{t})}\Big|_{\mathbf{t}=\mathbf{0}} = \frac{1}{n!} \int_{\mathbb{R}^{n}} \! \Big( \sum_{i=1}^{n} z_{i} \Big) \,\Delta_{n}^{2}(z)\, \prod_{j=1}^{n} \rho(z;\mathbf{0}) \, dz_{i} = 0\,,
\]
which follows from the skew-symmetry of the argument of the integral under reflections $z \to - z$. 
As studied in~\cite{HermitianPRE,BONORA1992453,Jurkiewicz:1991sj,Senechal}, if the Hamiltonian $H(M;\mathbf{t})$ is of the form
\begin{equation}
\label{Ham_even}
H(M;\mathbf{t})\Big|_{\underset{\forall i\in \mathbb{N}}{t_{2i-1} = 0}}  = \Tr \Bigg(\frac{M^{2}}{2} - \sum_{k=1}^{\infty} t_{2k}\, M^{2k} \Bigg),
\end{equation}
i.e.\ it contains even power interaction terms only, the condition $a_{n}({\bf t}) = 0$ holds for all $\mathbf{t}$ and the variables $b_{n}(\mathbf{t})$ are specified by the even flows, i.e.\ the flows associated with the ``times" $t_{2k}$, of the Toda hierarchy. In particular, the first non-trivial flow associated with the coupling constant $t_{2}$ of the hierarchy gives the celebrated Volterra lattice equation~\cite{Volterra,vanMoerbeke1975}\begin{equation}
\label{Volterraeq}
\partial_{t_2}{B_{n}} = B_{n} (B_{n+1} - B_{n-1})\,,
\end{equation}
with $B_{n}(\mathbf{t}) = b_{n}(\mathbf{t})^{2}$. As observed in~\cite{HermitianPRE}, the whole hierarchy can be written in the form
\begin{equation}
\label{Volterra_hierarchy}
\partial_{t_{2k}}{B_{n}} = B_{n} \left(V_{n+1}^{(2 k)} - V_{n-1}^{(2 k)} \right),
\end{equation}
where $V_{n}^{(2 k)}(\mathbf{t})$ is a function of $B_{n}(\mathbf{t})$ and its first $(k-1)$-th nearest neighbours. For example, for the first three flows we have
\begin{align*}
V^{(2)}_n &= \,B_n  \\
V^{(4)}_n &= V^{(2)}_n \left (V^{(2)}_{n-1}+V^{(2)}_n+V^{(2)}_{n+1} \right) \\
V^{(6)}_n &= \,V^{(2)}_n \left (V^{(2)}_{n-1}V^{(2)}_{n+1} + V^{(4)}_{n-1} + V^{(4)}_n + V^{(4)}_{n+1} \right).
\end{align*}
The partition function of the ensemble of order $n$ is interpreted as a $\tau$-function of the semi-infinite Toda chain evaluated at the $n$-th site of the lattice. However, it is natural to ask how the properties of an ensemble of a given order $n$ change depending on the couplings ${\bf t}$. The answer is provided by the following important result~\cite{vanM_Toda_orthogonal, Adler_KP, vanMoerbekenotes} (the theorem below reports a portion of a more extensive statement that is of specific interest for our discussion below):

\begin{theorem}[\cite{vanMoerbekenotes}, Theorem 3.1]
\label{AvM2:theo}
    The function $\tau_{n}({\bf t})$, given by the expression~\eqref{taun}, is a $\tau$-function of the KP hierarchy, i.e.\ it satisfies the equations
\begin{equation}
\label{tauHirota}
    \left({s}_{k+4} (\tilde{\partial} ) - \frac{1}{2}\, \partial_{t_1} \partial_{t_{k+3}} \right) \tau_{n} \circ \tau_{n} = 0\,, \qquad k\ge 0\,,
\end{equation}
where $\tilde{\partial} := \left (\partial_{t_1},\frac{1}{2} \partial_{t_2},\frac{1}{3} \partial_{t_3},\dots \right)$ denotes the Hirota derivatives 
\[
\partial_{t_{\ell}}^{k}\,\left( f \circ g\right):=  (\partial_{\hat{t}_{\ell}} - \partial_{t_{\ell}})^{k} \,f\!\left(\hat{t}_{\ell}\right) g\!\left(t_{\ell}\right) \Big|_{\hat{t}_{\ell}=t_{\ell}}
\]
and ${s}_{\ell}(\tilde \partial)$ are the elementary Schur polynomials\footnote{\label{foot:schur} Elementary Schur polynomials ${s}_{\ell}({\bf x})$, with ${\bf x} = (x_{1},x_{2},\dots)^{\top}$ are defined via $ \sum\limits_{\ell=0}^{\infty} s_{\ell}({\bf x})\, z^{\ell}=\exp \sum\limits_{i=1}^{\infty}x_{i}\, z^{i}$.}.

\end{theorem}
It is important to note that equations~\eqref{tauHirota} contain a single $\tau_{n}$, for a given $n$, from the sequence~$\{\tau_{n}\}_{n \in \mathbb{N}}$, unlike the Volterra equations~\eqref{Volterraeq} and the Toda  equations~\eqref{LaxToda} which combine multiple elements from the sequence of $\tau$-functions. The first non-trivial equation of the hierarchy involves the first three parameters $t_{1}$, $t_{2}$, $t_{3}$ and reads as follows
\begin{equation}
\label{KPHirota}
\left(  \partial_{t_1}^{4} + 3\, \partial_{t_2}^{2} -  4\, \partial_{t_1} \partial_{t_3}\right) \tau_n \circ \tau_n = 0.
\end{equation}
Introducing the function
\[
u_{n}({\bf t}) = 2\,\partial_{t_1}^2 \log \tau_{n}({\bf t}),
\]
the equation~\eqref{KPHirota} implies that $u_n({\bf t})$ satisfies the KP (specifically KPII) equation
\[
\partial_{t_{1}}\!\left(  \partial_{t_{1}}^{3}\! u_{n} + 6 u_n \; \partial_{t_{1}}\! u_{n} - 4 \, \partial_{t_{3}}\! u_{n} \right ) + 3\, \partial_{t_{2}}^{2}u_{n}= 0.
\]
In the following, we show that a similar result holds for the Volterra hierarchy. We note that this result cannot be inferred directly from the above Theorem~\ref{AvM2:theo}, since the KP hierarchy~\eqref{KPHirota} contains derivatives with respect to the variable $t_{1}$ which is absent in the Volterra hierarchy.\\
Hence, we prove the following
\begin{lemma}
Let $\{B_{n}(\mathbf{t}) \}_{n \in \mathbb{N}}$ be a solution of the Volterra lattice~\eqref{Volterraeq}. For a given $n \in \mathbb{N}$, let us denote $\phi(\mathbf{t}) = B_{n}(\mathbf{t})$ and 
$\psi(\mathbf{t}) = B_{n-1}(\mathbf{t})$. Then $\phi(\mathbf{t})$ and $\psi(\mathbf{t})$ fulfil the following compatible systems of $(1+1)$-dimensional conservation laws
\begin{subequations}\label{eq:phipsi_both}
\begin{align}
\label{phipsi_system1}
&\begin{cases}
\phi_{y} =\left(\phi^{2} + 2 \phi \psi + \phi_{x} \right)_{x} \\[.5ex]
\psi_{y} =\left(\psi^{2} + 2 \phi \psi - \psi_{x} \right)_{x} 
\end{cases} \\[1ex]
\label{phipsi_system2}
&\begin{cases}
\phi_{t} = \left(\phi^{3} + 3 (\psi + 2 \phi) \phi \psi + 3 (\phi + \psi) \phi_{x} + \phi_{xx} \right)_{x} \\[.5ex]
\psi_{t} = \left(\psi^{3} + 3 (\phi + 2 \psi) \phi \psi - 3 (\phi + \psi) \psi_{x} + \psi_{xx} \right)_{x} 
\end{cases}
\end{align}
\end{subequations} 
with the notation $x = t_{2}$, $y = t_ 4$ and $t = t_6$. 
\end{lemma}
\noindent We note that the equations~\eqref{eq:phipsi_both} are invariant under the transformation 
\begin{align*}
(\phi,\psi) \to (-\psi,-\phi), \quad 
(x,y,t) \to (x,-y,t).
\end{align*}

\begin{proofC}
Let us first observe that writing Volterra's equations~\eqref{Volterraeq} for $\partial_{t_{2}} {B_{n-2}} $, $\partial_{t_2} {B_{n-1}}$, $\partial_{t_2} {B_{n}}$ and $\partial_{t_2} {B_{n+1}}$, one can express the variables $B_{n-3}, B_{n-2},B_{n+1}, B_{n+2}$ in terms of the variables $B_{n-1}, B_{n}$ and their derivatives with respect to $t_{2}$. Substituting these expressions into the equation~\eqref{Volterra_hierarchy} for the flows $\partial_{t_4} B_{n-1}$ and $\partial_{t_4} B_{n}$, we obtain the system~\eqref{phipsi_system1}.
 Substituting the above mentioned expressions into the equation~\eqref{Volterra_hierarchy} and proceeding in a similar way for the flows $\partial_{t_{6}}B_{n-1}$ and $\partial_{t_6} B_{n}$, we obtain the system~\eqref{phipsi_system2}.
\end{proofC}

The above result allows us to prove that the variable $B_{n}(\mathbf{t})$ for a given $n$ satisfies, as a function of~$t_{2}, t_{4}, t_{6}$, the mKP equation. In particular, we have the following

\begin{theorem}\label{thm:mKP_volterra}
Let $\{B_{n}(\mathbf{t}) \}_{n \in \mathbb{N}}$ be a solution of the Volterra lattice~\eqref{Volterraeq}. For a given $ n \in \mathbb{N}$ let us denote~$\phi(\mathbf{t}) = B_{n}(\mathbf{t})$. Then, $\phi$ fulfils the modified Kadomtsev-Petviashvili (mKP) equation
\begin{equation}
\label{eqmKPsys}
\begin{cases}
4 \phi_t = 6\phi_x (\xi - \phi^2)+ \phi_{xxx} + 3 \xi_y \\[.5ex]
\phi_y = \xi_x
\end{cases} 
\end{equation}
with the notation $x = t_{2}$, $y = t_ 4$ and $t = t_6$.
\end{theorem}
\begin{proofC}
Let us write the equations for $\phi$ in~\eqref{phipsi_system1} and~\eqref{phipsi_system2} in potential form as follows
\begin{subequations}
\begin{align}
\label{dphiy}
\partial_{y} (\partial^{-1}_x\phi) &=\phi^{2} + 2 \phi \psi + \phi_{x}  \\
\label{dphit}
\partial_{t} (\partial^{-1}_x \phi) &= \phi^{3} + 3 (\psi + 2 \phi) \phi \psi + 3 (\phi + \psi) \phi_{x} + \phi_{xx},
\end{align}
\end{subequations} 
and evaluate the quantity
\[
3 \partial_{y}^{2} (\partial^{-1}_x \phi) - 4 \partial_{x} \partial_{t} (\partial^{-1}_x \phi) = 3 \partial_{y}^{2} (\partial^{-1}_x \phi) - 4 \partial_{t} \phi
\]
by differentiating both sides of the equations~\eqref{dphiy} and \eqref{dphit} with respect to $y$ and $x$ respectively. Using the equations~\eqref{phipsi_system1} and~\eqref{phipsi_system2}, we obtain 
\begin{equation}
\label{mKPpot}
3 \partial_{y}^{2} (\partial^{-1}_x \phi) - 4 \partial_{t} \phi = - 6 \partial_{y} (\partial^{-1}_x \phi) \phi_{x} + 6 \phi^{2} \phi_{x} - \phi_{xxx}.
\end{equation}
Introducing the variable $\xi = \partial_{y} (\partial^{-1}_x \phi)$, the equation~\eqref{mKPpot} is identified with the mKP equation~\eqref{eqmKPsys}. Hence, the statement is proved.
\end{proofC}\\

\subsection{Thermodynamic limit}
\label{sec:therm_hermitian}
Let us consider the unitary ensemble specified by the partition function~\eqref{taun}. Rescaling the variables~$z_{i}$ and $t_{k}$ as follows 
\[
z_{i} \to \sqrt{N}\; z_{i}\,, \qquad t_{k} \to N^{1-k/2}\,t_{k} \,,
\]
where $N$ is a (large) scale parameter,  the partition function~\eqref{taun} reads as 
\begin{equation}
\label{taunbis}
\tau_{n}(\mathbf{t}) = {\frac{N^{n^{2}/2}}{\,n!}} \int_{\mathbb{R}^{n}} \Delta_{n}^2(z)\, \prod_{i=1}^{n} \rho(z_i\,; \mathbf{t};N)\, d z_{i},  
\end{equation}
with 
\begin{equation}    
    \rho(z\,; \mathbf{t};N) = \exp \bigg[ {N \bigg(\!-\frac{z^2}{2} + \sum_{k=1}^{\infty} t_{k}\, z^{k} \bigg) \bigg] }\,.
\end{equation}
We are interested in the behaviour of the unitary ensemble in the thermodynamic limit such that
\[
\frac{n}{N} = \mathcal{O}(1) \;\; \text{as }\;\; n,N\to \infty.
\]
In particular, we see that the thermodynamic limit corresponds to the continuous limit of the Toda lattice. The standard formal limit is performed by introducing the re-scaled interpolating functions
\begin{equation}
v(x\,,\mathbf{t}) = \sqrt{\varepsilon} \; a_{n}(\mathbf{t})\,,  \qquad u(x\,,\mathbf{t}) =  \sqrt{\varepsilon} \; b_{n}(\mathbf{t})\,, \qquad \textup{where} \quad x = \varepsilon n \quad \textup{and} \quad \varepsilon =\frac{1}{N}\,.
\end{equation}
It is important to note that the above scaling is consistent with the one of the initial conditions, for which we have
\begin{equation}
\label{uvinit}
v(x,\mathbf{0}) = 0, \qquad u(x,\mathbf{0}) = \sqrt{x}.
\end{equation}
Writing the Toda system~\eqref{Todat1} in terms of $u(x,\mathbf{t})$, $v(x,\mathbf{t})$, $u(x\pm \varepsilon,\mathbf{t})$ and $v(x\pm \varepsilon,\mathbf{t})$ and formally expanding in Taylor series as $\varepsilon \to 0$,
we obtain the equations
\begin{subequations}
\label{Todacont}
    \begin{align}
        \partial_{t_1}\!v &= \sum_{n=1}^{\infty} (-1)^{n+1} \,\varepsilon^{n-1} \sum_{k=0}^{n} \frac{\partial_x^\ell u}{\ell!} \, \frac{ \partial_x^{(n-\ell)} u}{ (n-\ell)!}  \,\\
        \partial_{t_1}\!u &= \frac{1}{2}\, u \sum_{n=1}^{\infty} \varepsilon^{n-1} \, \frac{\partial_x^n v}{n!} \,,
    \end{align}
\end{subequations}
which, up to $O(\varepsilon^3)$, read as
\begin{subequations}
\label{Todacontbis}
    \begin{align}
        \partial_{t_1}\!v &= 2\, u \,\partial_x u - \varepsilon \left(u \,\partial_x^2u + (\partial_x u)^{2} \right) +\varepsilon^{2} \left(\frac{u \,\partial_x^3u}{3} + \partial_xu \, \partial_x^2u \right) + \mathcal{O}(\varepsilon^{3})\\[1ex]
        \partial_{t_1}\!u &=\frac{1}{2}\, u\, \partial_x v + \frac{\varepsilon}{2\cdot 2!}\, u\, \partial_x^2 v + \frac{\varepsilon^{2}}{2 \cdot 3!}\, u\, \partial_x^3 v + \mathcal{O}(\varepsilon^{3}).
    \end{align}
\end{subequations}
Setting $\varepsilon\!=\!0$ gives the continuum Toda lattice equations, also known as dispersionless Toda lattice~(dTL) system
\begin{subequations}
\label{dToda}
    \begin{align}
        \partial_{t_1}\!v &= 2 \,u \,\partial_x u\,\\
        \partial_{t_1}\!u &=\frac{1}{2}\, u \,\partial_x v\,.
    \end{align}
\end{subequations}
The above dTL system is a completely integrable hyperbolic system of hydrodynamic type~\cite{Dubrovin_Novikov}, solvable by the method of characteristics. In fact, proceeding in a similar fashion for the higher flows of the Toda hierarchy, one obtains the higher flows of the dTL hierarchy.

\noindent Introducing {the Riemann invariants} 
\[
r^{1} = \frac{v}{2} +  u\,, \qquad r^{2} =  \frac{v}{2} - u \,,
\]
{ (standardly denoted with upper indices)}, the dTL system takes the following diagonal form
\begin{equation}
\label{dtodariemann}
\partial_{t_1} r^{i} = \lambda^{i} \,\partial_{x} r^{i}\,, \qquad i \in \{1,2\}\,,
\end{equation}
where $\lambda^{1}=- \lambda^{2}= (r^{1} - r^{2})/2$ are the {associated} characteristic speeds.
The general solution is given by the hodograph formula
\begin{equation}
\label{dTodahod}
x + \lambda^{i}\, t_1  = \partial_i \Omega\,, \qquad i=1,2,
\end{equation}
where $\partial_i := \partial/\partial r^i$ and $\Omega(r^{1},r^{2})$ satisfies the Euler-Poisson-Darboux equation
\[
\partial_{1}\partial_2\Omega + \frac{1}{2 (r^{1} - r^{2})} \left(\partial_1 \Omega - \partial_2 \Omega \right) =0\,.
\]
Hydrodynamic systems such as the dTL, and the hodograph formulae of the form~\eqref{dTodahod}, have been studied in greater generality~\cite{Tsarev_1990,Dubrovin_Novikov}. A typical phenomenon associated with this class of systems is that {generic} solutions develop gradient catastrophe singularities in finite time. 

We notice that the random matrix ensemble described by the partition function~\eqref{taunbis} is a multidimensional system where the pair of order parameters~$(a_{n}(\mathbf{t}),b_{n}(\mathbf{t}))$ (risp. $\big(v(x; \mathbf{t}),u(x; \mathbf{t})\big)$ in the continuum limit) solve the Toda (resp.\ dTL) hierarchy. The state of the system is, at least in principle, reconstructed by evolving the initial condition~\eqref{abinit} (resp.\!~\eqref{uvinit}), associated with the GUE, with respect to all flows of the Toda (resp.\ dTL) hierarchy. In practice, one shall specify a model by fixing a finite number of interacting terms so that the integral~\eqref{taunbis} is convergent, i.e.\ truncate the exponent in~\eqref{taunbis} to a given even coupling $t_{2k}$ for a certain~$k$.

The dTL equation~\eqref{dToda} provides the leading order approximation away from the singularity. In the vicinity of the gradient catastrophe, higher order terms in the expansion~\eqref{Todacontbis} are no longer negligible and a renormalisation procedure is required in order to evaluate the leading order contribution to the solution. This problem has been addressed in~\cite{Dubrovin_2006,DubrovinGravaKlein,Moro_Grava_Dubrovin} where it was shown that the critical asymptotic behaviour near the point of gradient catastrophe for a generic Hamiltonian perturbation of a hyperbolic system of hydrodynamic type is universal and it is given by a particular solution of a Painlev\'e equation. This result was proven for the KdV~\cite{GravaClaeys} and NLS~\cite{BertolaTovbis} equations and in~\cite{Moro_Grava_Dubrovin} conjectured to be true for any two-component Hamiltonian system of hydrodynamic type (subject to suitable genericity conditions). In~\cite{Moro_Grava_Dubrovin} it was pointed out that the conjecture also applies to the continuum limit of the Toda lattice system. 
The thermodynamic properties of the random matrix ensemble~\eqref{taunbis} with even powers in relation to the continuum limit of the Volterra hierarchy have been studied in~\cite{HermitianPRE}. For this reduction, only the order parameter~$b_{n}(\mathbf{t})$ survives, yielding at the leading order the Burgers-Hopf hierarchy
\[
\partial_{t_{2k}}\!u = c_{k}\, u^{k} \, \partial_x u\,.
\]
{where $c_k$ {are some constants  which can be removed by a suitable rescaling of $t_{2k}$}}. Solutions of the Burgers-Hopf hierarchy also develop gradient catastrophe in finite time. In the vicinity of the gradient catastrophe, the order parameter $u(x,\mathbf{t})$ exhibits universal behaviour described by a particular solution of the the Painlev\'e $\text{P}_{\text{I}}^{(2)}\!$ equation followed by the onset of a dispersive shock~\cite{JURKIEWICZ1990178,Jurkiewicz:1991sj,Senechal,BONORA1992453,HermitianPRE}. We emphasise that there is a significant difference between the problem formulated for weights with even powers only and the analogous problem where the weights contain also odd powers (see e.g.~\cite{Ercolani2008,Ercolani_Pierce_2012}).

\section{Orthogonal ensemble and integrable differential identities}
\label{sec:SME_gen}
In this section we analyse the orthogonal ensemble and its relationship with the Pfaff lattice. In~\cite{vanM_Pfaff_skew,adler2002, vanMoerbekenotes} the authors derived a formula in terms of Pfaffians for the $\tau$-functions~$\{\tau_{2n}^{(1)}(\mathbf{t})\}_{n \in \mathbb{N}}$ of ensembles of symmetric matrices of even order. We recall some of the results mentioned above~\cite{vanM_Pfaff_skew,adler2002, vanMoerbekenotes,benassi2021symmetric} and further build upon our previous findings from~\cite{benassi2021symmetric}, concerned with a reduction of the Pfaff lattice hierarchy restricted to the even flows {(even Pfaff lattice)}. In this case, the Pfaff Lax matrix is sparse and we use the method of integrable differential identities to characterise the order parameters. Furthermore, we prove that the Lax equations can be written equivalently as a double infinite semi-discrete dynamical chain, that is Main Result~\ref{res:pfaff_lattice_proof}. Moreover, Main Result~\ref{res:orthogonal_ensemble} follows from the relationship between the Pfaff lattice and skew-orthogonal polynomials~\cite{vanM_Pfaff_skew}, which can be mapped onto orthogonal polynomials (related to the Toda lattice) via the transformation found in~\cite{Adler2000_skew,adler2002}. A key step of the method requires the explicit calculation of the Pfaff Lax matrix at $\mathbf{t}=\mathbf{0}$.
 {For the sake of simplicity, in the following we drop the upper index $\beta =1$ and set $Z_{n} :=Z_{n}^{(1)}$, $\tau_{n} :=\tau_{n}^{(1)}$  and similarly for all the quantities specified by $\beta$.} 

\subsection{Pfaff and even Pfaff hierarchies, and ensemble observables }
\label{sec:SME_Pfaff}

In this section we study the differential identities satisfied by the partition function of the orthogonal ensemble.
The underlying integrable structure for the orthogonal ensemble is based on a splitting of the Lie algebra $\mathfrak{gl}(\infty)$ considered in~\cite{Ad,Kos,Sy,STS,Algebraic_Integrability_Lie_Algebras}\footnote{
The splitting acting on a semi-infinite matrix $A \in \mathfrak{gl}(\infty)$, with $\mathfrak{gl}(\infty) = \mathfrak{t} +{\mathfrak{n}}$, is such that the projections $(A)_\mathfrak{t}$ and $(A)_\mathfrak{n}$ are expressed in terms of $A_0$, i.e.\ the projection  of $A$ on its   $2\times 2$ block diagonal part, the upper triangular part $A_+$ and the lower triangular part $A_-$ with respect to $2 \times 2$ block diagonal component. The form of the projection $(A)_\mathfrak{t}$ is given in~\eqref{eq:t_projection}.\label{foot:pfaff}}. 

Let us start by recalling the connection between the orthogonal ensemble and the Pfaff lattice hierarchy as introduced in~\cite{vanM_Pfaff_skew} and studied in~\cite{Adler2000_skew,adler2002,KodamaPierce}.
The probability distribution is defined on the non-compact symmetric space $SL(n,\mathbb{R})/SO(n)$, formally in the same way as for the unitary ensemble, by the expression~\eqref{intro:probdensity} with partition function 
\begin{equation}
\label{eq:HSME}
Z_n(\mathbf{t}) = \int_{\mathcal{S}_n} {\rm e}^{-H(M; \mathbf{t})}dM\,, \qquad H(M; \mathbf{t}) = {\rm Tr}\Bigg( \frac{M^2}{2} - \sum_{k\geq 1} t_{k} \, M^{k} \Bigg),
\end{equation}
where the integral is performed over the space of real symmetric matrices of order $n$, i.e.\ $\mathcal{S}_n$.  
Similarly to the unitary ensemble, the partition function can be reduced to an integral over $\mathbb{R}^{n}$ in the eigenvalues of the form 
 \begin{equation}
     Z_n(\mathbf{t}) = \frac{C_n}{n!}\int_{\mathbb{R}^n}\left|\Delta_n(z)\right|\,\prod_{i=1}^n \rho(z_i; \mathbf{t}) \, dz_i\,,
 \end{equation}
with $\rho(z; \mathbf{t})$ given as in~\eqref{rho}.
At $\mathbf{t}\!=\!\mathbf{0}$, one recovers the GOE. 

In the following, we focus on the $\tau$-functions for even order matrices 
\begin{equation}
\label{eq:tau_pfaff_evensized}
    \tau_{2n}(\mathbf{t}) = \frac{1}{(2n)!}\int_{\mathbb{R}^{2n}}\left|\Delta_{2n}(z)\right|\,\prod_{i=1}^{2n} \rho(z_i; \mathbf{t}) \, dz_i\,,
\end{equation}
such that $\tau_{2n}(\mathbf{t}) = Z_{2n}(\mathbf{t})/C_{2n}$. In~\cite{vanM_Pfaff_skew,adler2002,vanMoerbekenotes} it was proved that the sequence $\{ \tau_{2n}(\mathbf{t})\}_{n\in \mathbb{N}}$ coincides with a $\tau$-function of the Pfaff lattice hierarchy. 
This result follows from the observation that~$\tau_{2n}(\mathbf{t})$ can be expressed as the Pfaffian of an appropriately defined moments matrix.

We concisely summarise the argument and the result. Introducing the skew-symmetric scalar product
\begin{align} \label{eq:skew_prod}
    \langle\, f \,,\,g \,\rangle_{\mathbf{t}} = - \langle\, g \,,\,f \,\rangle_{\mathbf{t}} =  \dfrac{1}{2} \int_{\mathbb{R}^2} \,  f(x)\,g(y)\,\,\text{sgn}(y-x) \,\rho(x; \mathbf{t})\,\rho(y; \mathbf{t})\, dx\,dy \,,
\end{align}
one can define the skew-symmetric moments matrix
 \[
m_{2n}(\mathbf{t})=\left (\langle\,x^{i}\,,y^{j}\,\rangle_\mathbf{t} \right )_{0 \le i,j \leq 2n-1} \,.
\]
A direct calculation shows that the $\tau$-function~\eqref{eq:tau_pfaff_evensized} is given by the formula 
\begin{equation}
\label{taupfaffdet}
\tau_{2n}(\mathbf{t}) = 2^n\,\textup{pf} \left( m_{2n} (\mathbf{t}) \right),
\end{equation}
where $\textup{pf}(A)$ denotes the Pfaffian of the even order matrix $A$, i.e.\ $\textup{pf}(A) = \sqrt{\det(A)}$.For the sake of comparison with similar formulae found in the literature, we point out that the factor $2^{n}$ is consistent with the definition of skew-symmetric scalar product~\eqref{eq:skew_prod} featuring a factor $\frac{1}{2}$, see e.g.~\cite{Adler2000_skew}.
Let us now consider the semi-infinite formal extension of the moments matrix and its unique factorisation of the form
\begin{equation}
m_{\infty}(\mathbf{t})= \left( S(\mathbf{t})^{-1} \right) J \left( S(\mathbf{t})^{-1}\right)^{\top}   \,, 
\label{skew_decomposition_intro}
\end{equation} 
where $J$ is the semi-infinite skew-symmetric matrix such that $J^2\!=\!-I$, and $S(\mathbf{t})$ is a semi-infinite lower triangular matrix. Given $S(\mathbf{t})$ and the shift matrix $\Lambda\!=\! \big\{\delta_{i,j-1} \big\}_{i,j=1}^{\infty}$, with $\delta_{i,j}$ the Kronecker delta, one can construct the Lax  matrix 
\begin{equation} 
L(\mathbf{t})= S(\mathbf{t}) \,\Lambda \,S(\mathbf{t})^{-1}, \,
\label{eq:l_t_intro}
\end{equation} 
which satisfies the compatible hierarchy of equations
\begin{equation}
    \frac{\partial L}{\partial t_{k}} = \left[\,-(L^{k})_{\mathfrak{t}}\,, L\,\right]{ ,}
    \label{eq:hamiltonian_L_intro}
\end{equation} 
{ referred to as the Pfaff lattice.} The subscript $\mathfrak{t}$ denotes the projection of the given matrix over the space of lower triangle matrices with  $2\times2$ diagonal blocks along the diagonal evaluated as follows:
\begin{equation}\label{eq:t_projection} 
A_\mathfrak{t} = A_- - J(A_+)^{\top} J + \frac{1}{2}\left(A_0 - J (A_0)^{\top} J \right), \, \qquad A \in \mathfrak{gl}(\infty)\,,  
\end{equation}
where $A_{\pm}$ denote, respectively, the upper and lower triangular part of~$A$, with all $2\times 2$ blocks along the diagonal equal to zero, and $A_{0}$ is obtained projecting $A$ on its $2 \times 2$ blocks along the diagonal.
As shown in~\cite{vanM_Pfaff_tau_function,Adler1999ThePL}, the partition function of the orthogonal ensemble is a particular solution of the Pfaff lattice hierarchy.

In the following, we focus on the study of the sequence $\{\tau_{2n}(\mathbf{t}) \}_{n\in \mathbb{N}}$ under the assumption that the Hamiltonian contains even powers of $M$ only, as in the equation~\eqref{Ham_even}. 
This choice leads us to the introduction of the reduced even Pfaff hierarchy which can be viewed as an infinite component analogue of the Volterra hierarchy obtained as a reduction of the Toda hierarchy's even flows. 
As observed in~\cite{benassi2021symmetric}, the even flows of the Pfaff hierarchy
\begin{equation}
    \frac{\partial L}{\partial t_{2k}} = \left[\,-(L^{2k})_{\mathfrak{t}}\,, L\,\right],
    \label{eq:even_Pfaff_Lax_equation}
\end{equation}
admit a reduction that is compatible with a Lax matrix of the form
\begin{equation}
    L(\mathbf{t})\Big|_{\underset{\forall i\in \mathbb{N}}{t_{2i-1} = 0}}  =   \left(
\begin{array}{ccccccc}
 ~~0~~ & ~~1~~ & ~~0~~ & ~~0~~ & ~~0~~ & ~~0~~ & \dots\\[1ex]
w^{-1}_1& ~~0~~ & w^{0}_{1}\hspace*{-0.5ex}& 0 & 0 & 0 &\ddots\\[1ex]
 0 & w^{1}_{1}\hspace*{-0.5ex}& 0 & 1 & 0 & 0 & \ddots\\[1ex]
 w^{-2}_{1}\hspace*{-0.5ex}& 0 & w^{-1}_2& 0 & w^{0}_{2}\hspace*{-0.5ex}& 0 & \ddots \\[1ex]
 0 & w^{2}_{1}\hspace*{-0.5ex}& 0 & w^{1}_{2}\hspace*{-0.5ex}& 0 & 1 & \ddots \\[1ex]
\vdots& \ddots &\ddots& \ddots & \ddots& \ddots & \ddots\\[1ex]
\end{array}
\right).
\label{eq:even_Pfaff_Lax_matrix}
\end{equation} 
This reduction corresponds to the request that the Lax matrix preserves the structure attained at~$\mathbf{t} = \mathbf{0}$, i.e.\ all lower diagonals in even position (counted starting from first diagonal below the main one) remain zero for any value of the couplings (the times of the hierarchy)~$t_{2i}$, when~$t_{2i-1} =0$ for all $i \in \mathbb{N}$.

 In the present formulation, for each given $\ell \in \mathbb{Z}$, the sequences $\{w^{\ell}_n(\mathbf{t})\}_{n\in\mathbb{N}}$ are interpreted as order parameters. Hence, the order parameters are particular solutions of the system of differential equations~\eqref{eq:even_Pfaff_Lax_equation} with initial condition specified by the GOE (i.e.\ the partition function~\eqref{eq:HSME} evaluated at~$\mathbf{t}\!=\!\mathbf{0}$).
The strategy we adopt to investigate the orthogonal ensemble in the thermodynamic limit is the study of the entries~$\{w^{\ell}_n(\mathbf{t})\}^{\ell\in\mathbb{Z}}_{n\in\mathbb{N}}$ as solutions of an integrable hierarchy in the continuum limit via a suitable asymptotic interpolation. This is similar to the approach proposed in~\cite{Witten} for the unitary ensemble, and developed in a consequent extensive literature, see e.g.\ \cite{Adler_KP,deift_Toda,KodamaPierceToda,HermitianPRE}. However, the case of symmetric matrices exhibit important conceptual and technical differences, for the associated differential identities are realised by an integrable chain of hydrodynamic type~\cite{chain}. Hydrodynamic chains are indeed systems of a higher level of complexity if compared with the Hopf equation or the dispersionless Toda system arising in the study of the unitary ensemble.
The evolution equations for~$\{w^{\ell}_n(\mathbf{t})\}^{\ell \in\mathbb{Z}}_{ n\in\mathbb{N}}$ and the associated initial datum are given below in Section \ref{sec:Pfaff_equations_and_initial_datum}.

 As observed {in~\cite{adler2002}}, the factorisation~\eqref{eq:l_t_intro}, given in terms of the lower triangular matrix $S(\mathbf{t})$, holds for the matrix $L(\mathbf{t})$. The entries of $S(\mathbf{t})$ are combinations of $\{\tau_{2n}(\mathbf{t})\}_{n\in \mathbb{N}}$, and their derivatives with respect to $t_{n}$ written in terms of elementary Schur polynomials $s_k(-\tilde{\partial})$ (see {Theorem~\ref{AvM2:theo} for the definitions of $\tilde{\partial}$ and $s_k$}).

Based on Theorem 0.1 {in~\cite{adler2002}}, we first prove the following
\begin{corollary}
The entries of the matrix $S(\mathbf{t})$ in~\eqref{eq:l_t_intro} with $t_{2i-1}=0$ for all $i\in \mathbb{N}$ are given by the following formulae
 \begin{equation}
     \begin{split} 
    S_{2n,2n-k}(\mathbf{t})\Big|_{\underset{\forall i\in \mathbb{N}}{t_{2i-1} = 0}} & = \left.\dfrac{ \sqrt{2} \, s_{k}(-\tilde{\partial}\,)\, \tau_{2n}(\mathbf{t})}{\sqrt{\tau_{2n}(\mathbf{t})\, \tau_{2n+2}(\mathbf{t})}}\right|_{\underset{\forall i\in \mathbb{N}}{t_{2i-1} = 0}}  \, \\[.5ex]
    S_{2n+1,2n-k+1}(\mathbf{t})\Big|_{\underset{\forall i\in \mathbb{N}}{t_{2i-1} = 0}} &= \left.\dfrac{ \sqrt{2}\left(\partial_{t_1} s_{k-1}(-\tilde{\partial}\,) + s_{k}(-\tilde{\partial}\,)\right)\, \tau_{2n}(\mathbf{t})}{\sqrt{\tau_{2n}(\mathbf{t})\, \tau_{2n+2}(\mathbf{t})}} \right|_{\underset{\forall i\in \mathbb{N}}{t_{2i-1} = 0}} \,,
    \end{split} 
\end{equation}
with $s_{-1}=0$, $s_0 = 1$ and $\tau_0(\mathbf{t}) = 1$.
\end{corollary}
The proof and explicit examples are provided in Appendix~\ref{app:L_elements_in_tau}. The above result allows us to evaluate, via a direct calculation, the elements of the sequences $\{w^{\ell}_{n}(\mathbf{t}) \}_{n \in \mathbb{N}}$, with $\ell \in \mathbb{Z}$,  in terms of $\tau_{2n}(\mathbf{t})$ and their derivatives restricted to $t_{2i-1} = 0$, with $i \in \mathbb{N}$. For example, we have
\begin{subequations}
\label{wn:expression}
\begin{align}
\label{w0n:expression}
    w^0_n&=\left.\sqrt{\dfrac{\tau_{2(n+1)}(\mathbf{t})}{\tau_{2(n-1)}(\mathbf{t})}}\dfrac{s_0(-\tilde{\partial}\,) \,\tau_{2(n-1)}(\mathbf{t})}{s_0(-\tilde{\partial}\,)\, \tau_{2n}(\mathbf{t})}  \right|_{\underset{\forall i\in \mathbb{N}}{t_{2i-1} = 0}} =\left.\frac{\sqrt{\tau_{2(n-1)}(\mathbf{t}) \,\tau_{2(n+1)}({\mathbf{t}})}}{\tau_{2 n}(\mathbf{t})}\right|_{\underset{\forall i\in \mathbb{N}}{t_{2i-1} = 0}} \\[1ex]
    \label{w1n:expression}
    w^1_n &=  \left. \dfrac{\left( s_2(-\tilde{\partial}\,) + s_2(\tilde{\partial}\,) \right) \tau_{2n}(\mathbf{t})}{\sqrt{\tau_{2(n-1)}(\mathbf{t})\tau_{2(n+1)}(\mathbf{t})}}\right|_{\underset{\forall i\in \mathbb{N}}{t_{2i-1} = 0}}\, =\left. \frac{\partial_{t_1}^2 \tau_{2n}(\mathbf{t})}{\sqrt{\tau_{2(n-1)}(\mathbf{t})\,\tau_{2(n+1)}(\mathbf{t})}}\right|_{\underset{\forall i\in \mathbb{N}}{t_{2i-1} = 0}} 
\end{align}
\begin{equation}
\begin{split} 
\label{wm1n:expression}
    w^{-1}_n &= \left. - \dfrac{s_2(-\tilde{\partial}\,)\,\tau_{2n}(\mathbf{t})}{\tau_{2n}(\mathbf{t})}- \dfrac{s_2(\tilde{\partial}\,)\,\tau_{2(n-1)}(\mathbf{t})}{\tau_{2(n-1)}(\mathbf{t})}
   \right|_{\underset{\forall i\in \mathbb{N}}{t_{2i-1} = 0}} \\[.5ex] 
  &= \left. \frac{\left(\partial_{t_2}-\partial^2_{t_1}\right)\tau_{2n}(\mathbf{t})}{2 \tau_{2n}(\mathbf{t})}-\frac{\left(\partial_{t_2}+\partial_{t_1}^2\right)\tau_{2(n-1)}(\mathbf{t})}{2\tau_{2(n-1)}(\mathbf{t})}\right|_{\underset{\forall i\in \mathbb{N}}{t_{2i-1} = 0}} \,. 
  \end{split} 
\end{equation}
\end{subequations}
Explicit formulae for $w^{\pm k}_n(\mathbf{t})$ with $k >1$ are obtained in a similar manner, although, as shown in Appendix \ref{app:L_elements_in_tau}, their expressions become increasingly more complex. 

Before proceeding with the derivation of the evolution equations for the elements of the sequences $\{w^{\ell}_n(\mathbf{t})\}_{ n\in\mathbb{N}}$, with $\ell \in\mathbb{Z}$, we briefly discuss their statistical mechanical interpretation in relation to the orthogonal ensemble. 

Since here we consider the orthogonal ensemble where the Hamiltonian contains even couplings only, we adapt the definition of free energy~\eqref{PHI} and chemical potential to this specific case as follows
\[
 \Phi_{n}(t_{2},t_{4},\dots) := \log \tau_{2n}(\mathbf{t}) \Big|_{\underset{\forall i\in \mathbb{N}}{t_{2i-1} = 0}} \;,
\]
and 
\[
\mu_{n}(t_{2},t_{4},\dots) = \Phi_{n+1} - \Phi_{n} = \log \tau_{2n+2}(\mathbf{t}) - \log \tau_{2n}(\mathbf{t})\Big|_{\underset{\forall i\in \mathbb{N}}{t_{2i-1} = 0}} \; .
\]
Using the expression~\eqref{w0n:expression}, we find that the variation of the chemical potential $\Delta \mu_{n} = \mu_{n+1} - \mu_{n}$ is expressed in terms of the entries of $w^{0}_{n}(\mathbf{t})$ as 
\begin{equation}
\label{chemicpot2}
\Delta \mu_{n} = 2 \log w^{0}_{n+1}\;.
\end{equation}
Moreover, from the expression~\eqref{w1n:expression} we obtain
\begin{equation}
\label{corr}
\mathbb{E}_{2n} \left(\sum_{i,j=1}^{2 n} z_{i} z_{j} \right) = w^{0}_{n} w^{1}_{n} \; ,
\end{equation}
where the expectation values $\mathbb{E}(\varphi)$ are defined as
\begin{equation*}
\mathbb{E}_{n}(\varphi) = \left . \frac{\displaystyle\int_{\mathbb{R}^{n}} \varphi \; \left|\Delta_n(z)\right| \, \prod_{i=1}^{n} \rho(z_i; \mathbf{t})\,dz_{i}}{\displaystyle\int_{\mathbb{R}^{n}} \left|\Delta_n(z)\right|\, \prod_{i=1}^{n} \rho(z_i; \mathbf{t})\, dz_{i}}\, \right |_{ \underset{\forall i\in \mathbb{N}}{t_{2i-1} = 0}} \; .
\end{equation*}
Similarly, the expressions~\eqref{wn:expression} imply
\begin{equation}
\label{corrvar}
\mathbb{E}_{2n} \left(\sum_{i=1}^{2 n} z_{i}^{2} \right)-\mathbb{E}_{2n-2} \left(\sum_{i=1}^{2 n-2} z_{i}^{2} \right) = 2 w^{-1}_{n} + w^{0}_{n}w^{1}_{n} + w^{0}_{n-1}w^{1}_{n-1} \; .
\end{equation}
The above formulae~\eqref{chemicpot2},~\eqref{corr} and~\eqref{corrvar} show that certain statistical mechanical observables of the orthogonal ensemble with even couplings can be expressed explicitly in terms of the matrix elements~$w^{\ell}_{n}(\mathbf{t})$. In the following, we show that the entries $w^{\ell}_{n}(\mathbf{t})$ can be obtained as a solution to a double infinite nonlinear dynamical chain fulfilling a specific initial condition, and the thermodynamic limit is described by the continuum limit of the chain obtained by a suitable singular asymptotic expansion.

\subsection{Even Pfaff hierarchy as a double infinite chain}
\label{sec:Pfaff_equations_and_initial_datum}
In this section, we focus on the first Lax equation of the reduced even Pfaff hierarchy 
\begin{equation}
    \der{L}{t_{2}} = \left[\,-(L^2)_{\mathfrak{t}}\,, L\,\right],
    \label{eq:even_Pfaff_t2}
\end{equation}
and in particular on the initial value problem as specified by the orthogonal ensemble.
The approach we adopt for the calculation of the Lax matrix entries~$w^{\ell}_{n}(\mathbf{t})$ relies on a result proved in~\cite{vanM_Pfaff_skew},
specifically on the observation that the Lax operator $L(\mathbf{t})$ admits eigenvectors (see~\cite{vanM_Pfaff_skew, vanMoerbekenotes})
\[
q(z; \mathbf{t}) = (q_0 (z; \mathbf{t}), q_1(z; \mathbf{t}),\dots, q_n(z; \mathbf{t}),\dots)^{\top}
\]
such that
\begin{equation}
L(\mathbf{t})\,q(z; \mathbf{t}) = z \,q(z; \mathbf{t})\,, 
\label{eq:pfaff_eigenvalues}
\end{equation}
where the $q_n(z; \mathbf{t})$ are $n$-th degree skew-orthonormal polynomials with respect to the skew-symmetric inner product~\eqref{eq:skew_prod}
\begin{equation}
\begin{aligned}
    \langle q_{2m}(z; \mathbf{t}),q_{2n+1}(z; \mathbf{t})  \rangle_{\mathbf{t}} &= -\langle q_{2n+1}(z; \mathbf{t}),q_{2m}(z; \mathbf{t})\rangle_{\mathbf{t}} = \delta_{m n} \, \\[1ex]
     \langle q_{2m+1}(z; \mathbf{t}),q_{2n+1}(z; \mathbf{t})\rangle_{\mathbf{t}} &= \langle q_{2m}(z; \mathbf{t}),q_{2n}(z; \mathbf{t}) \rangle_{\mathbf{t}} =0\,.
\end{aligned}
\end{equation}

Theorem \ref{theo:discrete_chain} below incorporates a result previously obtained heuristically in~\cite{benassi2021symmetric}, giving the explicit form of the Pfaff lattice as a double infinite discrete chain for the lattice variables $\{w^{\ell}_{n}(\mathbf{t}) \}_{n \in \mathbb{N}}$, with~$\ell \in \mathbb{Z}$ (part $(\ref{thm:semi-discrete_syst})$ of the statement). Moreover, exploiting the eigenvalue problem~\eqref{eq:pfaff_eigenvalues} evaluated at~$\mathbf{t}=\mathbf{0}$, we provide the explicit expression of the Lax matrix $L(\mathbf{0})$ ( part $(\ref{thm:initial_datum})$ of the statement).
{ Since in the rest of the paper we consider the reduction to the even hierarchy involving even couplings $t_{2i}$ only, with a slight abuse of notation we shall denote $\mathbf{t} = (t_{2},t_{4},\dots)$ as all odd couplings are set equal to zero, i.e.\ $t_{2i-1}=0$}. Hence, we prove the following

\begin{theorem}
\label{theo:discrete_chain} 
Let $\{w^{\ell}_{n}(\mathbf{t}) \}^{\ell \in \mathbb{Z}}_{n \in \mathbb{N}}$ be the entries of the Lax matrix~\eqref{eq:even_Pfaff_Lax_matrix} of the orthogonal ensemble, evaluated at {$\mathbf{t} = (t_{2},t_{4},\dots)$}, solution of the equation~\eqref{eq:even_Pfaff_t2}. Then 
\begin{enumerate}[$(i)$]
    \item \label{thm:semi-discrete_syst}
    $w^\ell_n(\mathbf{t})$ satisfy the following double infinite chain:
\begin{subequations}  \label{pfaff_lattice}
\begin{align}
\begin{split} 
\partial_{t_2} w^{-k}_{n} &= \frac{w^{-k}_{n} }{2}  \Big(w^{0}_{n}\, w^{1}_{n} - w^{0}_{n-1}\, w^{1}_{n-1} +  w^{0}_{n+k-1}\, w^{1}_{n+k-1}  - w^{0}_{n+k-2}\, w^{1}_{n+k-2} \Big) \,  \\[1ex]
&~~+w^{-k+1}_{n+1}\, w^{0}_{n}- w^{-k+1}_{n}\, w^{0}_{n+k-2}+ w^{-k-1}_{n}\, w^{0}_{n+k-1} - w^{-k-1}_{n-1}\, w^{0}_{n-1}, \,  \quad -k \leq -2
\end{split} \\[1.5ex]
\partial_{t_2} w^{-1}_{n}&= w^{-1}_{n}\Big( w^{0}_{n}\, w^{1}_{n} - w^{0}_{n-1}\, w^{1}_{n-1} \Big) + w^{0}_{n}\Big(w^{0}_{n} + w^{-2}_{n}  \Big)- w^{0}_{n-1}\Big(w^{0}_{n-1} + w^{-2}_{n-1}  \Big) \, \\[1.2ex]
\partial_{t_2} w_{n}^{0} &= \frac{w^0_n}{2}  \Big( w^{0}_{n+1}\, w^{1}_{n+1} - w^{0}_{n-1}\, w^{1}_{n-1} \Big)+w^0_n \Big(w^{-1}_{n+1} - w^{-1}_{n}\Big) \, \\[1.5ex]
\partial_{t_2} w_{n}^{1} &= \frac{w^{1}_{n}}{2}  \Big( w^{0}_{n-1}\, w^{1}_{n-1}  - w^{0}_{n+1} \,w^{1}_{n+1}\Big) + w^{0}_{n+1}\, w^{2}_{n} - w^{0}_{n-1}\, w^{2}_{n-1} \, \\[1.5ex]
\begin{split} 
\partial_{t_2} w^{k}_{n} &= \frac{w^{k}_{n}}{2}  \Big(w^{0}_{n-1}\, w^{1}_{n-1} -w^{0}_{n}\, w^{1}_{n} + w^{0}_{n+k-1}\, w^{1}_{n+k-1}   - w^{0}_{n+k}\, w^{1}_{n+k} \Big)  \\[1ex]
&~~ + w^{k+1}_{n}\, w^{0}_{n+k} -w^{k+1}_{n-1}\, w^{0}_{n-1}+ w^{k-1}_{n+1}\, w^{0}_{n} - w^{k-1}_{n}\,w^{0}_{n+k-1}, \,   \hspace{12ex} k \geq 2 \,;
\end{split} 
\end{align}
\end{subequations} 
    \item \label{thm:initial_datum} with initial condition   
    {
    \begin{subequations}
\label{eq:w_initial}
\begin{align}
        w^{-k}_n(\mathbf{0}) &= 0 \,, &\quad&-k <- 2\, \label{eq:w_-k_initialcondition} \\[1ex]
        w^{-2}_n(\mathbf{0})&= - \frac{1}{2}\sqrt{2n(2n-1)}  \,\label{eq:w_-2_initialcondition} \\[1ex]
        w^{-1}_n(\mathbf{0}) & =\dfrac{1}{2} \,\label{eq:w_-1_initialcondition} \\[1ex]
        w^0_n(\mathbf{0}) &= \frac{1}{2}\sqrt{2n(2n-1)}   \, \label{eq:w_0_initialcondition} \\[1ex]
        w^k_n(\mathbf{0}) &= 2^{k+1}\,\frac{(k+n-1)!}{(n-1)!} \, \left(\frac{(2n-2)!}{(2k+2n-2)!}\right)^{\!1/2} \label{eq:w_k_initialcondition}
        \, &\quad &\;\;\;\;k > 0.
\end{align}
\end{subequations}}
\end{enumerate}

\end{theorem}

\begin{proofC} \begin{enumerate}[($i$)]
\item  Let us represent the generic element of the matrix $L$ defined in~\eqref{eq:even_Pfaff_Lax_matrix} as follows
\begin{equation} \label{eq:pfaff_t2_generic_element}
\begin{aligned} 
    L_{ij}&=\varphi(i) \,\sigma(j) \left( \delta_{1,j-i} \,w^0_{\frac{j-1}{2}} +\vartheta(i-j) \, w^{-\frac{i-j+1}{2}}_{\frac{j+1}{2}} \right)\! 
    +\varphi(j) \,\sigma(i) \left(\delta_{1,j-i} +\vartheta(i-j)\,  w^{\frac{i-j+1}{2}}_{\frac{j}{2}}\right) \;,
\end{aligned} 
\end{equation}
where $\delta_{i,j}$ is the Kronecker symbol and, for the sake of simplicity, we have dropped the explicit dependence of $w^{\ell}_n(\mathbf{t})$ on $\mathbf{t}$. The symbols $\sigma(i)$, $\varphi(i)$ and $\vartheta(i)$, are defined  for $i\in\mathbb{Z}$ as
\begin{equation}\label{eq:even_odd}
    \varphi(i)=\begin{cases}
        1\,, \quad |i| \text{ even}  \\
        0\,, \quad |i| \text{ odd}
    \end{cases} \, \qquad \sigma (i)=\begin{cases}
        1\,, \quad |i| \text{ odd}  \\
        0\,, \quad |i| \text{ even}
    \end{cases} \, \qquad \vartheta(i)=\begin{cases}
        1\,, \quad i \ge 0  \\
        0\,, \quad i < 0
    \end{cases} \,. 
\end{equation}
Below, we exploit the following properties of $\varphi$ and $\sigma$ with respect to their arguments
\begin{align*}
    \varphi(i+n)&=\begin{cases}
        \varphi(i)\,, \quad |n| \text{ even} \\
        \sigma(i),\, \quad |n| \text{ odd}
    \end{cases}\, \qquad \sigma (i+n)=\begin{cases}
        \sigma(i)\,, \quad |n| \text{ even}  \\
        \varphi(i)\,, \quad |n| \text{ odd}
    \end{cases} \, \\[1ex]
    \begin{split}   &\varphi(i+j)=\varphi(i)\,\varphi(j)+\sigma(i)\,\sigma(j),\, \qquad \sigma(i+j)= \varphi(i)\,\sigma(j)+\sigma(i)\,\varphi(j) \, \\[1.5ex]
    &\varphi(i)^{m}=\varphi(i),\, \qquad \sigma(i)^{m}=\sigma(i),\,  \qquad \varphi(i)\,\sigma(i)=0,\, \quad m \in \mathbb{N}\,.
\end{split} 
\end{align*}
Using the notations and properties above, the proof follows from a direct calculation of the right hand side of the equation
\begin{equation}
\der{L_{ij}}{t_{2}} = -\left[(L^2)_{\mathfrak{t}}, L \right]_{ij}.
\label{eq:even_Pfaff_elements}
\end{equation}
In particular we have (see Appendix \ref{app:pfaff_lattice_explicit} for full details)
\begin{equation} \label{pfaff_lattice_in_W}
	\begin{aligned}
	\left[ (L^2)_{\mathfrak{t}}\,, L \right]_{ij} &=\, \varphi(j)\,W_{\varphi}(i,j) + \sigma(j)\,W_{\sigma}(i,j)   + \varphi(i)\,\sigma(j)\,W_{\varphi\sigma}(i,j)+\sigma(i)\,\varphi(j)\,W_{\sigma\varphi}(i,j)\,,
	\end{aligned} 
\end{equation}
where 
\begin{align}
	\begin{split} 
	W_{\varphi}(i,j) &:=  \tfrac{1}{2}\,\delta_{i,j-1} \left (w^{-1}_{\frac{j}{2}}+ w^0_{\frac{j}{2}}\,w^1_{\frac{j}{2}}\right )\,
	\end{split} 
	\\[1ex]
	\begin{split} 
	W_{\sigma}(i,j) &:= \tfrac{1}{2}\,\delta_{i,j-1} \left ( w^{-1}_{\frac{j-1}{2}} \,w^0_{\frac{j-1}{2}}+ w^0_{\frac{j-3}{2}}\,w^1_{\frac{j-3}{2}}\,w^0_{\frac{j-1}{2}}\right )\,,  
	\end{split} 
\end{align}
and the expressions for $W_{\varphi\sigma},W_{\sigma\varphi}$, due to their length, are reported in Appendix \ref{app:pfaff_lattice_explicit}.
The equations for~$w^{\ell}_n(\mathbf{t})$ are obtained by comparing the left and right hand side of the equation~\eqref{eq:even_Pfaff_elements}. To this purpose, it is useful to introduce the index $k=(i-j+1)/2$, and analyse the following three separate cases \\

\noindent {\bf Case 1:} $i-j=-1$ ($k=0$, first upper diagonal). If $i = 2n$, we have $\varphi(i)\,\sigma(j)=1$ and
\begin{equation*}
\begin{split} 
        \partial_{t_2} L_{2n,2n+1} &= -W_{\sigma}(2n,2n+1)-W_{\varphi\sigma}(2n,2n+1)\,,
\end{split} 
\end{equation*}
which gives the sought equation for $w^{0}_{n}$, i.e.
\begin{equation*}
\begin{split} 
\partial_{t_2} w^0_n &= \tfrac{1}{2} \, w^0_n \left( w^{0}_{n+1} w^{1}_{n+1} - w^{0}_{n-1} w^{1}_{n-1} \right)+w^0_n \left(w^{-1}_{n+1} - w^{-1}_{n}\right) \,.
\end{split} 
\end{equation*}
We also note that the case $i =2n-1$ is trivial as $L_{2n-1,2n} =1$ and, consistently, we get 
\[
\left[(L^2)_{\mathfrak{t}},L\right]_{2n-1,2n} =0.
\]
\noindent {\bf Case 2:} $i-j=1$ ($k=1$, first lower diagonal). If $i = 2 n$, we have $\varphi(i)\,\sigma(j)=1$ and then
 \begin{equation*}
    \begin{split}
        \partial_{t_2} L_{2n,2n-1} &= -W_{\sigma}(2n,2n-1)-W_{\varphi\sigma}(2n,2n-1) \,,
    \end{split} 
 \end{equation*} 
 which gives the sought equation for $w^{-1}_{n}$, i.e.
  \begin{equation*}
    \begin{split}
        \partial_{t_2} w^{-1}_{n}&= w^{-1}_{n}\left( w^{0}_{n} w^{1}_{n} - w^{0}_{n-1} w^{1}_{n-1} \right) + w^{0}_{n}\left(w^{0}_{n} + w^{-2}_{n}  \right)+ w^{0}_{n-1}\left(w^{0}_{n-1} + w^{-2}_{n-1}  \right)\,.
    \end{split} 
 \end{equation*} 
If $i = 2n{ +}1$, we have $\varphi(j)\,\sigma(i)=1$  and
\begin{equation*}
        \begin{split}
        \partial_{t_2} L_{2n+1,2n} &= -W_{\varphi}(2n+1,2n)-W_{\sigma\varphi}(2n+1,2n)\,,
        \end{split} 
        \end{equation*}
which gives the sought equation for $w^{1}_{n}$, i.e.
\begin{equation*}
        \begin{split}
            \partial_{t_2} w_{n}^{1} &= \tfrac{1}{2} \,w^{1}_{n} \left( w^{0}_{n-1} w^{1}_{n-1}  - w^{0}_{n+1} w^{1}_{n+1}\right) + w^{0}_{n+1} w^{2}_{n} - w^{0}_{n-1} w^{2}_{n-1}  \,.
        \end{split} 
\end{equation*}
\noindent {\bf Case 3:} $i-j>1$ ($k>1$). The only non-trivial equations are obtained from the analysis of the diagonals in odd position, as the entries of even diagonals identically vanish. Therefore, considering the lower diagonal in position $2k-1$, if $j=2n-1$, we have  $\varphi(i)\,\sigma(j)=1$ and
\begin{equation*}
        \begin{split} 
        \partial_{t_2} L_{2(n+k)+2,2n-1} &= -W_{\sigma}(2(n+k)+2,2n-1)-W_{\varphi\sigma}(2(n+k)+2,2n-1)\,,
        \end{split}
\end{equation*}
which gives the sought equation for $w^{-k}_{n}$, i.e.\    
\begin{equation*}
        \begin{split} 
            \partial_{t_2} w^{-k}_{n} &= \tfrac{1}{2} \, w^{-k}_{n} \left(w^{0}_{n} w^{1}_{n} - w^{0}_{n-1} w^{1}_{n-1} +  w^{0}_{n+k-1} w^{1}_{n+k-1}  - w^{0}_{n+k-2} w^{1}_{n+k-2} \right) \\[1ex]
&+w^{-(k-1)}_{n+1} w^{0}_{n}- w^{-(k-1)}_{n} w^{0}_{n+k-2} + w^{-(k+1)}_{n} w^{0}_{n+k-1} - w^{-(k+1)}_{n-1} w^{0}_{n-1}  \,.
        \end{split}
        \end{equation*}
If $j=2n$, we have $\varphi(j)\,\sigma(i)=1$ and
\begin{equation*}
        \begin{split}
        \partial_{t_2} L_{2(n+k)+1,2n} &= -W_{\varphi}(2(n+k)+1,2n)-W_{\sigma\varphi}(2(n+k)+1,2n)\,,
        \end{split} 
\end{equation*}
which gives the sought equation for $w^{k}_{n}$, i.e.\ 
\begin{equation*}
        \begin{split}
            \partial_{t_2} w^{k}_{n} &= \tfrac{1}{2}\, w^{k}_{n} \left(w^{0}_{n-1} w^{1}_{n-1} -w^{0}_{n} w^{1}_{n} + w^{0}_{n+k-1} w^{1}_{n+k-1}   - w^{0}_{n+k} w^{1}_{n+k} \right) \\[1ex]
& + w^{k+1}_{n} w^{0}_{n+k} -w^{k+1}_{n-1} w^{0}_{n-1} + w^{k-1}_{n+1} w^{0}_{n} - w^{k-1}_{n}w^{0}_{n+k-1} \,.
        \end{split} 
\end{equation*}
As mentioned above, the commutator vanishes for all remaining entries. 

\item The proof is based on the map between the Toda and Pfaff lattice discovered in~\cite{Adler2000_skew, adler2002} and based on a mapping between {the} symmetric inner product $(\,\cdot\,\,,\,\cdot\,)_{\mathbf{0}}^{{(2)}}$ ({defined below in equation~\eqref{eq:Hermiteweight}}) and {the above} skew-symmetric inner product $\langle\,\cdot\,\,,\,\cdot\,\rangle_{\mathbf{0}}$. {This map  further} implies a mapping between a class of monic orthogonal polynomials and skew-orthogonal polynomials. In particular, this map allows us to relate the orthonormal polynomials~$p_n(z;\mathbf{0})$ in~\eqref{eq:orthonormal_poly}, i.e.\ the eigenfunctions of the Toda lattice Lax matrix, and the skew-orthonormal polynomials $q_n(z;\mathbf{0})$, i.e.\ the eigenfunctions of the Pfaff lattice Lax matrix.  

Let us first consider the eigenvalue problem~\eqref{eq:pfaff_eigenvalues} at $\mathbf{t} = \mathbf{0}$, i.e.
\begin{equation}\label{eq:eigenvalue_pfaff_zero_time}
    L(\mathbf{0})\,q(z;\mathbf{0}) = z\, q(z;\mathbf{0})\,.
\end{equation}
For the sake of simplicity, until the end of the proof, all quantities, $q_{n}$ and~$w^{\ell}_{n}$ are understood to be evaluated at $\mathbf{t} = \mathbf{0}$.
Writing the equations~\eqref{eq:eigenvalue_pfaff_zero_time} in components, we have
\begin{subequations}
\begin{align}  
    z\, q_{2n} (z) &= q_{2n+1}(z) + \sum_{k=0}^{n-1} w^{n-k}_{k+1} \, q_{2k+1}(z) \,
    \label{eq:eigenvalue_pfaff_components_even}\\
    z\, q_{2n-1} (z) &= w^0_n \, q_{2n}(z) + \sum_{k=0}^{n-1} w^{-(n-k)}_{k+1}\, q_{2k}(z) \,. \label{eq:eigenvalue_pfaff_components_odd}
\end{align}
 \end{subequations}
Given the skew-orthogonal polynomials $q_{n}(z)$, we introduce the associated monic skew-orthogonal polynomial $Q_{n}(z)$ with respect to the inner product $\langle \,\cdot\,\,, \,\cdot\, \rangle_{\mathbf{0}}$ defined in~\eqref{eq:skew_prod} such that
\begin{equation*}
\begin{aligned}
    \langle Q_{2n}(z)\,,Q_{2m+1}(z) \rangle_{\mathbf{0}} &= -\langle Q_{2m+1}(z)\,,Q_{2n}(z) \rangle_{\mathbf{0}} = \nu_n \, \delta_{m n} \,  \\[1ex]
    \langle Q_{2n+1}(z)\,,Q_{2m+1}(z) \rangle_{\mathbf{0}} &= 0 = \langle Q_{2n}(z)\,,Q_{2m}(z) \rangle_{\mathbf{0}} 
    \,,
\end{aligned}
\end{equation*}
where $\nu_n = \sqrt{\pi}(2n)!/2^{2n}$.

In particular, $q_n(z)$ and $Q_{n}(z)$ are related as follows
\begin{equation} \label{eq:q_orthonorm}
    q_n(z) = \begin{cases}
    \dfrac{Q_n(z)}{\sqrt{\strut{\nu_{n/2}}}} & \text{ if }n\text{ is even} \\[4ex] 
    \dfrac{Q_n(z)}{\sqrt{\strut{\nu_{(n-1)/2}}}} & \text{ if }n\text{ is odd} \; .
    \end{cases} \,
\end{equation}
Hence, using the above relations~\eqref{eq:q_orthonorm}, we can re-write~\eqref{eq:eigenvalue_pfaff_components_even} and~\eqref{eq:eigenvalue_pfaff_components_odd} in terms of the monic skew-orthogonal polynomials $Q_n(z)$ as follows 
\begin{subequations}
\label{eqs:Qandw}
\begin{align}
    z\, \dfrac{Q_{2n}(z)}{\sqrt{{\nu_n}}}  &= \dfrac{Q_{2n+1}(z)}{\sqrt{{\nu_n}}} + \sum_{k=0}^{n-1} w^{n-k}_{k+1} \, \dfrac{Q_{2k+1}(z)}{\sqrt{{\nu_k}}} \, \\[1ex]
    z\, \dfrac{Q_{2n-1} (z)}{\sqrt{\nu_{n-1}}} &= w^0_n \, \dfrac{Q_{2n}(z)}{\sqrt{\nu_n}} + \sum_{k=0}^{n-1} w^{-(n-k)}_{k+1} \, \dfrac{Q_{2k}(z)}{\sqrt{\nu_k}} \,.
\end{align}
\end{subequations}
According to~\cite{Adler2000_skew}, we have that, for Gaussian weights,
\begin{subequations}
\label{eq:hermite_skew_orthoQ}
\begin{align} 
        Q_{2n}(z) &= P_{2n}(z) \, \label{eq:hermite_skew_ortho_even}\\[1ex] 
        Q_{2n+1}(z) &= P_{2n+1}(z) - n\,P_{2n-1}(z) \, \qquad n \in \mathbb{N}_{0}, \label{eq:hermite_skew_ortho_odd} 
\end{align}
\end{subequations}
where $P_{m}(z)$ are the monic orthogonal polynomials\footnote{We note that $P_n(z)$ are {Hermite polynomials} related to the standard Hermite polynomials $H_n(z)$ as follows
\begin{equation*}
\begin{split} 
    P_n(z) = \dfrac{1}{2^n} H_n(z),{\text{ with }}\,\, H_n(z) = \left(-1\right)^n \exp(x^2) \, \dfrac{d^n}{dx^n}\, \exp(-x^2), \quad n \in \mathbb{N}_{0} \,.
\end{split}     
\end{equation*}} with respect to the symmetric inner product\footnote{Note that this definition of scalar product slightly differs from the one in the inner product defined in~\eqref{intro:inner_product}, therefore, to distinguish the two of them, here we use the superscript in $(\cdot,\cdot)^{(2)}_{\mathbf{0}}$ and $\rho^{(2)}(x,\mathbf{0})$.}
\begin{equation} 
\label{eq:Hermiteweight}
    \left(\, f\,,g \,\right)^{{(2)}}_{\mathbf{0}} = \int_{\mathbb{R}}  f(x) \, g(x) \,\rho^{(2)}(x;\mathbf{0})\, dx \,, \qquad \rho^{(2)}(x;\mathbf{0})=\exp(-x^2)
\end{equation}
such that for all $m,n \in \mathbb{N}_{0}$
\begin{equation} \label{eq:poly_ortho_zero}
    \left( P_n(z)\,,P_m(z) \right)_{\mathbf{0}}^{(2)} = \int_{\mathbb{R}} P_n(x)\, P_m(x) \, \rho^{(2)}(x;\mathbf{0}) \, dx = \gamma_n\, \delta_{m n} \,,
\end{equation}
where $\gamma_n = \sqrt{\pi}\,n!/2^n$. 
Using the relations~\eqref{eq:hermite_skew_orthoQ} into the equations~\eqref{eqs:Qandw} we have
\begin{subequations}
\label{eq:skew_even_odd}
\begin{align}
\label{eq:skew_even} 
&z\, \dfrac{P_{2n}(z)}{\sqrt{{\nu_n}}}  = \dfrac{P_{2n+1}(z)-n\,P_{2n-1}(z)}{\sqrt{{\nu_n}}} + \sum_{k=0}^{n-1} w^{n-k}_{k+1} \, \dfrac{P_{2k+1}(z)-k\,P_{2k-1}(z)}{\sqrt{{\nu_k}}} \, \\
\label{eq:skew_odd}
    &z\, \dfrac{P_{2n-1}(z)-(n-1)P_{2n-3}(z)}{\sqrt{\nu_{n-1}}} = w^0_n \, \dfrac{P_{2n}(z)}{\sqrt{\nu_n}} + \sum_{k=0}^{n-1} w^{-(n-k)}_{k+1} \, \dfrac{P_{2k}(z)}{\sqrt{\nu_k}} \,.
\end{align} 
\end{subequations}
Recalling that orthogonal {Hermite polynomials satisfy a three-term} recurrence relation
\begin{equation*}
\label{eq:three_Hermite}
    z \, P_n(z) = P_{n+1}(z) + \dfrac{n}{2} \, P_{n-1}(z) \,, \quad P_0(z)=1, \quad P_{-1}(z)=0,
\end{equation*}
{separating even and odd degree we have}, respectively,
\begin{subequations}
\label{eq:threepointsplit}
\begin{align}
    &z P_{2n}(z) = P_{2n+1}(z) + n P_{2n-1}(z)\label{eq:threepointeven}\,\\
    &z P_{2n-1}(z) = P_{2n}(z) + \Big(n-\frac{1}{2}\Big)P_{2n-2}(z)\,.\label{eq:threepointodd}  
\end{align}
\end{subequations}
Substituting now the  expressions~\eqref{eq:threepointsplit} into the equations \eqref{eq:skew_even_odd}, we obtain two relations involving, respectively, only odd and even degree polynomials $P_{m}(z)$, i.e.
\begin{subequations}
\label{eq:Pw}
\begin{align}
     &\frac{ P_{2n+1}(z) + n P_{2n-1}(z)}{\sqrt{\nu_n}} =   \dfrac{P_{2n+1}(z)-n\,P_{2n-1}(z)}{\sqrt{{\nu_n}}} + \sum_{k=0}^{n-1} w^{n-k}_{k+1} \, \dfrac{P_{2k+1}(z)-k\,P_{2k-1}(z)}{\sqrt{{\nu_k}}}\,,\\
   &\frac{ P_{2n}(z) + \frac{1}{2}P_{2n-2}(z) - (n-1) 
    \left(n-\frac{3}{2}\right)P_{2n-4}(z)}{\sqrt{\nu_n}}= w^0_n \, \dfrac{P_{2n}(z)}{\sqrt{\nu_n}} + \sum_{k=0}^{n-1} w^{-(n-k)}_{k+1} \, \dfrac{P_{2k}(z)}{\sqrt{\nu_k}}\,.
\end{align}
\end{subequations}
For any $m\in\mathbb{N}_{0}$, we can thus project both the left and right hand sides of~\eqref{eq:Pw} onto $P_{m}(z)$ with respect to the inner product~\eqref{eq:poly_ortho_zero}. This gives a set of identities leading to the expressions~\eqref{eq:w_initial} (see Appendix \ref{app:field_variables} for full details). Hence, the theorem is proven.
\end{enumerate}
\end{proofC}

The chain~\eqref{pfaff_lattice} is equivalent to the equation~\eqref{eq:even_Pfaff_Lax_equation}, i.e.\ the first flow of the {even Pfaff hierarchy}. Although the reduction procedure described above is similar to the derivation of the Volterra hierarchy~\eqref{Volterraeq} from the Toda hierarchy in the case of the unitary ensemble, the result is significantly different. Indeed, the chain~\eqref{pfaff_lattice} constitutes a system in infinitely many components, whilst the Volterra hierarchy is a family of scalar equations. {Higher equations of the even Pfaff hierarchy can be written following a similar procedure although calculations are much more involved}.

As shown in Section~\ref{sec:SME_Pfaff}, the entries $w^{\ell}_n(\mathbf{t})$ of the Pfaff lattice Lax matrix~\eqref{eq:even_Pfaff_Lax_matrix} can be interpreted as observables of the orthogonal ensemble calculated as a particular solution of the set of differential equations~\eqref{pfaff_lattice}. The required particular solution is fixed by the initial condition, i.e.\ the values~$ w^{\ell}_n(\mathbf{0})$. Hence, the explicit evaluation of the initial condition (associated with the orthogonal ensemble) allows us to completely specify $w^{\ell }_n(\mathbf{t})$ as the solution to the initial value problem for the even Pfaff hierarchy. As discussed below, the explicit form of the initial condition is crucial for the construction of the singular asymptotic expansion required to characterise the solution in the thermodynamic limit. This will be the focus of Section \ref{sec:thermolimit}.

We emphasise that the mapping between Toda and Pfaff lattice~\cite{Adler2000_skew} specified by the relations~\eqref{eq:hermite_skew_orthoQ} also holds in general for $\mathbf{t} \neq \mathbf{0}$. Specifically, the polynomials $q_n(z; \mathbf{t})$, skew-orthogonal with respect to the scalar product~\eqref{eq:skew_prod}, can be expressed in terms of orthogonal polynomials with scalar product 
\begin{equation*}
\left(\,f\,,\,g\,\right)^{(2)}_{2\mathbf{t}} = \int_\mathbb{R} f(x)\,g(x)\,\rho^{(2)}(x;\mathbf{t})\,dx\,, \qquad \rho^{(2)}(x;\mathbf{t})=\exp\Big(-x^2 + \sum_{j\geq 1}2\,t_j \,x^j \Big).
\end{equation*} 
At $\mathbf{t}=\mathbf{0}$, the explicit expressions for $w^{\ell}_{n}$ result from this remarkable mapping, as the orthogonal polynomials reduce to the standard Hermite polynomials. Calculating $w^{\ell}_{n}(\mathbf{t})$, at $\mathbf{t}\neq \mathbf{0}$, requires the study of the differential equation~\eqref{eq:even_Pfaff_t2} with initial condition
{\small
\begin{equation} 
\label{eq:Lax_red_init}
L(\mathbf{0}) =   \left(
\begin{array}{ccccccc}
 0 & 1 & 0 & 0 & 0 & 0 & \dots  \\[1.8ex]
 \dfrac{1}{2}& 0 & w^{0}_{1}(\mathbf{0})\hspace*{-0.5ex}& 0 & 0 & 0 & \ddots \\[1.8ex]
 0 & w^{1}_{1}(\mathbf{0})\hspace*{-0.5ex}& 0 & 1 & 0 & 0 & \ddots \\[1.8ex]
 -w^{0}_{1}(\mathbf{0})\hspace*{-0.5ex}& 0 & \dfrac{1}{2}& 0 & w^{0}_{2}(\mathbf{0})\hspace*{-0.5ex}& 0 & \ddots  \\[1.8ex]
 0 & w^{2}_{1}(\mathbf{0})\hspace*{-0.5ex}& 0 & w^{1}_{2}(\mathbf{0})\hspace*{-0.5ex}& 0 & 1 & \ddots  \\[1.8ex]
 0 & 0 & -w^{0}_{2}(\mathbf{0})\hspace*{-0.5ex}& 0 & \dfrac{1}{2}& 0 & \ddots\\[1.8ex]
 0 & w^{3}_{1}(\mathbf{0})\hspace*{-0.5ex}& 0 & w^{2}_{2}(\mathbf{0})\hspace*{-0.5ex}& 0 & w^{1}_{3}(\mathbf{0})\hspace*{-0.5ex}& \ddots \\[1.8ex]
 \vdots & \ddots & \ddots & \ddots & \ddots & \ddots & \ddots 
\end{array}
\right) \,. 
\end{equation}}

\noindent 
Note that $w^{-k}_n(\mathbf{0}) = 0$ for all integers $k\!>\!2$. 
In Section \ref{sec:reductions}, we introduce a further reduction which preserves this property for all $\mathbf{t}$, i.e.\ $w^{-k}_n(\mathbf{t}) = 0$ for $k>2$, maintaining the form of the initial condition~\eqref{eq:w_initial}. As we shall see, such a reduction exists for both the semi-discrete even Pfaff lattice~\eqref{pfaff_lattice}  and for its asymptotic expansion in the limit $n\to 
\infty$ as detailed in Section \ref{sec:thermolimit}. 

\begin{remark}
Based on the result of Theorem \ref{theo:discrete_chain}$(\ref{thm:initial_datum})$, for any fixed integer $k> 0$, the elements of the sequence $\{ w^k_n(\mathbf{0}) \}_{n \in \mathbb{N}}$ can be written explicitly as follows (see Appendix~\ref{app:field_variables})
\begin{equation}
    w^k_n(\mathbf{0}) = n \sqrt{\frac{\nu_{n-1}}{\nu_n}}\,w^{k-1}_{n+1}(\mathbf{0}) = 2\,\frac{(k+n-1)!}{(n-1)!}\left(\prod_{\ell = 1}^{k+n-1}w^0_\ell(\mathbf{0})\right)^{\!\!-1} \,.
\end{equation}
\end{remark}

\section{Thermodynamic limit of the orthogonal ensemble}\label{sec:thermolimit}
In Section~\ref{sec:therm_hermitian} we have shown how, upon introducing a suitable scale $N$, the entries $a_{n}(\mathbf{t})$ and $b_{n}(\mathbf{t})$ of the Lax matrix for the Toda lattice can be interpolated yielding the system of partial differential equations~\eqref{Todacont} as a formal power series in $\varepsilon = 1/N \ll 1$. At the leading order in $\varepsilon$, the equations~\eqref{Todacont} reduce to a system of hydrodynamic type, i.e.\ the dTL equations~{\eqref{dToda}}, whose solutions develop gradient catastrophe singularities in the space of couplings. We have also seen that if the Hamiltonian defining the probability distribution contains only even power interaction terms (see expression~\eqref{Ham_even}) there exists a compatible reduction of the Toda lattice hierarchy restricted to even flows only, namely the Volterra hierarchy, satisfied by the sequence $\{b_{n}(\mathbf{t})\}_{n\in\mathbb{N}}$ whilst $a_{n}(\mathbf{t})\!=\!0$ for all $n\in\mathbb{N}$. In the continuum limit, the Volterra lattice admits a regular expansion as $\varepsilon \to 0$ that is of the same form as the asymptotic expansion of the initial datum. At the leading order, the continuum equations coincide with the Hopf hierarchy. 

In the present section, we follow a similar approach to study the orthogonal ensemble with Hamiltonian containing even couplings only, and {the even Pfaff lattice as} its related integrable lattice. We interpolate the dependent variables of the reduced even Pfaff lattice, i.e.\ the Lax matrix entries $w^{\ell}_{n}(\mathbf{t})$. These variables, as we shall see, define a sequence of order parameters for the orthogonal ensemble. However, a key difference, compared to the unitary ensemble, is that the asymptotic expansion of the initial datum with respect to the scaling parameter $\varepsilon$ is singular. Hence, the asymptotic expansion for interpolating functions is also required to be singular. As detailed below, the required singular expansion induces a resonance between different orders, which is realised by the appearance of a non-homogeneous term in the resulting system of equations. This resonance can however be resolved by imposing a specific constraint on higher order terms. Remarkably, the procedure is consistent with the initial datum and leads to a completely integrable chain of hydrodynamic type that belongs to a class studied in~\cite{chain, Pavlov2003IntegrableHC,Pavlov2006Class,Odesskii_2010}. This analysis yields our Main Result \ref{res:reduced_pfaff_continuum_limit}.

\subsection{Reduced even Pfaff hierarchy: continuum limit}{\label{sec:recap_old_chain}}

We now build up on Theorem~\ref{theo:discrete_chain} to study the behaviour of the order parameters in the thermodynamic limit for the orthogonal ensemble where the Hamiltonian contains even power interactions only. The thermodynamic limit is realised via the continuum limit of the reduced even Pfaff lattice, i.e.\ the system of equations~\eqref{pfaff_lattice}. { We recall that, as detailed in Theorem \ref{theo:discrete_chain}, we work here with null odd couplings and $\mathbf{t} = (t_2, \,t_4,\,\dots, t_{2n},\dots)$.} 

Similarly to the case of the unitary ensemble, let us  introduce the variable $x\!=\!\varepsilon n$, with $\varepsilon\!=\!1/N\!\ll\!1$, such that $x$ remains finite as both $n, N \to \infty$. Hence, we introduce the interpolating functions $U^\ell(x,\mathbf{t};\varepsilon)$, {with upper indices $\ell \in \mathbb{Z}$}, such that
\begin{equation} \label{eq:continuum_limit}
U^\ell(x,\mathbf{t};\varepsilon):= w^\ell_{n}(\mathbf{t}),  \qquad U^\ell(x \pm j \varepsilon ,\mathbf{t};\varepsilon):= w^\ell_{n \pm j}(\mathbf{t}) \qquad \textup{for} \quad x = \varepsilon n, \quad j \in \mathbb{N}.
\end{equation} 
Therefore, the initial condition on the Pfaff Lax matrix entries~$w^{\ell}_n(\mathbf{0})$ (see equations~\eqref{eq:w_initial}) implies the following asymptotic expansions of the initial conditions for the interpolating functions~\eqref{eq:continuum_limit} as~$\varepsilon \to 0$:
\vspace*{-2ex}
\begin{subequations}\label{eq:initial_datum}
\begin{align}
&U^{-k}(x,\mathbf{0};\varepsilon) =0, \,  &-k < -2\,\\[.1ex]
 & U^{-2}(x,\mathbf{0};\varepsilon) = -\frac{x}{\varepsilon} + \frac{1}{4} + \mathcal{O}(\varepsilon)\,  &\label{eq:cont_initial_datum_1}\\[.1ex]
 & U^{-1}(x,\mathbf{0};\varepsilon) =\frac{1}{2} & \\[.1ex]
 & U^{0}(x,\mathbf{0};\varepsilon) = \frac{x}{\varepsilon} - \frac{1}{4} + \mathcal{O}(\varepsilon)\,  &\label{eq:cont_initial_datum_2}\\[.1ex]
  &  U^{k}(x,\mathbf{0};\varepsilon) = 2 + \frac{k}{2x}\,\varepsilon+\mathcal{O}(\varepsilon^2),\,  &k> 0 \,.
\end{align}
\end{subequations}
Consistently with equations~\eqref{eq:initial_datum}, we consider an asymptotic expansion for the interpolating functions~$U^{\ell}(x,\mathbf{t}; \varepsilon)$ of the same form, i.e.\ 
{  \begin{equation} \label{eq:continuum_var}
U^\ell(x,\mathbf{t};\varepsilon)= \begin{cases}
         \displaystyle \sum_{i\geq 0} U^\ell_i(x, \mathbf{t})\varepsilon^i = U^\ell_0(x, \mathbf{t})+ \varepsilon\,U^\ell_1(x, \mathbf{t}) + \mathcal{O}(\varepsilon^2) \,  & \qquad \ell \in \mathbb{Z}\backslash \{0\,,-2\}\,  \\[3ex]
         \displaystyle \sum_{i\geq -1} U^\ell_i(x, \mathbf{t})\varepsilon^i= \dfrac{1}{\varepsilon} \,U^{\ell}_{-1}(x, \mathbf{t}) + U^\ell_0(x, \mathbf{t})+ \mathcal{O}(\varepsilon) \, & \qquad \ell \in \{0\,,-2\}\,.
    \end{cases} 
\end{equation} }
We now proceed with the derivation of the equations for the terms of the expansion~\eqref{eq:continuum_var} at the orders~$\mathcal{O}(\varepsilon^{-1})$ and $\mathcal{O}(\varepsilon^{0})$. To enhance the readability of the equations below, we introduce a simplified notation so that all quantities of interest depend on one index only. 
Let us set
\begin{equation}\label{eq:definition_scalar}
    v(x, \mathbf{t}) := U^0_0(x, \mathbf{t})\,,
\end{equation}
and introduce the infinite vector
\begin{equation}
\label{eq:upm}
    \mathbf{u}:=
    \mathbf{u}^- +
    \mathbf{u}^+\,,
\end{equation} 
where $\mathbf{u}^{-}$ has non-zero entries only in negative positions, and $\mathbf{u}^{+}$ has non-zero entries in non-negative positions, i.e.\ $ \mathbf{u}^-=\left(\,\dots, u^{-k}(x, \mathbf{t}), \dots, u^{-1}(x, \mathbf{t}),0, 0,\dots\right)^{\top}$ and $ \mathbf{u}^+=\left(\,\dots,0,0,u^0(x, \mathbf{t}), \dots, u^k(x, \mathbf{t}), \dots\,\right)^{\top}$, such that
\begin{subequations}\label{eq:definition_variables} 
     \begin{align} 
           & u^\ell(x, \mathbf{t}) := U^\ell_0(x, \mathbf{t}) \,,\quad \text{for }\ell \in \mathbb{Z}\backslash \{0\,,-2\} \\[1ex]
           & u^{0}(x, \mathbf{t}) := U^0_{-1}(x, \mathbf{t})\,\\[1ex] 
          &  u^{-2}(x, \mathbf{t}) := U^{-2}_0(x, \mathbf{t}) + U^0_0(x, \mathbf{t})\,.
        \end{align}
\end{subequations}
We can now prove the following:  
\begin{theorem}\label{thm:chain_mixed} 
Let $\{U^{\ell}(x,\mathbf{t};\varepsilon)\}_{\ell \in \mathbb{Z}}$ be the interpolation functions defined in~\eqref{eq:continuum_limit}, with asymptotic expansion of the form~\eqref{eq:continuum_var}, and notations~\eqref{eq:definition_scalar} and~\eqref{eq:definition_variables}. Then, the Pfaff lattice equations~\eqref{pfaff_lattice} imply that
\begin{enumerate}[$(i)$]
\item 
    \begin{equation}\label{eq:constraint_thm_1}
        U^{-2}_{-1}(x, \mathbf{t})= -U^0_{-1}(x, \mathbf{t}) \,;
    \end{equation}
\item the vector $\mathbf{u}(x, \mathbf{t})$ satisfies the following quasilinear system of infinitely many PDEs
\begin{equation}
\label{eq:chain_system}
    \partial_{t_2} \mathbf{u} = A(\mathbf{u})\,\partial_x \mathbf{u}\,,
\end{equation}
where $A(\mathbf{u}) = \left(A_{ij} \right)_{i,j \in \mathbb{Z}}$ is the infinite matrix:
\begin{equation}
\label{eq:A+-}
    A:=A^{-}+A^{+}\,,
\end{equation}
with
\vspace*{-3ex}

{\small
\begin{equation*} \arraycolsep=4.8pt
A^-\!=\! \left(  \begin{array}{c c | c  c c c} 
    \ddots & \vdots & \vdots & \vdots & \vdots & \iddots \\[1ex]
    \cdots & \!\!\left(A^{-}\right)^{-2}_{-1} & \left(A^{-}\right)^{-2}_{0} & \left(A^{-}\right)^{-2}_{1} & 0 & \cdots \\[1.5ex]
    \cdots & \!\!\left(A^{-}\right)^{-1}_{-1} & \left(A^{-}\right)^{-1}_{0} &  \left(A^{-}\right)^{-1}_{1} & 0 & \cdots \\[1.5ex] \hline 
   \cdots & 0\Tstrut  & 0 & \!\!0 & 0 & \cdots \\[1ex]
   \iddots & \vdots &\vdots & \vdots & \vdots & \ddots
    \end{array}
    \right) ~~~
A^+\!=\!\left(  \begin{array}{c c | c  c c c} 
    \ddots & \vdots & \vdots & \vdots & \vdots & \iddots \\[.7ex]
    \cdots & ~0~ & 0 &  0 & 0 & \cdots \\[1ex] \hline 
   \cdots & 0  & \left(A^{+}\right)^0_{0}\Tstrut & \left(A^{+}\right)^{0}_{1} & 0 & \cdots \\[1.5ex]
   \cdots & 0 &\left(A^+\right)^{1}_{0} & \left(A^+\right)^{1}_{1} & \left(A^+\right)^{1}_{2} & \cdots \\[.7ex]
   \iddots & \vdots &\vdots & \vdots & \vdots & \ddots
    \end{array}
    \right),
\end{equation*} }

\vspace{1ex}

where the non-zero entries of $A^{-}$ are 
\begin{align*}
    &\left(A^{-}\right)^{-k}_0=\begin{cases}
    u^{-k} u^1+k \,u^{-(k+1)} -(k-2) u^{-(k-1)}\, \qquad &k > 2 \\[.5ex]
    u^{-2}u^{1}+ 2\,u^{-3}\, \hspace{27.7ex} &k=2 \\[.5ex] 
        u^{-1}u^{1}+ u^{-2} \, \hspace{34.2ex} &k=1
    \end{cases} \\[1.5ex]
    \begin{split} 
    &\left(A^{-}\right)^{-k}_1 = u^{-k}u^0,\, \hspace{5ex} k>0\,\\[1.5ex]
    &\left(A^{-}\right)^{-k}_{-(k+1)} = u^0, \hspace{5.3ex} k>0\, 
    \end{split} \hspace{15ex}
    \left(A^{-}\right)^{-k}_{-(k-1)} = \begin{cases}
        u^0\, \qquad &k>2\, \\[.5ex]
        2u^0\, \qquad &k=2\,. 
    \end{cases}
\end{align*}
and the non-zero entries of $A^{+}$ are
\begin{align*}
    \left(A^{+}\right)^k_0 &=\begin{cases}
        u^0\,u^1\, \hspace{34.2ex} &k=0 \\[.5ex]
        (k+1)u^{k+1}-(k-1)u^{k-1}-u^k u^1\, \qquad\qquad &k>0 
    \end{cases} \\[1.5ex]
    \left(A^{+}\right)^k_{1} &= 
        -u^ku^0,\, \quad k\ge 0  \hspace{6ex}
    \left(A^{+}\right)^k_{k+1} = 
        u^0, \, \quad k\ge 1 \,\hspace{6ex}
    \left(A^{+}\right)^k_{k-1} = 
        u^0, \, \quad k> 1 \,;
\end{align*}
\item the function $v(x, \mathbf{t})$, defined in~\eqref{eq:definition_scalar}, satisfies the following non-homogeneous quasilinear PDE
\begin{equation} \label{eq:outsider}
    \begin{split}
       \partial_{t_{2}}v&=  \partial_x\!\left( u^0\,u^{1} \, v\right) +  u^{0} \,\partial_x u^{-1} + u^0 \,\partial_x \!\left(u^0 \,U^1_{1} \right) \,,
    \end{split}
\end{equation}
where $u^{-1}$, $u^{0}$, $u^{1}$ are entries of the solution $\mathbf{u}$ of the system~\eqref{eq:chain_system} and  the function $U^1_1(x, \mathbf{t})$ is undetermined.

\end{enumerate} 
\end{theorem}
\begin{proofC} Using the definition~\eqref{eq:continuum_limit} together with the asymptotic expansions~\eqref{eq:continuum_var} into the Pfaff lattice equations~\eqref{pfaff_lattice} and expanding in Taylor series with respect to $\varepsilon$, we extract the equations at the leading and next-to-leading order. 
\begin{enumerate}[$(i)$]
\item  At the leading order $\mathcal{O}(\varepsilon^{-1})$, the equations~\eqref{pfaff_lattice} for $\ell\in \{-3,-2,-1,0\}$  yield the following set of conditions for the coefficients $U^{-2}_{-1}(x, \mathbf{t})$, $U^{0}_{-1}(x, \mathbf{t})$ and $U^1_0(x, \mathbf{t})$: 
    \begin{subequations}
\begin{align}
    &U^0_{-1}(x, \mathbf{t})\,\partial_x U^{-2}_{-1}(x, \mathbf{t}) = U^{-2}_{-1}(x, \mathbf{t})\,\partial_x U^{0}_{-1}(x, \mathbf{t}) \, \label{eq:epsilon_1_k3}  \\[1ex] 
   & \partial_{t_2} U^{-2}_{-1}(x, \mathbf{t})  = U^{-2}_{-1}(x, \mathbf{t}) \,\partial_x\!\left(  U^{1}_{0}(x, \mathbf{t})\, U^{0}_{-1}(x, \mathbf{t})   \right) \, \label{eq:epsilon_1_k2}  \\[1ex]
   & \partial_x\!\left( U_{-1}^{-2}(x, \mathbf{t})\,U_{-1}^0(x, \mathbf{t}) + (U^{0}_{-1}(x, \mathbf{t}))^2  \right)  =0  \, \label{eq:epsilon_1_k1}  \\[1ex]
    &\partial_{t_2} U^{0}_{-1}(x, \mathbf{t}) =   U^{0}_{-1}(x, \mathbf{t}) \,\partial_x\!\left(U^{1}_{0}(x, \mathbf{t})\,U^{0}_{-1}(x, \mathbf{t})\right) \,.\label{eq:epsilon_1_k0} 
\end{align}
\end{subequations}
Observing that, consistently with the initial conditions~\eqref{eq:cont_initial_datum_1} and~\eqref{eq:cont_initial_datum_2}, $U^{-2}_{-1}(x, \mathbf{t})$ {and} $U^{0}_{-1}(x, \mathbf{t})$ cannot be constant or identically vanishing, direct integration with respect to $x$ of the equation~\eqref{eq:epsilon_1_k3} gives
\begin{equation} \label{eq:constr_1}
    |U^{-2}_{-1}(x, \mathbf{t})| = g_1(t_{2},t_{4},\dots) |U^{0}_{-1}(x, \mathbf{t})|  \,
\end{equation}
with $g_1(t_2, t_4, \dots)$ a positive arbitrary function of its arguments.
Comparing equations~\eqref{eq:epsilon_1_k2} and~\eqref{eq:epsilon_1_k0}, we have
\begin{equation}
    U^{0}_{-1}(x, \mathbf{t}) \,\partial_{t_2} U^{-2}_{-1}(x, \mathbf{t}) =U^{-2}_{-1}(x, \mathbf{t})\,\partial_{t_2} U^{0}_{-1}(x, \mathbf{t})  \,,
\end{equation}
and then
\begin{equation} \label{eq:constr_2}
    |U^{-2}_{-1}(x, \mathbf{t})| = g_2(x,\, t_4,t_6,\dots) |U^{0}_{-1}(x, \mathbf{t})|  \,
\end{equation}
where $g_2(x, t_4, \dots)$ is also a positive arbitrary function.
Hence, equations~\eqref{eq:constr_1} and~\eqref{eq:constr_2} imply that
\[
g_1(t_{2},t_{4},\dots) \equiv g_2(x,\, t_4,t_6,\dots) 
=: g(t_{4},t_{6},\dots),
\]
i.e.\ $g_{1}$ and $g_{2}$ coincide and do not depend on $x$ and $t_{2}$.
Therefore
\begin{equation} \label{eq:constr_3}
    |U^{-2}_{-1}(x, \mathbf{t})|=g\big(t_4,t_6,\dots\big)\,|U^{0}_{-1}(x, \mathbf{t})| \,.
\end{equation}
Finally, integrating the constraint~\eqref{eq:epsilon_1_k1} with respect to\ $x$ we find 
\begin{equation}
\label{eq:constr_5}
    U^{-2}_{-1}(x, \mathbf{t})\,U^{0}_{-1}(x, \mathbf{t}) + \left( U^0_{-1}(x, \mathbf{t}) \right)^2 = f(t_{2},t_{4},\dots)\,.
\end{equation}
Combining~\eqref{eq:constr_3} and~\eqref{eq:constr_5}, we get
\begin{equation}\label{eq:constr_4}
    |U^0_{-1}(x, \mathbf{t})|^2 \left( 1+ g(t_4, t_6, \dots) \,\text{sgn}(U^{-2}_{-1}(x, \mathbf{t}))\, \text{sgn}(U^{0}_{-1}(x, \mathbf{t})) \right) = f(t_{2},t_{4},\dots) \,.
\end{equation}
Since $U^{0}_{-1}(x, \mathbf{t})$ depends on $x$, the above equation~\eqref{eq:constr_4} implies that simultaneously
\begin{align*}
f(t_{2},t_{4},\dots) &=0 \,  \\
 1+ g(t_4, t_6, \dots)\,\text{sgn}(U^{-2}_{-1}(x, \mathbf{t}))\, \text{sgn}(U^{0}_{-1}(x, \mathbf{t})) &=0, \, 
\end{align*}
Hence $g(t_4,\,t_6,\,\dots)$ can take values $\pm 1$ only. Moreover, consistently with~\eqref{eq:constr_2}, $g(t_4,\,t_6,\,\dots)> 0$, therefore $g(t_4,\,t_6,\,\dots)=1$. Finally, we have \[
U^{-2}_{-1}(x, \mathbf{t}) = - U^{0}_{-1}(x, \mathbf{t}),
\] 
which is consistent with the initial datum~\eqref{eq:cont_initial_datum_1} and~\eqref{eq:cont_initial_datum_2}.
\item  A direct calculation of conditions at the order $\mathcal{O}(\varepsilon^0)$, using definitions~\eqref{eq:upm} and~\eqref{eq:definition_variables}, leads to the closed system of quasilinear PDEs~\eqref{eq:chain_system}.
\item Conditions at the order $\mathcal{O}(\varepsilon^{0})$ also yield the quasilinear PDE~\eqref{eq:outsider} for the variable $v(x, \mathbf{t})$ defined in~\eqref{eq:definition_scalar}. This equation contains the function $U^{1}_{1}(x, \mathbf{t})$ which remains undetermined, being a contribution at the order $\mathcal{O}(\varepsilon)$. 
\end{enumerate}
This completes the proof.
\end{proofC}

We notice that the singular expansion in $\varepsilon$~\eqref{eq:continuum_var} is responsible for a resonance between the terms at the order $\mathcal{O}(\varepsilon^{0})$ and higher order corrections $\mathcal{O}(\varepsilon)$, leading to the non-homogeneous equation~\eqref{eq:outsider} where $U^{1}_{1}(x, \mathbf{t})$ is undetermined. In fact, obtaining the evolution equation for $U^{1}_{1}(x, \mathbf{t})$ would require considering the equations at the order $\mathcal{O}(\varepsilon)$, {leading to further undetermined functions}. A similar resonance mechanism occurs as one proceeds to higher orders, so that the resulting system of PDEs is not closed. In order to obtain a determined closed system of PDEs at the order of interest, in this case~$\mathcal{O}(\varepsilon^{0})$, we fix the function $U^{1}_{1}(x, \mathbf{t})$ in a way that the asymptotic expansion~\eqref{eq:continuum_var} is consistent with the prescribed initial condition. This is achieved by the following:
\begin{corollary} \label{cor:quasilinear_system}
If
\begin{equation}
\label{eq:U11}
U^{1}_{1} = \frac{1}{2 u^{0}(x, \mathbf{t})} \, ,
\end{equation}
the equation~\eqref{eq:outsider} is determined and the asymptotic expansion~\eqref{eq:continuum_var} is consistent with the initial condition~\eqref{eq:initial_datum}. Moreover, the combined system of equations~\eqref{eq:chain_system} and~\eqref{eq:outsider} constitutes a closed system of quasilinear PDEs.
\end{corollary}
\begin{proofC}
Evaluating equation~\eqref{eq:U11} at $\mathbf{t} = \mathbf{0}$, we immediately see that it is compatible with the initial condition
\[
u^{0}(x,\mathbf{0}) = x, \qquad U^{1}_{1}(x,\mathbf{0}) = \frac{1}{2 x}.
\]
Hence, substituting the expression~\eqref{eq:U11} into the equation~\eqref{eq:outsider} the equations~\eqref{eq:chain_system} and~\eqref{eq:outsider} result into a closed system of quasilinear PDEs for the variables $\mathbf{u}$ and $v$.
\end{proofC}

Table~\ref{tab:entries} provides a summary of the notation for the Pfaff Lax matrix entries and the associated interpolating functions emerging at each step of the procedure described above leading to the chain of equations at the order $\mathcal{O}(\varepsilon^0)$, that is subject matter of the section below.

\begin{table}[h]
    \centering \footnotesize
    \begin{tblr}{
        hlines, vlines,
        colspec = {c c c c},
        row{1} = {font=\bfseries, c},
        rowsep = 0.3ex,
        colsep = 0pt,
    }
    & {~~Even Pfaff lattice~~} & {~~Initial datum (GOE)~~} & {~~Continuum limit~~} & {~~Chain at $O(\varepsilon^0)$~~} \\
    
    & \SetCell[r=4]{c}
    \begin{tabular}{@{}l@{}}
    $w^{\ell}_n(\mathbf{t})$, $n \in \mathbb{N}$, $\ell \in \mathbb{Z}$
    \end{tabular}
    
    & \SetCell[r=4]{m}
    \begin{tabular}{@{}l@{}}
    $w^\ell_n(\mathbf{0})=0$, for $\ell < -2$ \\[1ex]
    $w^{-2}_n(\mathbf{0})=-w^{0}_n(\mathbf{0})$
    \end{tabular}
    
    & \SetCell[r=4]{m}
    \begin{tabular}{@{}l@{}}
    ~~$U^\ell(x,\mathbf{t};\varepsilon):= w^\ell_{x/\varepsilon}(\mathbf{t})$, for $x = \varepsilon n$ ~~\\[2ex]
    ~~$U^\ell = \begin{cases}
    \sum\limits_{i \ge 0} U^\ell_i(x, \mathbf{t}) \, \varepsilon^i, & \ell \notin \{0,-2\} \\
    \sum\limits_{i \ge -1} U^\ell_i(x, \mathbf{t}) \, \varepsilon^i, & \ell \in \{0,-2\}~~
    \end{cases}$
    \end{tabular}
    
    & \SetCell[r=4]{m}
    \begin{tabular}{@{}l@{}}
    $~~u^\ell(x, \mathbf{t})$, $x \in \mathbb{R}$, $\ell \in \mathbb{Z}$ \\[0.5ex]
    $~~u^\ell := U^\ell_0$ \quad $\ell \notin \{0,-2\}$~~ \\[0.5ex]
    $~~u^0 := U^0_{-1}$ \quad $v := U^0_0$ \\[0.5ex]
    $~~u^{-2} := U^{-2}_0 + U^0_0$
    \end{tabular} \\
    & & & \\
    & & & \\
    & & & \\
    \end{tblr}
    \caption{{Summary of the notations adopted for order parameters of the Pfaff lattice, their initial conditions and asymptotic expansions for the associated interpolating functions.}}
    \label{tab:entries}
\end{table}

\subsection{Integrable chains of hydrodynamic type}\label{sec:integrability_of_hydro_chain}
In this section we focus on the infinite hydrodynamic system~\eqref{eq:chain_system}. We prove that this is an integrable hydrodynamic chain according to the definition introduced in~\cite{chain}. We now recall some  definitions given in~\cite{chain} that are key to state the main result of this section (i.e.\ Main Result \ref{res:reduced_pfaff_continuum_limit}).
\begin{definition}[Hydrodynamic chain]
\label{def:hydro_chain}
A chain of hydrodynamic type is an infinite system of quasilinear PDEs of hydrodynamic type of the form
\begin{equation}\label{eq:def_chain}
    \partial_{t} \mathbf{v} = K(\mathbf{v})\,\partial_x \mathbf{v}\,,
\end{equation} 
where $\mathbf{v}\!=\!(\dots,v^{-1},v^{0},v^{1},\dots)^{\top}$ and $K(\mathbf{v}) = (K^{i}_{j}(\mathbf{v}))_{i,j\in \mathbb{Z}}$ is an infinite matrix, with elements depending on the entries of $\mathbf{v}$ only. The  system~\eqref{eq:def_chain} is said to be a chain of class $C$ if
\begin{enumerate}
    \item Each row of $K(\mathbf{v})$ contains a finite number of non-zero elements;
    \item Each element of $K(\mathbf{v})$ depends on a finite number of variables $v^i$.
\end{enumerate}
\end{definition}
{The} above definition is a straightforward generalisation of the definition of hydrodynamic chain given in~\cite{chain} where the matrix $K(\mathbf{v})$ is semi-infinite.
\begin{definition}[Haantjes tensor]
Given an infinite matrix $K(\mathbf{v})$ associated with a hydrodynamic chain of class $C$ of the form~\eqref{eq:def_chain}, the Haantjes tensor ${\cal H}^i_{jk}$ of $K(\mathbf{v})$ is defined as follows
\begin{equation}
\label{def:Hantjes}
 {\cal H}^i_{jk} := {\cal N}^i_{pr}\,K^p_j\, K^r_k-{\cal N}^p_{jr}\,K^i_p\, K^r_k -{\cal N}^p_{rk}\, K^i_p\, K^r_j + {\cal N}^p_{jk}\, K^i_r\, K^r_p\,, 
\end{equation}
where ${\cal N}^i_{jk}$ are the components of the Nijenhuis tensor
\begin{equation}
    {\cal N}^i_{jk}:= K^p_j\, \partial_p K^i_k -K^p_k\, \partial_p K^i_j - K^i_p\left(\partial_j K^p_k - \partial_k K^p_j\right), 
    \label{eq:Nij}
\end{equation}
with the notation $\partial_\ell := \partial_{v^{\ell}}$, and summation over repeated indices is understood.
\end{definition}
\begin{definition}[Diagonalisability]
\label{def:haantjes_chain}
A hydrodynamic chain of class $C$ of the form~\eqref{eq:def_chain} is said to be diagonalisable if all components of the Haantjes tensor ${\cal H}^i_{jk}(\mathbf{v})$ of $K(\mathbf{v})$ vanish identically. 
\end{definition}\noindent
In order to define the integrability of a hydrodynamic chain of the form~\eqref{eq:def_chain} we look for solutions
\begin{equation}
    \mathbf{v} = \mathbf{v}\!\left(r^1, \dots, r^N\right), 
    \label{eq:reduction}
\end{equation}
where $\mathbf{v}$ depends on $x$ and $\mathbf{t}$ via the $N$ variables $r^i = r^i(x, \mathbf{t})$ {with upper indices $i=1,\dots,N$}. These are referred to as Riemann invariants, and satisfy the system of diagonal equations
\begin{equation}
    \partial_{t}\, r^i = \lambda^i \! \left(r^1, \dots, r^N\right)\partial_x r^i\,,
    \label{eq:Riemann}
\end{equation}
where $\lambda^i\left(r^1,\dots, r^N\right)$ are the characteristic speeds. We assume that the characteristic speeds satisfy the system of equations
\begin{equation}
   \partial_k \! \left( \frac{\partial_j \lambda^i}{\lambda^j - \lambda^i}\right) =  \partial_j \!\left( \frac{\partial_k \lambda^i}{\lambda^k - \lambda^i}\right) \;\;\;\;i\neq j \neq k\,,
   \label{eq:semiham}
\end{equation}
where $\partial_k := \partial_{r^k}$ denotes differentiation with respect to the $k$-th Riemann invariant. A system for which the characteristic speeds fulfill the condition~\eqref{eq:semiham} is said to be semi-Hamiltonian~\cite{Tsarev_1985}. Conditions~\eqref{eq:semiham} ensure that equations~\eqref{eq:Riemann} constitute a system of conservation laws~\cite{Sevennec} and it is solvable by the generalised hodograph method~\cite{Tsarev_1990, Dubrovin_Novikov}. 
Taking into account~\eqref{eq:reduction}-\eqref{eq:semiham}, the system~\eqref{eq:def_chain} implies
\begin{equation}\label{eq:recursive_for_GT}
    \lambda^i \, \partial_i v^k = K^k_{\ell}(\mathbf{v})\, \partial_i v^{\ell}\,, \qquad i = 1, \dots, N\,, 
\end{equation}
where sum over the index $\ell$ is understood, whereas $i$ is fixed. 
As discussed in~\cite{chain}, the analysis of the compatibility conditions of the system of the form \eqref{eq:recursive_for_GT} leads to an overdetermined system of equations referred to as Gibbons-Tsarev system. Consistently with the definition below, when the Gibbons-Tsarev system is automatically in involution, the sought hydrodynamic reductions exist in any number of components, and the associated hydrodynamic chain of class $C$ is said to be integrable.

\begin{definition}[Integrability in the sense of Ferapontov~\cite{chain}]\label{def:int_chain}
    A hydrodynamic chain of class $C$ is integrable if it admits $N$-component reductions of the form~\eqref{eq:reduction} for arbitrary $N$.
\end{definition}
We emphasise that the diagonalisability of the chain is a necessary (but not sufficient) condition for integrability in the sense of Ferapontov, as pointed out in~\cite{chain}.

Let us now  consider the system of infinitely many PDEs~\eqref{eq:chain_system}, which is indeed a chain of hydrodynamic type, and study its integrability according to  Definition~\ref{def:int_chain}. The evolution equations for the variables $u^{-k}(x, \mathbf{t})$ with $-k < 0$ are written explicitly as follows
\begin{equation}
\begin{aligned} 
    \partial_{t_2} u^{-k} &=\big(k \,u^{-(k+1)} -(k-2) u^{-(k-1)}+u^{-k} u^1 \big)\partial_x u^0 + u^0\, u^{-k}\,\partial_x u^1 
    \\[1ex]  
    &~~+ u^0 \big(\partial_x u^{-(k-1)} + \partial_x u^{-(k+1)} \big), \,  \hspace{20ex} -k < - 2  \\[1ex] 
     \partial_{t_2} u^{-2}&= \big( u^{-2}u^{1}+ 2\,u^{-3}  \big) \partial_x u^{0} +  u^{0}\,u^{-2} \, \partial_x u^{1} +  u^{0} \,\partial_x u^{-3} + 2 u^{0} \,\partial_x u^{-1}\,  \\[1ex] 
      \partial_{t_2} u^{-1}&= \big( u^{-1}u^{1}+ u^{-2}   \big) \partial_x u^{0}  +  u^{0}u^{-1}\, \partial_x u^{1} + u^{0} \partial_x u^{-2}  \,,   
      \label{eq:chainnegative}
\end{aligned}
\end{equation}
while the equations for $u^k(x, \mathbf{t})$ with $k\geq 0$ read as
\begin{equation}
    \label{eq:chainpositive}
    \begin{aligned}
    \partial_{t_2} u^{0}&=  u^{0}u^{1} \, \partial_x u^{0} + \left( u^0\right)^2 \partial_x u^{1}\,  \\[1ex]
    \partial_{t_2} u^1 &= \big(2u^2 - (u^1)^2\big)\partial_x u^0 - u^0 u^1 \,\partial_x u^1 + u^0 \,\partial_x u^2 \, \\[1ex]
    \partial_{t_2} u^k &= \big((k+1) u^{k+1} - (k-1) u^{k-1} - u^k u^1 \big) \partial_x u^0 -u^0 u^k \,\partial_x u^1\\[1ex]
      &~~ 
    + u^0 (\partial_x u^{k+1} + \partial_x u^{k-1} ), \hspace{30ex}  k > 1 \,. 
    \end{aligned}
    \end{equation}
We observe that for $k\neq 0$, each equation in~\eqref{eq:chainnegative} and~\eqref{eq:chainpositive} contains only the nearest neighbours $u^{k+1}$, $u^{k-1}$ and the variables $u^0$ and $u^1$. Moreover, we note that the evolution equation for $u^0$ does not contain $u^{-1}$, therefore the non-negative half of the chain can be solved independently of the negative one.

\begin{remark}
We recall that in~\cite{benassi2021symmetric} we derived an alternative continuum limit for the chain~\eqref{pfaff_lattice} where the interpolating functions, say~$\{v^\ell(x; \mathbf{t})\}_{\ell\in \mathbb{Z}}$ 
admit a non-singular asymptotic power series with the same leading-order behaviour in $\varepsilon$. Up to the rescaling $t=\varepsilon\, t_2$, 
the leading order of the expansion in $\varepsilon \to 0$ reads as~\cite[eq.~(2.1)]{benassi2021symmetric}

\vspace*{-2ex}

\begin{equation} \label{eq:old_chain}
\begin{split} 
\hspace*{-1ex}
\partial_{t} v^{-k} &=\big( k v^{-(k+1)} -(k-2) v^{-(k-1)} + v^{1} v^{-k} \big) \partial_x v^{0} + v^{0} v^{-k}\, \partial_x v^{1} 
\\[1ex]
&~~+ v^{0}\big(\partial_x v^{-(k+1)} + \partial_x v^{-(k-1)}\big), \hspace{20ex}  -k < 0,
\\[.5ex]
\partial_{t} v^{0} &=v^{0} v^{1}\, \partial_x v^{0} + \left( v^{0} \right)^{\!2} \partial_x v^{1} + v^{0} \,\partial_x v^{-1}  \\[1ex] 
\partial_{t} v^{1} &=\big( 2 v^{2} - \big( v^{1} \big)^{\!2} \big) \partial_x v^{0} - v^{0} v^{1}\, \partial_x v^{1} + v^{0} \,\partial_x v^{2} \\[1.2ex]
\partial_{t} v^{k} &=\big( (k+1) v^{k+1} - (k-1) v^{k-1} - v^{1} v^{k} \big) \partial_x v^{0} - v^{0} v^{k}\, \partial_x v^{1} 
\\[1ex]
&~~+v^{0}(\partial_x v^{k-1} + \partial_x v^{k+1}), \hspace{27.8ex}   k> 1.
\end{split}
\end{equation}
\noindent 
Interestingly, in this limit the non-negative part of the chain does not decouple from the negative one unlike in~\eqref{eq:chainnegative}-\eqref{eq:chainpositive}.
\end{remark}
We now prove the necessary condition for integrability as per~\cite{chain}. In particular, we have the following:
\begin{proposition}\label{prop:Haantjes} The Haantjes tensor of the hydrodynamic chain~\eqref{eq:chain_system} identically vanishes.
\end{proposition}
\begin{proofC}
First, let us note that, by definition, the Nijenhuis tensor~\eqref{eq:Nij} is antisymmetric in its lower indices, i.e.\ $\mathcal{N}^i_{jk} = - \mathcal{N}^i_{kj}$ for all $i,\,j,\,k \in \mathbb{Z}$. It is straightforward to show via a direct calculation that $\mathcal{N}^0_{jk} = 0$ for all $j$ and $k$. 
If $i\neq 0$, given the explicit form of the hydrodynamic chain~\eqref{eq:chain_system}, by direct inspection, we have that the only non-zero elements of the Nijenhuis tensor are
$$
\mathcal{N}^{i}_{0\,1},\;\;\; \mathcal{N}^i_{0\,i},\;\;\;\mathcal{N}^i_{0\,i\pm 1},\;\;\;\mathcal{N}^i_{1\,i\pm1},$$
and their counterparts with swapped lower indices. The explicit expressions for generic values of $i$ can be found in Appendix \ref{app:Nij}. 
From the form of $\mathcal{N}^i_{jk}$ and of the  matrix $A(\mathbf{u})$ for the chain~\eqref{eq:chain_system}, it follows that for a given $i$ the components of the antisymmetric tensor $\mathcal{H}^i_{jk}$ are evidently zero apart from the cases $j,\,k\in\{0,\,1,\,2,\,3,\,i,\,i\pm1,\,i\pm2,\, i\pm 3\}$. In these cases, {a direct, although cumbersome, calculation gives} $\mathcal{H}^i_{jk}=0$ for all listed values.
\end{proofC}

We now establish the integrability of the chain by verifying the existence of hydrodynamic reductions through the compatibility of the associated Gibbons-Tsarev system. In particular, we have the following:
\begin{theorem}\label{thm:SMEintegrability} The hydrodynamic chain~\eqref{eq:chain_system} is integrable in the sense of Ferapontov.
\end{theorem}
\begin{proofC}
In order to check Definition \ref{def:int_chain}, we look for solutions of the form $\mathbf{u}=\mathbf{u}(r^1, \dots, r^N)$ for arbitrary $N$, where the Riemann invariants $r^i$ satisfy the equations~\eqref{eq:Riemann} with the condition~\eqref{eq:semiham}.
Let us consider the equation~\eqref{eq:recursive_for_GT} for $K(\mathbf{u}) = A(\mathbf{u})$ defined for the system~\eqref{eq:chain_system}. In particular, we can express the derivatives $\partial_i u^\ell$, for, say, $\ell\in \{-2,\, 1,\,2,\,3\}$, as follows
\begin{subequations}\label{eq:diu}
    \begin{align}
      \partial_i u^{-2} &= \frac{ \lambda^i}{u^0}\,\partial_i u^{-1}-\frac{u^{-2}u^0 + u^{-1}\lambda^i}{(u^0)^2}\,\partial_i u^0\,\\[.5ex]
        \partial_i u^1 &=\frac{\lambda^i - u^1 u^0}{(u^0)^2}\,\partial_i u^0\, \label{eq:diu1}\\[.5ex]
        \partial_i u^2 &=\frac{(\lambda^i)^2 - 2 u^2 (u^0)^2}{(u^0)^3}\,\partial_i u^0\,\\[.5ex]
        \partial_i u^3 &= \frac{(u^0)^3\left(2u^1-3u^3\right)-(u^0)^2\left(1+u^2\right)\lambda^i+(\lambda^i)^3}{(u^0)^4}\,\,\partial_i u^0\,.
    \end{align}
\end{subequations}
The compatibility conditions
\begin{equation}
    \partial_i \partial_j u^{1} = \partial_j \partial_i u^{1},\;\;\;\;\;  \partial_i \partial_j u^{2} = \partial_j \partial_i u^{2},\;\;\;\;\;  \partial_i \partial_j u^{3} = \partial_j \partial_i u^{3}
\end{equation}
give
\begin{subequations}
\label{eq:GT_SMEpositive}
\begin{align}
    \partial_j \lambda^i &= \frac{4 (u^0)^2 - \lambda^i\lambda^j}{u^0 (\lambda^i - \lambda^j)}\partial_j u^0\\[.5ex] 
   \partial_i\partial_j u^0 &= \frac{-8 (u^0)^2 + (\lambda^i)^2 + (\lambda^j)^2}{
u^0 (\lambda^i - \lambda^j)^2}\partial_i u^0 \partial_j u^0, 
    \end{align}
\end{subequations}
for $i\neq j$. Moreover, the compatibility conditions $\partial_i\partial_j u^{-2} = \partial_j \partial_i u^{-2}$ imply
\begin{equation}
\label{eq:GT_SMEextra}
\begin{aligned}
\partial_i\partial_j u^{-1} &= \frac{-4 u^0 +(\lambda^i)^2}{u^0(\lambda^i - \lambda^j)^2}\partial_i u^0 \partial_j u^{-1} +\frac{-4 u^0 +(\lambda^j)^2}{u^0(\lambda^i - \lambda^j)^2}\partial_j u^{0} \partial_i u^{-1}   \,.
    \end{aligned}
\end{equation}
Equations~\eqref{eq:GT_SMEpositive} and~\eqref{eq:GT_SMEextra} constitute the sought Gibbons-Tsarev system for the chain~\eqref{eq:chain_system}. We also note that all the other compatibility conditions of the form $\partial_i\partial_j u^{k} = \partial_j \partial_i u^{k}$ are automatically satisfied modulo equations~\eqref{eq:GT_SMEpositive} and~\eqref{eq:GT_SMEextra}.
One can verify by direct calculation that equations~\eqref{eq:GT_SMEpositive} and~\eqref{eq:GT_SMEextra} are an involution, i.e.
\begin{equation}
    \partial_k \partial_j \lambda^i = \partial_j \partial_k \lambda^i, \;\;\;\;\partial_k \partial_i\partial_j u^0 = \partial_i \partial_k \partial_j u^0,\;\;\;\;\partial_k \partial_i\partial_j u^{-1} = \partial_i \partial_k \partial_j u^{-1},\;\;\;\;\;i\neq j\neq k,
\end{equation}
for all permutations of the indices $i,j,k\in\{1,\dots,N\}$, modulo equations~\eqref{eq:GT_SMEpositive} and~\eqref{eq:GT_SMEextra}. This completes the proof.
\end{proofC}

\subsection{On the classification of the non-negative chain}\label{sec:OS}

As mentioned in the previous section, the components of the non-negative part of the chain~\eqref{eq:chain_system}, associated with the evolution of the variables $u^\ell$ with~$\ell\geq 0$, evolve independently of those corresponding to $\ell<0$, as it is apparent from the explicit form of the equations~\eqref{eq:chainnegative} and~\eqref{eq:chainpositive}. 

In this section we focus on the semi-infinite chain with non-negative indices, i.e.\ the system~\eqref{eq:chainpositive}. In particular, we note that Proposition \ref{prop:Haantjes} and Theorem \ref{thm:SMEintegrability} hold for this system as well, which is therefore diagonalisable and integrable in the sense of Ferapontov. The corresponding Gibbons-Tsarev systems is given by the equations~\eqref{eq:GT_SMEpositive}.

We note that the form of the system \eqref{eq:chainpositive} suggests that $u^1$ plays a special role, similarly to $u^0$, as the evolution equations for $u^k$ feature both $u^0$ and $u^1$. However, only $u^{0}$ appears in the Gibbons-Tsarev system~\eqref{eq:GT_SMEpositive}. Below, we show that, in fact, the chain~\eqref{eq:chainpositive} can be recast by a simple change of variables into a form where $u^1$ is absorbed in the dependent variables, and only $u^0$ appears in all equations of the chain. Indeed, defining 
\begin{subequations}
\begin{align} 
\psi^0(x, \mathbf{t}) &:= u^0(x, \mathbf{t}) \\[.5ex]
\psi^k(x, \mathbf{t}) &:= u^k(x, \mathbf{t}) \,u^0(x, \mathbf{t}), \qquad k > 0\,, 
\end{align} 
\end{subequations}
the non-negative chain reads as
\begin{equation}
\label{eq:hydrochain_1seed}
\begin{aligned}
    \partial_{t_2} \psi^k &= k\left(\psi^{k+1}-\psi^{k-1}\right)\partial_x \psi^0 + k\,\psi^0 \left(\partial_x\psi^{k+1}+\partial_x\psi^{k-1}\right) \qquad k\geq 0.
\end{aligned}
\end{equation}
For the sake of completeness, we include here also the negative chain~\eqref{eq:chainnegative} in terms of the variables 
\[
\psi^{-k}(x, \mathbf{t}) = u^{-k}(x, \mathbf{t}) \,u^0(x, \mathbf{t}) \qquad k > 0,
\]
that is
\begin{equation*}
\begin{aligned}
    \partial_{t_2}\psi^{-1} &= 2\psi^{-1} \,\partial_x \psi^1_x + \psi^ 0\, \partial_x \psi^{-2}\\[.7ex]
    \partial_{t_2} \psi^{-2} &= k\left(\psi^{-3}-2\psi^{-1}\right)\partial_x \psi^0 +2 \psi^{-2} \partial_x \psi^1 + \psi^0 \left(\partial_x\psi^{-3}+2 \partial_x\psi^{-1}\right) \\[.5ex]
     \partial_{t_2} \psi^{-k} &= (k-1)\left(\psi^{-(k+1)}-\psi^{-(k-1)}\right)\partial_x \psi^0 +2 \psi^{-k}\, \partial_x \psi^1+ \psi^0 \left(\partial_x\psi^{-(k+1)}+\partial_x\psi^{-(k-1)}\right)\,, \;\;\;\;k > 2.
    \end{aligned}
\end{equation*}
We notice that both $\psi^0$ and $\psi^1$ appear in all equations of the negative chain. However, since the non-negative chain is partially decoupled from the negative one, in principle $\psi^0$ and $\psi^1$ can be thought as fixed functions of $(x, \mathbf{t})$ obtained from the solution of~\eqref{eq:hydrochain_1seed}.

\begin{remark}
    It is worth comparing the Gibbons-Tsarev system~\eqref{eq:GT_SMEpositive}-\eqref{eq:GT_SMEextra} for the chain~\eqref{eq:chainnegative}-\eqref{eq:chainpositive} and the one for the chain~\eqref{eq:old_chain} obtained in~\cite{benassi2021symmetric} for the ``na\"{i}ve" continuum limit, that is 
\begin{subequations}\label{eq:GT_system_old}
\begin{align} 
\partial _{j} \lambda ^{i}&= \frac{4 (v^{0})^{2} - \lambda ^{i} \lambda ^{j}}{v^{0} (\lambda ^{i} - \lambda ^{j})} \,\partial _{j} v^{0} \\[.5ex] 
\partial _{i} \partial _{j} v^{0}&= \frac{(\lambda ^{i})^{2} + (\lambda ^{j})^{2} - 8 (v^{0})^{2}}{v^{0} (\lambda ^{i} - \lambda ^{j})^{2}} \, \partial _{i} v^{0} \partial _{j} v^{0} \\[.5ex]  
\partial _{i} \partial _{j} v^{1}&= - \frac{(\lambda ^{j} - 2 \lambda ^{i}) \lambda ^{j} + 4 (v^{0})^{2}}{v^{0} (\lambda ^{i} - \lambda ^{j})^{2}} \partial _{i} v^{0} \partial _{j} v^{1} - \frac{(\lambda ^{i} - 2 \lambda ^{j}) \lambda ^{i} + 4 (v^{0})^{2}}{v^{0} (\lambda ^{i} - \lambda ^{j})^{2}} \,\partial _{j} v^{0} \partial _{i} v^{1}, \label{eq:GT_system_old_last}
\end{align} 
\end{subequations}
(see~\cite[eq.\ (2.12)]{benassi2021symmetric}). In particular, the first two equations in~\eqref{eq:GT_system_old} coincide with the equations~\eqref{eq:GT_SMEpositive}, while equations~\eqref{eq:GT_system_old_last} and~\eqref{eq:GT_SMEextra}, although different, share the same form, i.e.\ they are both semi-linear in {$v^{1}$ and $u^{-1}$}, respectively.
\end{remark}

The rest of this section is devoted to the classification of the hydrodynamic chain~\eqref{eq:hydrochain_1seed} according to the framework introduced by Odesski and Sokolov in~\cite{Odesskii_2010}.
Their classification leads to six normal forms of Gibbons-Tsarev systems and corresponding classes. Systems in the same class are equivalent up to  transformations of the form
\begin{equation}
u^k\rightarrow f^k(u^0, u^1, \dots, u^k)\qquad  \text{with} \quad \frac{\partial f^k}{\partial u^k} \neq 0\,.
\end{equation}
Let us briefly summarise the main necessary steps for the classification. Let $\boldsymbol\psi\!=\!(\psi^0, \psi^1, \dots)^{\top}$ denote the semi-infinite vector of the dependent variables of the integrable hydrodynamic chain of the form 
\begin{equation}
    \partial_t\boldsymbol{\psi} = B(\boldsymbol{\psi})\,\partial_x \boldsymbol{\psi}\,, 
\end{equation} 
where the matrix $B(\boldsymbol{\psi})$ is such that $B^{i}_{j}(\boldsymbol\psi) = 0$ if $j> i+1$, $B^{i}_{i+1} (\boldsymbol{\psi}) \neq 0$ and $B^{i}_{j}(\boldsymbol\psi)$ depend on $\psi^0$, \dots, $\psi^{j+1}$.
In~\cite{Odesskii_2010}, the authors show that the general form of a Gibbons-Tsarev system for an integrable chain of the form above is

\begin{subequations}\label{eq:ROdesskii}
\begin{align}
\partial_i \lambda^j &= \left(\frac{R(\lambda^j)}{\lambda^i - \lambda^j} +\lambda^i \Big( (\lambda^j)^2\,z_4 +  \lambda^j\,z_5 + z_6\Big) +  (\lambda^j)^3\,z_4 +  (\lambda^j)^2\, z_3 + \lambda^j\,z_7 + z_8 \right)\partial_i \psi^0
\\[1ex]
    \begin{split} 
    \partial_i \partial_j \psi^0 &= \left(\frac{2\,  (\lambda^i)^2(\lambda^j)^2\,z_4 + \lambda^i\, \lambda^j (\lambda^i + \lambda^j)\,z_3+ ((\lambda^i)^2 + (\lambda^j)^2)\,z_2+(\li + \lj)\,z_1+2 z_0}{(\li-\lj)^2} + z_9\right)\partial_i \psi^0 \partial_j \psi^0,
    \end{split} 
\end{align}
\end{subequations}

\noindent
where 
\[
R(\lambda) =  \lambda^4\,z_4 +  \lambda^3\,z_3 +  \lambda^2\,z_2 + \lambda\,z_1 + z_0,
\]
and the coefficients $z_i=z_i(\psi^0, \,\psi^1)$, with $i\in\{1,\dots,9\}$, fulfil suitable differential conditions. 

Under a change of variables and reparametrisations of the characteristic speeds, the polynomial~$R(\lambda)$ can be reduced to one of six canonical forms~\cite[Prop.\ 1]{Odesskii_2010}. The fourth of them is $R(\lambda) = \lambda$, and the following proposition shows that the Gibbons-Tsarev system associated with the hydrodynamic chain~\eqref{eq:hydrochain_1seed} belongs to this canonical form.

\begin{proposition}\label{prop:classification_chain}
    The hydrodynamic chain~\eqref{eq:hydrochain_1seed} belongs to the {\textnormal{Case 4}} of the classification in Proposition 1 from \textup{\cite{Odesskii_2010}}. Hence, the normal form of the Gibbons-Tsarev system~\eqref{eq:GT_SMEpositive} corresponds to the case $R(\lambda) = \lambda$.
\end{proposition}
\begin{proofC}
The proof closely follows Section 4 of~\cite{Odesskii_2010}, and we refer readers to this paper for further details.
Let us first note that the equation~\eqref{eq:diu1} in terms of the variables $\psi^k$, reads as
\begin{equation}
\partial_i \psi^1 = \frac{\lambda^i}{\psi^0}\,\partial_i \psi^0.
\label{eq:dipsi1}
\end{equation}
A preliminary step in~\cite{Odesskii_2010} involves normalising the chain via a change of variables $\psi^k\mapsto  \hat{\psi}^k $ such that, e.g.
\begin{equation}
\partial_i \hat{\psi}^1 = \li \, \partial_i \hat{\psi}^0.  
\end{equation}
Given~\eqref{eq:dipsi1}, it is straightforward to check that the required change of variables is
\begin{equation}\label{eq:transf_psihat}
  \psi^0 = \exp(\hat{\psi}^0), \qquad \psi^k = \hat{\psi}^k,\, \quad k>0.   
\end{equation}
Hence, in these new variables, the Gibbons-Tsarev system~\eqref{eq:GT_SMEpositive} reads as
\begin{subequations} \label{eq:GTnewseed}
    \begin{align}
        \partial_i \lambda^j &= \frac{- 4\exp(2\hat{\psi}^0) +\lambda^i\lambda^j}{\lambda^i - \lambda^j} \,\partial_i \hat{\psi}^0 \label{eq:GTnewseed1}\\[.5ex]
        \partial_i \partial_j \hat{\psi}^0 &= \frac{- 8\exp(2\hat{\psi}^0)  + 2\lambda^i \lambda^j}{(\lambda^i-\lambda^j)^2}\,\partial_i \hat{\psi}^0\, \partial_j \hat{\psi}^0\,. 
    \end{align}
\end{subequations} 
Comparing~\eqref{eq:GTnewseed} and~\eqref{eq:ROdesskii}, we have 
\begin{equation}
    R(\lambda) = \lambda^2 -4 \exp({2 \hat{\psi}^0}),
    \label{eq:Rquad}
\end{equation}
i.e.\ $R(\lambda)$ is quadratic -- as noted above $R(\lambda)$ is a polynomial in $\lambda$ or degree up to four. In this case, the coefficients $z_i(\hat{\psi}^0,\hat{\psi}^1)$ in~\eqref{eq:ROdesskii} are
\begin{equation*}
\begin{aligned}
    z_0 &= -4\exp(2\hat{\psi}^0),\qquad z_2 = 1,\qquad z_7=1,\qquad z_9 =1,\qquad z_1 = z_3 = z_4 = z_5 = z_6 = z_8 =0.
\end{aligned}    
\end{equation*} 
In order to reduce the polynomial $R(\lambda)$ to its canonical form, we apply the following reparametrisation of the characteristic speeds, as detailed in~\cite[eq.\ (4.8)]{Odesskii_2010}:
\begin{equation}\label{eq:unimodular_transf}
    \lambda^i = \frac{a\bar{\lambda}^i + b}{\bar{\lambda}^i - \xi},
\end{equation}
where $a$, $b$, $\xi$ are in principle functions of both $\hat{\psi}^0$ and $\hat{\psi^1}$, but in our case they are assumed to depend on~$\hat{\psi}^0$ only. These functional parameters can be suitably chosen to reduce $R(\lambda)$ to a normal form.
Differentiating the expression~\eqref{eq:unimodular_transf} to calculate $\partial_i \,\bar{\lambda}^j$, and substituting into equation~\eqref{eq:GTnewseed1}, we obtain
\begin{equation}
\label{eq:lambabar}
\partial_i\, \bar{\lambda}^j = \left((\bar{\lambda}^j - \xi)^2 \frac{R(\bar{\lambda}^j)}{\bar{\lambda^i}-\bar{\lambda^j}}
+\bli \big( (\blj)^2\bar{z}_4 +  \blj\,\bar{z}_5 + \bar{z}_6 \big) + (\blj)^3\bar{z}_4  + (\blj)^2\bar{z}_3  + \blj\,\bar{z}_7 + \bar{z}_8 \right)\partial_i \hat{\psi}^0,
\end{equation}
with suitably defined coefficients $\bar{z}_j$.
Choosing $a=2\exp(\hat{\psi}^0)$, after some simplifications we find
{
\begin{equation}
    R(\bar{\lambda}^j)=\frac{b+2\exp(\hat{\psi}^0)\big(2 \bar{\lambda}^j - \xi\big) }{b+ 2\xi\exp(\hat{\psi}^0)},
\end{equation}
}
and
{
\begin{equation*}
\begin{split}
\bar{z}_3 &= 4\exp(\hat{\psi}^0),\;\;\;\;\bar{z}_7 = \frac{\partial b}{\partial \hat{\psi}^0} +  \bigg(\frac{\partial \xi}{\partial \hat{\psi^0}} - 3 \xi\bigg)2\exp(\hat{\psi}^0),\;\;\;\;\bar{z}_8 =  \frac{\partial \xi}{\partial \hat{\psi^0}}\,b - \frac{\partial b}{\partial \hat{\psi}^0} \,\xi + 2\xi^2 \exp(\hat{\psi}^0)\, ,\\
\bar{z}_4 &= \bar{z}_5 = \bar{z}_6 =0.
\end{split}
\end{equation*}
}
Finally, choosing 
{$b = 2\xi\exp(\hat{\psi}^0) $ and $\xi =1$ } we obtain the normal form $R(\bar{\lambda}) = \bar{\lambda}$.

\end{proofC}

\section{Even reduced Pfaff lattice: further reduction }\label{sec:reductions}
In this final section we construct a further reduction of the Pfaff lattice~\eqref{pfaff_lattice} and its continuum limit given by the hydrodynamic chain~\eqref{eq:chain_system}, incorporating the structure of the initial conditions, respectively,~\eqref{eq:w_initial} and~\eqref{eq:initial_datum}. We show that in both cases a further reduction exists giving, remarkably, the very same one-dimensional semi-discrete integrable system: a one-dimensional  lattice. This constitutes Main Result~\ref{res:reduced_pfaff_new_system}.

\subsection{Discrete case}
\label{sec:discretecase}
In Section~\ref{sec:Pfaff_equations_and_initial_datum}, we have seen that the Pfaff Lax matrix possesses a particularly simple structure at~$\mathbf{t} = \mathbf{0}$, as all entries $w^{-k}_n(\mathbf{0})$ with $-k<-2$ vanish and  $w^{-1}_n(\mathbf{0})$ does not depend on $n$ (see  initial conditions~\eqref{eq:w_initial}). We now demonstrate that, if this structure is preserved also for $\mathbf{t} \neq \mathbf{0}$, the even reduced Pfaff lattice admits a further non-trivial reduction whose solution characterises the associated orthogonal ensemble in terms of a  one dimensional integrable chain. The new chain  involves a considerably lower number of dependent variables. In particular, the entries $w^0_n(\mathbf{t})$ and $w^{-2}_n(\mathbf{t})$ turn out to be proportional to $w^{-1}_n(\mathbf{t})$, and each~$w^k_n(\mathbf{t})$ is expressed in terms of~$w^k_1(\mathbf{t})$. We prove the following

{
\begin{theorem}\label{thm:reduction_pfaff} Let $ \{w^{\ell}_{n}(\mathbf{t})\}_{n \in \mathbb{N}}^{\ell \in \mathbb{Z}}$ be a solution of the Pfaff lattice~\eqref{pfaff_lattice}. 
\noindent 
If, for all $k>2$ and $n \in \mathbb{N}$, $w^{-k}_n(\mathbf{t}) \equiv 0$  and $w^{-1}_{n}(\mathbf{t}) = w^{-1}(\mathbf{t})$ -- i.e.\ $w^{-1}_{n}(\mathbf{t})$ does not depend on~$n$ -- there exists a solution of~\eqref{pfaff_lattice} that is compatible with the initial conditions~\eqref{eq:w_initial}, such that
\begin{subequations}\label{eq:entries_wn_reduced}
    \begin{align}
      w^0_n(\mathbf{t}) &= - w^{-2}_n(\mathbf{t}) = c_n\,w^{-1}(\mathbf{t}), \, \qquad c_n = \sqrt{2n(2n-1)}\, \label{eq:w0_red_def}\\[1ex]
        w^{k}_n(\mathbf{t}) &=\binom{n+k-1}{k}\prod_{\ell=1}^k\frac{c_{\ell}}{c_{n+k-\ell}}\,w^{k}_1(\mathbf{t}),\, \qquad k > 0\,,\label{eq:wk_red_def}
    \end{align}
\end{subequations}
and $w^{-1}(\mathbf{t})$ and $w^{k}_{1}(\mathbf{t})$ satisfy the following chain:
\begin{subequations}\label{eq:wk_evol_all}
    \begin{align}
        \partial_{t_2}w^{-1}&= (w^{-1})^2\left(\,c_1\,w^1_1\,\right) \,,\label{eq:wm1_evol}\\[1.75ex]
        \partial_{t_2}w^{k}_1&=w^{-1}\!\left(\, c_{k+1}\,w^{k+1}_1 - c_1\,w^1_1\,w^k_1 + \frac{k(c_1)^2-(c_k)^2}{c_k}\,{w^{k-1}_1} \,\right) , \qquad k > 0 \,. \label{eq:wk_evol} 
    \end{align}
\end{subequations}
Moreover, introducing the rescaled variables
\begin{equation}\label{eq:rescaled_w}
W^{-1}(\mathbf{t}) := w^{-1}(\mathbf{t})\,, \qquad  W^k(\mathbf{t}) := \frac{w^k_1(\mathbf{t})}{F_k},  \,\quad \text{with } \; F_k = \frac{2^k\,k!}{\sqrt{(2k)!}}\,, \quad k > 0 \,, 
\end{equation}
the system~\eqref{eq:wk_evol_all} reads as
\begin{subequations}\label{eq:system_red_W}
\begin{align}
    \partial_{t_2} W^{-1} &= 2(W^{-1})^2 \, W^1\, \\[1ex]
    \partial_{t_2} W^k &= 2\,W^{-1} \left( (k+1)W^{k+1} - W^1\,W^k - (k-1)W^{k-1} \right) \qquad k > 0 \,. 
\end{align}
\end{subequations}
\end{theorem}
\begin{remark}
    The reduction presented in this theorem can be found under sightly weaker hypotheses, namely assuming that $w^{-k}_n(\mathbf{t}) \equiv 0$ for all $k>2$, and proving that necessarily $w^{-1}_{n}(\mathbf{t})$ is independent of~$n$. However, the above formulation of the theorem allows a more concise proof whilst preserving the key points.
\end{remark}

\begin{proofC}
First of all we observe that under the assumptions 
\begin{equation}\label{eq:red_assumptions}
    w^{-1}_{n}(\mathbf{t}) = w^{-1}(\mathbf{t}),\, \qquad w^{-k}_{n}(\mathbf{t}) = 0, \quad  \text{ for } \; k >2\,, 
\end{equation}
valid at all $\mathbf{t}$, 
the Pfaff lattice equations~\eqref{pfaff_lattice} for $w^{-k}_n(\mathbf{t})$ with $k > 3$ are trivially satisfied. The first non-trivial conditions arise from the equation for~$w^{-3}_n(\mathbf{t})$, that is
\begin{equation}
    \label{eq:wm2_constraint}
    w^{-2}_{n+1} \,w^0_n - w^{-2}_{n}\, w^0_{n+1}=0\,.
\end{equation}
This constraint is consistent with the initial conditions on $w^{-2}_n(\mathbf{t})$ and $w^{0}_n(\mathbf{t})$ given in~\eqref{eq:w_initial}. Let us now consider the equations for~$w^{-2}_n(\mathbf{t})$ and~$w^{0}_n(\mathbf{t})$ in~\eqref{pfaff_lattice} together with the assumptions~\eqref{eq:red_assumptions}. We have
\begin{align}
&\partial_{t_2} w^{-2}_{n}= \frac{w^{-2}_n}{2}  \left(  w^0_{n+1}\, w^1_{n+1} - w^0_{n-1}\, w^1_{n-1} \right) \label{eq:wminus2_red} \\[.5ex]
&\partial_{t_2} w^{0}_{n}= \frac{w^{0}_n}{2} \left(  w^0_{n+1}\, w^1_{n+1} - w^0_{n-1}\, w^1_{n-1} \right),\label{eq:w0_red}
\end{align}
which imply that
\begin{equation}
   \frac{\partial_{t_2} w^{-2}_n}{w^{-2}_n}= \frac{\partial_{t_2} w^{0}_n}{w^{0}_n} \,. 
\end{equation}

\noindent
Integrating this expression with respect to $t_2$ we have 
\begin{equation}
    |w^{-2}_n(\mathbf{t})| = g_{n}(t_4,t_6, \dots)\,|w^0_n(\mathbf{t})|,
\label{eq:wm2_constraint_bis}
\end{equation}
with $g_{n}$ an arbitrary function of its arguments.
Recalling that, according to the initial datum, we have  $w^{-2}_n(\mathbf{0})= w^0_n(\mathbf{0})$,  compatibility with the initial condition~\eqref{eq:initial_datum}  implies that $g_n(\mathbf{0}) =1$. Thus, we can choose $g_n\equiv 1$~for all $(t_4, t_6, \dots)$\;. Therefore, consistently with the initial condition and the constraint~\eqref{eq:wm2_constraint}, we can impose the first identity of~\eqref{eq:w0_red_def}, i.e.\ 
\begin{equation*}
    w^{-2}_n(\mathbf{t}) = - w^0_n(\mathbf{t})\,.
\end{equation*} 
Let us now consider the equation for $w^{-1}(\mathbf{t})$ together with the assumptions~\eqref{eq:red_assumptions}, that is
\begin{equation} \label{eq:wminus1_red}
    \partial_{t_2} w^{-1} = w^{-1} \!\left( w^0_n\, w^1_n - w^0_{n-1}\, w^1_{n-1} \right).
\end{equation}
Since $w^{-1}(\mathbf{t})$ is independent of $n$, so is the difference appearing on the right hand side of equation~\eqref{eq:wminus1_red}. Using this fact and rewriting equation~\eqref{eq:w0_red} as follows (adding and subtracting the quantity $w^0_{n}\, w^1_{n}$ on the right hand side) 
\begin{equation}
    \partial_{t_2} w^{0}_n = \frac{w^0_n}{2} \left(  w^0_{n+1}\, w^1_{n+1} - w^0_{n}\, w^1_{n}+ w^0_{n}\, w^1_{n} -w^0_{n-1}\, w^1_{n-1} \right),
\end{equation}
it implies, together with the equation~\eqref{eq:wminus1_red}, that
\begin{equation} \label{eq:constraint_w0}
    \frac{\partial_{t_2} w^0_n}{w^0_n} = \frac{\partial_{t_2}w^{-1}}{w^{-1}} \,. 
\end{equation}
\noindent 
Integrating, we have
\begin{equation}
    |w^0_n(\mathbf{t})| = h_{n}(t_4,t_6,\dots)\,|w^{-1}(\mathbf{t})|\,.
\end{equation}
According to the initial condition $w^0_n(\mathbf{0})\!=\!c_n w^{-1}(\mathbf{0})$, with $c_n\!:=\!\sqrt{2n(2n-1)}$, we can consistently set~$h_{n}(t_4, t_6, \dots) = c_n$ and impose that, for any $(t_2, t_4, t_6, \dots$), the condition~\eqref{eq:w0_red_def} holds, i.e.\ 
\begin{equation*}
    w^0_n(\mathbf{t}) = c_n\,w^{-1}(\mathbf{t}) \,, \qquad c_n = \sqrt{2n(2n-1)}\,. 
\end{equation*}
The above expression shows that the dependence on all the couplings $\mathbf{t}$ for the entry $w^0_n(\mathbf{t})$ is completely encoded in~$w^{-1}(\mathbf{t})$, and the dependence on the index $n$ has the simple form of the coefficients $c_n$. 

Now, we focus on the constraint~\eqref{eq:wminus1_red}, and observe that it can be recast as follows
\begin{equation}
\label{eq:w01rec}
\begin{split}
    w^{0}_n\,w^{1}_{n} &= w^{0}_{n-1}\,w^{1}_{n-1} + \frac{\partial_{t_2} w^{-1}}{w^{-1}} \,.
\end{split}
\end{equation}
This equation can be solved recursively for $w^0_n w^1_n$, leading to (given the boundary conditions $w^0_0(\mathbf{t})\equiv 0$ and  $w^1_0(\mathbf{t}) \equiv 0$)
\begin{equation}\label{eq:constraint_w0w1}
    w^{0}_n\,w^{1}_{n} = n\, \frac{\partial_{t_2} w^{-1}}{w^{-1}}\,,
\end{equation}
which, using equation~\eqref{eq:w0_red_def}, for $n=1$ yields the evolution equation \eqref{eq:wm1_evol} for the entry $w^{-1}(\mathbf{t})$ 
\begin{equation*}
    \partial_{t_2}w^{-1}= (w^{-1})^2\left(\,c_1\,w^1_1\,\right).
\end{equation*}
Observing that~\eqref{eq:constraint_w0w1} implies
\begin{equation} \label{eq:constr_w0w1}
    w^{0}_n\,w^{1}_{n} = n\, w^{0}_1\,w^{1}_{1}\,,
\end{equation}
and using the equation~\eqref{eq:w0_red_def} for $w^{0}_n(\mathbf{t})$, we can express $w^1_n(\mathbf{t})$ according to~\eqref{eq:wk_red_def} for $k=1$, i.e.\
\begin{equation*}
    w^1_n(\mathbf{t}) =n\,\frac{c_1}{c_n}\,w^1_1(\mathbf{t})\,,
    \label{eq:w1n}
\end{equation*}
which implies that the dependence on the couplings is encoded in the first entry $w^1_1(\mathbf{t})$. The evolution of~$w^1_n(\mathbf{t})$ is then completely determined by the evolution of $w^1_1(\mathbf{t})$. 

Let us proceed with the analysis of\ $w^2_n(\mathbf{t})$. The general result for $w^k_n(\mathbf{t})$ with $k>2$ can then be proven by induction. 
Let us first consider the evolution equation for $w^1_n(\mathbf{t})$ from the Pfaff lattice equations~\eqref{pfaff_lattice}, which, using~\eqref{eq:constr_w0w1} and~\eqref{eq:w0_red_def}, can be written as follows
\begin{equation}\label{eq:constr_w1}
    \partial_{t_2}w^1_n = w^{-1} \!\left( c_{n+1}\,w^2_n - c_{n-1}\,w^2_{n-1}-c_1\,w^1_n\,w^1_1 \right)\,. 
\end{equation}
Setting $n=1$ and using the boundary condition $w^2_{0}(\mathbf{t})\equiv0$, we obtain the evolution equation for $w^1_1(\mathbf{t})$ as given in~\eqref{eq:wk_evol} for $k=1$, i.e.\
\begin{equation*}
    \partial_{t_2}w^1_1 = w^{-1} \!\left( c_{2}\,w^2_1 -c_1(w^1_1)^2 \right)\,. 
\end{equation*}
Substituting expression \eqref{eq:w1n} in equation~\eqref{eq:constr_w1} and using the evolution equation for $\partial_{t_2}w^1_1(\mathbf{t})$ we obtain the following recursive relation for  $w^2_n(\mathbf{t})$
\begin{equation*}
    c_{n}\,c_{n+1}\,w^2_n = c_{n}\,c_{n-1}\,w^2_{n-1} +n\,c_1 c_2\,w^2_1\,. 
\end{equation*}
This equation is solved directly by iteration and, using the boundary condition $w^2_0(\mathbf{t})\equiv 0$, we get
\begin{equation*}
    c_{n}\,c_{n+1}\,w^2_n = \sum_{k=1}^n k\,c_1 c_2\,w^2_1 = \frac{n(n+1)}{2}\,c_1 c_2\,w^2_1\,,
\end{equation*}
or, equivalently, 
\begin{equation*}
    w^2_n = \binom{n+1}{2}\,\frac{c_1 c_2}{c_{n}\,c_{n+1}}\,w^2_1\,,
\end{equation*}
i.e.\ expression~\eqref{eq:wk_red_def} for $k=2$. Again, the dependence on the couplings for $w^2_n(\mathbf{t})$ is completely encoded in the entry~$w^2_1(\mathbf{t})$. 
A similar analysis for the generic entry $w^k_n(\mathbf{t})$ completes the proof by induction. All details are given in Appendix \ref{app:wk_induction}. 

Finally, let us consider the evolution equation~\eqref{eq:wk_evol} for the rescaled variables $W^k$ such that $w^k_1(\mathbf{t}) = F_k\,W^k(\mathbf{t})$ where $F_{k}$ is given in~\eqref{eq:rescaled_w}. We have \ 
\begin{equation}\label{eq:Wk_evol_Fk}
    F_k \,\partial_{t_2} W^k = W^{-1} \left( c_{k+1}\,F_{k+1}\,W^{k+1} - c_1\,F_1\,F_k\,W^1\,W^k + \frac{k(c_1)^2-(c_k)^2}{c_k}\,F_{k-1}\,W^{k-1} \right).
\end{equation}
Observing that the factor $F_k$ satisfies the following equations 
\begin{equation*}
    c_{k+1}\,F_{k+1}=  
    2(k+1)\,F_k  \,,\qquad 
        \frac{k(c_1)^2-(c_k)^2}{c_k}\,F_{k-1}\,=  
        -2(k-1)\,F_k \, , \qquad c_1 \, F_1 = 2 \,,
    \end{equation*}
it follows that equation~\eqref{eq:Wk_evol_Fk}  coincides with equations~\eqref{eq:system_red_W}. This completes the proof.
\end{proofC}
}
\begin{remark}
We note that the coefficient in~\eqref{eq:wk_red_def} can be calculated explicitly. In particular, we have
\begin{itemize}
    \item if $k$  is even 
    \begin{equation}
        \binom{n+k-1}{k} \prod_{\ell=1}^k \frac{c_{\ell}}{c_{n+k-\ell}} = \binom{n+k-1}{k}^{\!1/2} \left(\frac{(2n-3)!!}{(2n+k+1)!!}\right)^{\!1/2}
    \end{equation}
    \item if $k$ is odd
    \begin{equation}
         \binom{n+k-1}{k} \prod_{\ell=1}^k \frac{c_{\ell}}{c_{n+k-\ell}} = \binom{n+k-1}{k}^{\!1/2} \left(\frac{(2n-3)!!}{(2n+k)!!}\right)^{\!1/2}.
    \end{equation}
\end{itemize}
\end{remark}

\subsection{Continuum case}
In Section~\ref{sec:discretecase} we considered the reduction where $w^{-k}_n(\mathbf{t}) \equiv 0$ for all $k > 2$, and derived the corresponding further reduced even Pfaff lattice equations. As mentioned, this reduction is particularly important in connection with the orthogonal ensemble for it matches the initial conditions associated with the GOE.

We now consider the similar reduction for the continuum limit represented by the hydrodynamic chain~\eqref{eq:chainnegative}-\eqref{eq:chainpositive}.
The initial conditions on the dependent variables of the hydrodynamic chain are given by the continuum limit of the initial conditions~\eqref{eq:initial_datum} for the discrete lattice, consistently with the definition of $u^k(x, \mathbf{t})$ in terms of the interpolating functions in~\eqref{eq:definition_variables}. Specifically, we have
\begin{subequations} \label{eq:id_continuum}
    \begin{align}
        u^{-k} (x,\mathbf{0}) &= 0\,, \hspace{7.5ex}  -k < -1\,\\
        u^{-1}(x,\mathbf{0}) &= \frac{1}{2}\, \label{eq:id_continuumm1}\\
        u^0(x,\mathbf{0}) &= x\,\label{eq:id_continuum0}\\[.5ex]
        u^k(x,\mathbf{0}) &= 2\,, \qquad\qquad k > 0 \label{eq:id_continuumm_pos} \,. 
    \end{align}
\end{subequations}
In this section, similarly to the discrete case, we assume $u^{-k}(\mathbf{t}) \equiv 0$ for all $k>2$. This leads to a new reduction of the hydrodynamic chain, as specified in the following
\begin{theorem} \label{thm:reduction_cont}
If $u^{-k}(x, \mathbf{t})\!\equiv\!0$ for all $k\!>\!2$, then there exists a reduction of the chain~\eqref{eq:chainnegative}-\eqref{eq:chainpositive} such that
\begin{subequations}\label{eq:constraints_recap}
    \begin{align}
        u^{-2}(x, \mathbf{t}) &=0\,\label{eq:constraint-2}\\[.2ex]
        u^{-1}(x, \mathbf{t}) &= u^{-1}(\mathbf{t})\,\label{eq:constraint-1}\\[.2ex]
        u^{0}(x, \mathbf{t}) &= 2 x\, u^{-1}(\mathbf{t})\label{eq:constraint0}\,\\[.2ex]
        u^{k}(x, \mathbf{t}) &= u^k(\mathbf{t})\,, \hspace{8ex}  k > 0\,,\label{eq:constraintk}
    \end{align}
\end{subequations}
that is compatible with the initial condition~\eqref{eq:id_continuum}. Moreover, the functions $u^{-1}(\mathbf{t})$ and $u^{k}(\mathbf{t})$, with $k>0$, solve the system \eqref{eq:system_red_W}, i.e.
\begin{subequations} \label{eq:chain_ODEs}
\begin{align}
    \partial_{t_2} u^{-1} &= 2 (u^{-1})^2\, u^1 \label{eq:reduction-1}\\[.75ex]
    \partial_{t_2}u^k &= 2 u^{-1}\!\big((k+1) u^{k+1} - u^1\, u^k - (k-1)u^{k-1}  \big), \qquad k > 0\,.\label{eq:reductionk} 
    \end{align}
\end{subequations}
\end{theorem}

\begin{proofC}
Let us observe that the evolution equations of the hydrodynamic chain \eqref{eq:chainnegative}-\eqref{eq:chainpositive} for the variables $u^{-k}(x, \mathbf{t})$ with $k > 3$ are trivially satisfied under the assumption $u^{-k}(x, \mathbf{t})\equiv0$ for $k > 2$.
The first non-trivial condition emerges from the equation for $u^{-3}(x, \mathbf{t})$
in~\eqref{eq:chainnegative}, that is
\begin{equation}
    0 = - u^{-2} \,\partial_x u^0 + u^0 \,\partial_x u^{-2}\,. 
    \label{eq:eqconstraint_-3}
\end{equation}
This constraint is automatically satisfied by the condition~\eqref{eq:constraint-2},
which is indeed consistent with the initial datum. We note, incidentally, that the equation~\eqref{eq:eqconstraint_-3} can be integrated by separation under the assumption that $u^{0} u^{-2} \neq 0$, but this assumption is not compatible with the initial conditions~\eqref{eq:id_continuum}, as $u^0(x,\mathbf{0})$ depends on $x$, whilst $u^{-2}(x,\mathbf{0})= 0$. 

Let us now consider the equation for $u^{-2}$ in~\eqref{eq:chainnegative}. The assumption~$u^{-k}(x, \mathbf{t}) \equiv 0$ for $k > 2$ implies
\begin{equation}
\label{eq:constraintu0um1}
    0 = 2 u^0 \, \partial_x u^{-1}\,. 
\end{equation}
Since $u^0(x, \mathbf{0}) \neq 0$, as per the initial condition, we must have  $\partial_x u^{-1}(x, \mathbf{t}) \equiv 0$ giving the condition~\eqref{eq:constraint-1}. 

Let us now consider the equations for $u^{-1}(x, \mathbf{t})$ in~\eqref{eq:chainnegative} and $u^0(x, \mathbf{t})$ in~\eqref{eq:chainpositive}. Using above condition~\eqref{eq:constraint-2} in the evolution equation for $u^{-1}$, we have
\begin{subequations}
\label{eq:reduced_0-1}
\begin{align} 
\partial_{t_2}u^{-1} &= u^{-1}\,\partial_x\!\left(u^0 u^1\right)\label{eq:reduced_0-1a} \\
\partial_{t_2}u^{0} &= u^{0}\,\partial_x\!\left(u^0 u^1\right). \label{eq:reduced_0-1b}
\end{align}
\end{subequations}
For $u^{-1}(\mathbf{t}) \neq 0$, equation~\eqref{eq:reduced_0-1a} can be written as follows
\[
\partial_x(u^0 u^1) = \frac{\partial_{t_2}u^{-1}}{u^{-1}}\,,
\]
implying that, since $u^{-1} = u^{-1}(\mathbf{t})$, the quantity $\partial_{x}(u^0 u^1)$ on the left hand side does not depend on $x$. Hence, equations~\eqref{eq:reduced_0-1} together imply that
\[
|u^0(x, \mathbf{t})| = |u^{-1}(\mathbf{t})| \,F(x, t_4, t_6, \dots),
\]
where $F$ is an arbitrary positive function independent of $t_2$. We then choose $F = |2x|$, so that the above condition is consistent with~\eqref{eq:constraint0} and the initial datum~\eqref{eq:id_continuum}.

We note that the assumption~\eqref{eq:constraintk} is consistent with the initial condition~\eqref{eq:id_continuumm_pos}. This, alongside with the assumptions above, brings the equations~\eqref{eq:chainnegative}-\eqref{eq:chainpositive} to the form~\eqref{eq:chain_ODEs}. This completes the proof. 
\end{proofC}

The above reduction, leading to the system~\eqref{eq:chain_ODEs}, allows us to solve the hydrodynamic chain in terms of a one-dimensional dynamical chain which describes the orthogonal ensemble with even couplings.

\begin{remark}
    The assumption \eqref{eq:constraintk} can be further justified as follows. Let us consider the evolution equation for $u^0(x, \mathbf{t})$ from~\eqref{eq:reduced_0-1}. Taking into account~\eqref{eq:constraint0}, it can be recast as
    $$
    \partial_x\!\left(x u^1\right) = \frac{\partial_{t_2}u^{-1}}{(u^{-1})^2}\,. 
    $$
    Integrating with respect to $x$ we have
    $$
    u^1(x, \mathbf{t}) = \frac{\partial_{t_2}u^{-1}}{(u^{-1})^2} + \frac{g_1(\mathbf{t})}{x}\,,
    $$
    where $g_1(\mathbf{t})$ is an arbitrary function of $\mathbf{t}$. Consistency with the initial condition~\eqref{eq:id_continuum} implies that $g_1(\mathbf{0})=0$. Choosing $g_1(\mathbf{t}) \equiv0$ effectively removes the dependence of $u^1(x, \mathbf{t})$ on $x$. A similar argument can be applied to the higher variables $u^{k}$ and the corresponding equations~\eqref{eq:chainpositive} with $k> 1$.
\end{remark}

\begin{remark}Theorem \ref{thm:reduction_cont} is the analogue for the hydrodynamic chain~\eqref{eq:chain_system} of Theorem \ref{thm:reduction_pfaff} for the Pfaff lattice~\eqref{pfaff_lattice}. 
 In particular, the equations~\eqref{eq:chain_ODEs} coincide with the equations~\eqref{eq:system_red_W} for the variables $W^\ell(\mathbf{t})$. 
 We also note that the variable  $W^\ell(\mathbf{t})$ are obtained from the entries of the Pfaff Lax matrix by the rescaling~\eqref{eq:rescaled_w}, so that $W^\ell(\mathbf{t})$ and $u^\ell(\mathbf{t})$, $\ell\in \{-1, 1, 2, ...\}$ share the same initial conditions~\eqref{eq:id_continuumm1},~\eqref{eq:id_continuumm_pos}.
Remarkably the solutions of both discrete and (leading order) continuum cases are obtained in terms of the very same system.
\end{remark}

\section*{Acknowledgments}
The authors are grateful to P.A. Clarkson, E.V. Ferapontov, A. Hone, Y. Kodama, A. Loureiro, and K.T.-R. McLaughlin for insightful discussions. They also thank the Isaac Newton Institute for Mathematical Sciences, Cambridge, for its support and hospitality during the programme {\it Dispersive Hydrodynamics: Mathematics, Simulation and Experiments, with Applications in Nonlinear Waves} (Cambridge, July–December 2022), supported by EPSRC grant EP/R014604/1, as well as the satellite programme {\it Emergent Phenomena in Nonlinear Dispersive Waves} (Newcastle, July–August 2024), supported by EPSRC grant EP/V521929/1, during which part of this work was conceived and developed.

This research has been partially supported by the Leverhulme Trust (grant RPG-2017-228, PI: A.M.), the London Mathematical Society Research in Pairs grant 42374 (C.B. and M.D.), and the Emmy Noether Fellowship EN-2324-04 (PI: C.B.). The authors also gratefully acknowledge the GNFM – Gruppo Nazionale per la Fisica Matematica, INdAM (Istituto Nazionale di Alta Matematica), for supporting activities that contributed to the research presented in this paper.

\section*{Conflict of interest}
The authors declare that there are no conflicts of interest related to the content of this article.

\section*{Data availability statement}
No external datasets were used in this study. The authors confirm that all data and relevant details supporting the findings are included within the paper, including the Appendices.

\newpage 
\begin{appendices}
\section{Entries of the reduced even Pfaff Lax matrix in terms of the $\tau$-function}\label{app:L_elements_in_tau}
We consider here the construction of the Pfaff Lax matrix $L$ for the hierarchy reduced to even couplings and express its entries in terms of Pfaff Lattice $\tau$-function. We reproduce the specific expressions given in~\cite{vanMoerbekenotes} for two of the entries and provide the form of further entries, following an approach mainly based on~\cite{vanMoerbekenotes,vanM_Pfaff_skew,adler2002}. This is the key observation that allows to recognise `similar' entries in the lattice, justifying the notation $w^k_n(\mathbf{t})$ (see Section~\ref{sec:Pfaff_equations_and_initial_datum}). 
The matrix $L(\mathbf{t})$ in the reduction considered here takes the form 
{\small
\begin{equation*}
  L(\mathbf{t}) \Big|_{\underset{\forall i\in \mathbb{N}}{t_{2i-1} = 0}}  = \left(
\begin{array}{ccccccc}
 0 & 1 & 0 & 0 & 0 & 0 & \dots\\[1.8ex]
L_{2,1}& 0 & L_{2,3}\hspace*{-0.5ex}& 0 & 0 & 0 &\ddots\\[1.8ex]
 0 & L_{3,2}\hspace*{-0.5ex}& 0 & 1 & 0 & 0 & \ddots\\[1.8ex]
 L_{4,1}\hspace*{-0.5ex}& 0 & L_{4,3}& 0 & L_{4,5}\hspace*{-0.5ex}& 0 & \ddots \\[1.8ex]
 0 & L_{5,2}\hspace*{-0.5ex}& 0 & L_{5,4}\hspace*{-0.5ex}& 0 & 1 & \ddots \\[1.8ex]
\vdots& \ddots &\ddots& \ddots & \ddots& \ddots & \ddots\\[1.8ex]
\end{array}
\right)  =  \left(
\begin{array}{ccccccc}
 ~~0~~ & ~~1~~ & ~~0~~ & ~~0~~ & ~~0~~ & ~~0~~ & \dots\\[1.8ex]
w^{-1}_1& ~~0~~ & w^{0}_{1}\hspace*{-0.5ex}& 0 & 0 & 0 &\ddots\\[1.8ex]
 0 & w^{1}_{1}\hspace*{-0.5ex}& 0 & 1 & 0 & 0 & \ddots\\[1.8ex]
 w^{-2}_{1}\hspace*{-0.5ex}& 0 & w^{-1}_2& 0 & w^{0}_{2}\hspace*{-0.5ex}& 0 & \ddots \\[1.8ex]
 0 & w^{2}_{1}\hspace*{-0.5ex}& 0 & w^{1}_{2}\hspace*{-0.5ex}& 0 & 1 & \ddots \\[1.8ex]
\vdots& \ddots &\ddots& \ddots & \ddots& \ddots & \ddots\\[1.8ex]
\end{array}
\right)
\end{equation*}
}

We recall that $L(\mathbf{t})$ is built by conjugating the shift matrix $\Lambda$ via the matrix $S(\mathbf{t})$. The latter is the decomposition matrix of the semi-infinite moment matrix $m_{\infty}(\mathbf{t})$ defined for the skew-symmetric inner product \eqref{eq:skew_prod}, i.e.\ 
\begin{equation}
    m_{\infty}(\mathbf{t})\Big|_{\underset{\forall i\in \mathbb{N}}{t_{2i-1} = 0}} = \left.(S^{-1}) \, J \, (S^{-1})^{ \top} \right|_{\underset{\forall i\in \mathbb{N}}{t_{2i-1} = 0}}\,, \qquad L(\mathbf{t})\Big|_{\underset{\forall i\in \mathbb{N}}{t_{2i-1} = 0}}  = \left.S \, \Lambda \, S^{-1}\right|_{\underset{\forall i\in \mathbb{N}}{t_{2i-1} = 0}} \,
    \label{eq:m_infty_l_in_Q}
\end{equation}
where $J$ is the skew-symmetric matrix such that $J^2 = -1$.
The matrix $S$ evaluated at $\mathbf{t}=(0,t_2,0,t_4,\dots)$ takes the form
\begin{equation}
    S(\mathbf{t})\Big|_{\underset{\forall i\in \mathbb{N}}{t_{2i-1} = 0}} = \begin{pmatrix}
    S_{0,0} & 0 & 0 & 0 & 0 & 0 & 0 & \dots \\[2ex] 
    0 & S_{0,0} & 0 & 0 & 0 & 0 & 0 &\dots \\[2ex] 
    S_{2,0} & 0 & S_{2,2} & 0 & 0 & 0 &0 & \dots \\[2ex] 
    0 & S_{3,1} & 0 & S_{2,2} & 0 & 0 &0 & \dots \\[2ex] 
    S_{4,0} & 0 & S_{4,2} & 0 & S_{4,4} & 0 &0 & \dots \\[2ex] 
    0 & S_{5,1} & 0 & S_{5,3} & 0 & S_{4,4} &0 & \dots \\[2ex] 
    \vdots & \ddots & \ddots & \ddots & \ddots & \ddots & \ddots & \ddots \\[2ex] 
    \end{pmatrix}
    \label{eq:q_matrix}
\end{equation}
where the entries are defined in terms of the skew-orthogonal polynomials $q_{2n}$, $q_{2n+1}$~\cite{adler2002} as follows
\begin{equation}
     \begin{split}
            q_{2n}(\mathbf{t},z) &= \sum_{j=0}^{2n} S_{2n,j} \, z^j  =  \sqrt{2}\,  z^{2n} \, h_{2n}^{-1/2} \, \frac{\tau_{2n}\left( \mathbf{t} - [z^{-1}] \right)}{\tau_{2n}(\mathbf{t})} \,, \qquad \text{ with }  h_{2n}=\frac{\tau_{2n+2}(\mathbf{t})}{\tau_{2n}(\mathbf{t})}\,, \\[.5ex]
            q_{2n+1}(\mathbf{t},z) &= \sum_{j=0}^{2n+1} S_{2n+1,j} \, z^j = {\sqrt{2}}\, z^{2n} \, h_{2n}^{-1/2} \, \frac{1}{\tau_{2n}(\mathbf{t})} \left(z + \partial_1 \right) \tau_{2n}\left( \mathbf{t} - [z^{-1}] \right) \,,
        \end{split}
        \label{eq:skew_ortho_tau}
\end{equation}
with the Sato shift~\cite{dickey1991soliton} 
\begin{equation}
    \mathbf{t} - [z^{-1}] = \left\{ t_k - \dfrac{1}{k} z^{-k} \right\}_{k \ge 1}  \,.
\end{equation}
Following~\cite{WillSats,dickey1991soliton} and using the Schur polynomials (see footnote~\ref{foot:schur}), the expressions in~\eqref{eq:skew_ortho_tau} can be written formally as
\begin{equation}
    \tau_{2n}\left(\, \mathbf{t} - [z^{-1}] \right) = \sum_{k=0}^{\infty} s_{k}(-\tilde{\partial}) \, \tau_{2n}(\mathbf{t})\,z^{-k} \,,
    \label{eq:tau_sato_schur}
\end{equation}
with {$\tilde{\partial}$ and $s_{\ell}$ as in Theorem~\ref{AvM2:theo}}.
As it is evident from~\eqref{eq:q_matrix}, if the odd couplings are set to zero, the only non-zero entries of the matrix $S$ are 
\begin{equation}
    S_{2n,2j}(\mathbf{t})\Big|_{\underset{\forall i\in \mathbb{N}}{t_{2i-1} = 0}} \,, \qquad S_{2n+1,2j+1}(\mathbf{t})\Big|_{\underset{\forall i\in \mathbb{N}}{t_{2i-1} = 0}}, \,\qquad {n\geq j}
\end{equation}
{with  $S_{2n+1,2n+1}(\mathbf{t})\Big|_{\underset{\forall i\in \mathbb{N}}{t_{2i-1} = 0}} =  S_{2n,2n}(\mathbf{t})\Big|_{\underset{\forall i\in \mathbb{N}}{t_{2i-1} = 0}} $.} Substituting~\eqref{eq:tau_sato_schur} in~\eqref{eq:skew_ortho_tau} we have 
 \begin{equation}
 \begin{split} 
    S_{2n,2j}(\mathbf{t})\Big|_{\underset{\forall i\in \mathbb{N}}{t_{2i-1} = 0}}& = \dfrac{{ \sqrt{2}}\,s_{2n-2j}(-\tilde{\partial})\, \tau_{2n}(\mathbf{t})}{\sqrt{\tau_{2n}(\mathbf{t})\, \tau_{2n+2}(\mathbf{t})}} \bigg|_{\underset{\forall i\in \mathbb{N}}{t_{2i-1} = 0}} \, {\qquad n\geq j}\\[.5ex]
    S_{2n+1,2j+1}(\mathbf{t})\Big|_{\underset{\forall i\in \mathbb{N}}{t_{2i-1} = 0}}&= \dfrac{{ \sqrt{2}}\left(\partial_{t_1} s_{2n-2j-1}(-\tilde{\partial}) + s_{2n-2j}(-\tilde{\partial})\right)\, \tau_{2n}(\mathbf{t})}{\sqrt{\tau_{2n}(\mathbf{t})\, \tau_{2n+2}(\mathbf{t})}} \bigg|_{\underset{\forall i\in \mathbb{N}}{t_{2i-1} = 0}} \, {\qquad n > j\,,}
    \end{split} 
\end{equation}
with $s_{-1}=0$, $s_0 = 1$ and $\tau_0(\mathbf{t}) = 1$. The index $k=2n-2j$ identifies the $k$-th diagonal of $S$ in \eqref{eq:q_matrix}. $k=0$ corresponds to the main diagonal, $k > 0$ corresponds to the lower diagonals in~\eqref{eq:q_matrix}. Therefore we can reformulate the previous expressions in terms of the new index and get the entries of the $k$-th diagonal of $S$ as 
\begin{equation}
    \begin{split} 
    S_{2n,2n-k}(\mathbf{t})\Big|_{\underset{\forall i\in \mathbb{N}}{t_{2i-1} = 0}}& = \dfrac{{ \sqrt{2}}\,s_{k}(-\tilde{\partial}\,)\, \tau_{2n}(\mathbf{t})}{\sqrt{\tau_{2n}(\mathbf{t})\, \tau_{2n+2}(\mathbf{t})}} \bigg|_{\underset{\forall i\in \mathbb{N}}{t_{2i-1} = 0}} \, \\[.5ex]
    S_{2n+1,2n-k+1}(\mathbf{t})\Big|_{\underset{\forall i\in \mathbb{N}}{t_{2i-1} = 0}}&= \dfrac{{ \sqrt{2}}\left(\partial_{t_1} s_{k-1}(-\tilde{\partial}\,) + s_{k}(-\tilde{\partial}\,)\right)\, \tau_{2n}(\mathbf{t})}{\sqrt{\tau_{2n}(\mathbf{t})\, \tau_{2n+2}(\mathbf{t})}} \bigg|_{\underset{\forall i\in \mathbb{N}}{t_{2i-1} = 0}}\,.
    \end{split} 
\end{equation}
According to~\eqref{eq:m_infty_l_in_Q}, the elements of the matrix $L$ are expressed in terms the elements of $S$.
Given the index $\ell \in \mathbb{Z}$ introduced in Section \ref{sec:SME_Pfaff}, elements on the $(2 |\ell|-1)$-th lower diagonal ($\ell = 0$ corresponds to the first upper diagonal) depend on combinations of Schur polynomials of order $2|\ell|$. For example, we have\footnote{The entries $w^0_n$ and $w^{-1}_n$ have been already given explicitly in~\cite{vanMoerbekenotes}.}
\begin{itemize}
    \item $\ell = 0$ 
    \begin{align*}
    \begin{split} 
   w^0_n   &= \dfrac{s_0(-\tilde{\partial}\,) \,\tau_{2n-2}}{s_0(-\tilde{\partial}\,)\, \tau_{2n}} \sqrt{\dfrac{\tau_{2n+2}}{\tau_{2n-2}}} \bigg|_{\underset{\forall i\in \mathbb{N}}{t_{2i-1} = 0}} = \left.\dfrac{\sqrt{\tau_{2n-2}\, \tau_{2n+2}}}{\tau_{2n}} \right|_{\underset{\forall i\in \mathbb{N}}{t_{2i-1} = 0}} \, \end{split}
\end{align*}
\item $|\ell| = 1$ 
\begin{align*}
      w^{-1}_n &=  - \dfrac{s_2(\tilde{\partial}\,)\,\tau_{2n-2}}{\tau_{2n-2}}
  - \dfrac{s_2(-\tilde{\partial}\,)\,\tau_{2n}}{\tau_{2n}} \bigg|_{\underset{\forall i\in \mathbb{N}}{t_{2i-1} = 0}}\,,
   \qquad 
           w^1_n =  \dfrac{\left( s_2(-\tilde{\partial}\,) + s_2(\tilde{\partial}\,) \right) \tau_{2n}}{\sqrt{\tau_{2n-2}\tau_{2n+2}}}\bigg|_{\underset{\forall i\in \mathbb{N}}{t_{2i-1} = 0}}
\end{align*}
\item $|\ell| = 2$
{ 
\begin{align*} 
    \begin{split} 
        w^{-2}_n  &  =  \frac{1}{\sqrt{\tau_{2n-2}\,\tau_{2n+2}}\,
         \tau_{2n}\,\tau_{2n+2}}  \Bigg[  \tau_{2n} \Big( s_2(-\tilde{\partial}) \tau_{2n}\,s_2(-\tilde{\partial}) - \tau_{2n}\,s_4(-\tilde{\partial}) \Big)\tau_{2n+2} \\[1ex]
        &    \hspace{10ex} +\,\tau_{2n+2} \Big( s_2(-\tilde{\partial})\tau_{2n}\,s_2(\tilde{\partial})+\partial_1 s_3(-\tilde{\partial})+ s_4(-\tilde{\partial})\Big)\tau_{2n} 
        \Bigg] \Bigg|_{\underset{\forall i\in \mathbb{N}}{t_{2i-1} = 0}} \,
        \end{split}\\[2ex] 
        \begin{split} 
        w^2_n  &=  \frac{1}{\sqrt{\tau_{2n-2}\,\tau_{2n}\,\tau_{2n+2}\,\tau_{2n+4}}} \Big[ s_{2}(\tilde{\partial}\,) \tau_{2n} \big( s_{2}(\tilde{\partial}\,)   + s_{2}(-\tilde{\partial}\,) \big)   - \tau_{2n} \partial_1 s_3(-\tilde{\partial}\,) \Big] \tau_{2n+2}\Bigg|_{\underset{\forall i\in \mathbb{N}}{t_{2i-1} = 0}}\,
    \end{split}
\end{align*} }
\item $|\ell| = 3$ 
{ 
\begin{align*}
\begin{split} 
    w^{-3}_n &= \frac{1}{\sqrt{\tau_{2n-2}\,\tau_{2n}\,\tau_{2n+2}\,\tau_{2n+4}}\,\tau_{2n+2}\,\tau_{2n+4}} \times \\[1ex]
    &\times\Bigg[ \tau_{2 n+4}  \Big(\tau_{2 n} s_4(-\tilde{\partial}) \tau_{2n+2} -s_2(-\tilde{\partial}) \tau_{2n} \, s_2(-\tilde{\partial}) \tau_{2n+2} \Big) s_2(\tilde{\partial}) \tau_{2n+2}  \\[1ex]
    &+\tau_{2 n+2}\,\tau_{2 n+4} \Big(\partial_1 \tau_{2 n} \, s_5(-\tilde{\partial}) -s_2(-\tilde{\partial}) \tau_{2n}\big(\partial_1  \, s_3(-\tilde{\partial}) + s_4(-\tilde{\partial}) \big) +\tau_{2 n}\,  s_6(-\tilde{\partial}) \,\Big)\tau_{2n+2}\\[1ex]
    &+\tau_{2 n+2}^2 \Big(s_2(-\tilde{\partial}) \tau_{2n} \, s_4(-\tilde{\partial}) -\tau_{2 n}\, s_6(-\tilde{\partial}) \Big)\tau_{2n+4}\\[1ex]
    &+\tau_{2 n+2} \Big(\tau_{2 n}  s_4(-\tilde{\partial}) \tau_{2n+2}-s_2(-\tilde{\partial}) \tau_{2n} \, s_2(-\tilde{\partial}) \tau_{2n+2} \Big) \, s_2(-\tilde{\partial}) \tau_{2n+4}
         \Bigg] \Bigg|_{\underset{\forall i\in \mathbb{N}}{t_{2i-1} = 0}}
    \end{split} \\[2ex]
    \begin{split} 
    w^3_n &= \frac{1}{\sqrt{\tau_{2n-2}\,\tau_{2n}\,\tau_{2n+4}\,\tau_{2n+6}}\,\tau_{2n+2}} \times \\[1ex]
    &~~\times \Bigg[ s_2(-\tilde{\partial})\tau_{2 n+4}\, \Big(s_2(\tilde{\partial})\tau_{2 n} s_2(\tilde{\partial})-\tau_{2 n}\, (\partial_1 s_3(-\tilde{\partial})\tau_{2 n+2}+s_4(-\tilde{\partial}))\Big)\tau_{2 n+2} \\[1ex]
    &~~+ s_2(\tilde{\partial})\tau_{2 n}\, \Big(s_2(\tilde{\partial})\tau_{2 n+2}\, s_2(\tilde{\partial})-\partial_1 \tau_{2 n+2} \, s_3(-\tilde{\partial})\Big)\tau_{2 n+4} \\[1ex]
    &~~- \tau_{2 n}\, \Big( \big(\partial_1 s_3(-\tilde{\partial})+s_4(-\tilde{\partial})\big)\tau_{2 n+2}\,s_2(\tilde{\partial})+\partial_1 \tau_{2 n+2}\, s_5(-\tilde{\partial})\Big)\tau_{2 n+4} \Bigg] \Bigg|_{\underset{\forall i\in \mathbb{N}}{t_{2i-1} = 0}}\,.
    \end{split} 
\end{align*}
}

\end{itemize}

\newpage 
\section{Even reduced Pfaff lattice as a double infinite chain.}\label{app:pfaff_lattice_explicit}
Recalling the definition of $\varphi(i)$ and $\sigma(i)$ in~\eqref{eq:even_odd}, let us consider the generic element of the Pfaff Lax matrix $L$ \eqref{eq:even_Pfaff_Lax_matrix} written according to~\eqref{eq:pfaff_t2_generic_element}
\begin{equation*}
\begin{aligned} 
    L_{ij}&=\varphi(i) \,\sigma(j) \left( \delta_{1,j-i} \,w\big(0\,,\tfrac{j-1}{2}\big) +\vartheta(i-j) \, w\big(\tfrac{-(i-j+1)}{2} ,\tfrac{j+1}{2}\big) \right) \\[1ex]
    &~~+\varphi(j) \,\sigma(i) \left(\delta_{1,j-i} +\vartheta(i-j)\,  w\big(\tfrac{i-j+1}{2} ,\tfrac{j}{2}\big)\right)\,. 
\end{aligned} 
\end{equation*}
Here, it is convenient to use the notation $w(k,n):= w^k_n$. 
The expression for the generic element of $L^2$ is
\begin{equation*}
\begin{aligned}
    (L^2)_{ij}&=\sum_{p\,\ge 1} L_{ip}\,L_{pj} \\
    &=\varphi(i) \,\varphi(j) \Big\{ \delta_{i,j-2}\, w\big(0\,,\tfrac{j-2}{2}\big)+\vartheta(i-j+1) \left(w\big(\tfrac{i-j+2}{2} ,\tfrac{j}{2}\big) + w\big(0\,,\tfrac{i}{2}\big)\, w\big(\tfrac{i-j+2}{2},\tfrac{j}{2}\big) \right) \Big\} \\ 
    &~~+\varphi(i) \,\varphi(j) \, \vartheta(i-j) \,\sum _{p=j}^i  \sigma(p)\, w\big(\tfrac{-i+p-1}{2} ,\tfrac{p+1}{2}\big) \,w\big(\tfrac{-j+p+1}{2} ,\tfrac{j}{2}\big) \\
    &~~+\sigma(i) \,\sigma(j) \Big\{ \delta_{i,j-2}\, w\big(0\,,\tfrac{j-1}{2}\big)+\vartheta(i-j+1) \left(w\big(\tfrac{-i+j-2}{2} ,\tfrac{j+1}{2}\big) + w\big(0\,,\tfrac{j-1}{2}\big) \, w\big(\tfrac{i-j+2}{2} ,\tfrac{j-1}{2}\big) \right) \Big\} \\ 
    &~~+\sigma(i) \,\sigma(j) \, \vartheta(i-j) \,\sum _{p=j}^i  \varphi(p)\, w\big(\tfrac{i-p+1}{2} ,\tfrac{p}{2}\big) \,w\big(\tfrac{j-p-1}{2} ,\tfrac{j+1}{2}\big) \,.
\end{aligned}
\end{equation*}
The projection $\mathfrak{t}$, acting on a generic semi-infinite matrix $A$, is given by a composition of projections which select $2\times 2$ blocks diagonals, and the upper/lower $2\times 2$ block triangular part respectively,
\begin{equation}
    A_{\mathfrak{t}} = A_- - J (A_+)^{\top} J + \frac{1}{2}\left( 
 A_0 - J (A_0)^{\top}J \right)\,.
\end{equation}
The generic element of the matrix $J$ and the fundamental projections are
\begin{equation}
\begin{aligned}
    J_{ij}&= \varphi(j)\,\delta_{i,j-1} - \sigma(j)\,\delta_{i,j+1} \,, \qquad 
    (A_0)_{ij} = A_{ij}\Big(\delta_{i,j} + \sigma(j)\,\delta_{i,j+1} + \varphi(j)\,\delta_{i,j-1} \Big)\,, \\[1ex]
    (A_+)_{ij} &=  \vartheta(j-i)\Big( A_{ij} - (A_0)_{ij} \Big) \,, \qquad 
    (A_{-})_{ij} = \vartheta(i-j)\Big( A_{ij} - (A_0)_{ij} \Big)    \,.
\end{aligned} 
\end{equation}
Therefore, we obtain 
\begin{equation}
\begin{aligned} 
    (A_{\mathfrak{t}})_{ij} &= \vartheta(i-j) A_{ij} -\tfrac{1}{2} A_{ij}  \\[1ex]
    &~~+ \tfrac{1}{2}\,\varphi(j) \big( A_{ij} \,\delta_{i,j-1}- A_{j-1,j} \,\delta_{i,j-1}+ A_{j-1,j-1} \,\delta_{i,j}\big)\\[1ex]
    &~~+ \tfrac{1}{2} \, \sigma(j) \big(A_{j+1,j+1} \,\delta_{i,j}- A_{ij} \,\delta_{i,j+1}- A_{j+1,j} \,\delta_{i,j+1}\big)\\[1ex] 
    &~~+ \varphi(i) \,\sigma(j) \big(A_{j+1,i-1} \,\delta_{i,j+2}-\vartheta(i-j-2) \,A_{j+1,i-1}\big)\\[1ex]
    &~~+  \varphi(j) \,\sigma(i) \big(A_{j-1,i+1} \,\delta_{i,j-2}+A_{j-1,i+1} \,\delta_{i,j-1}-\vartheta(i-j+2) \,A_{j-1,i+1}\big) \\[1ex]
    &~~+\varphi(i) \,\varphi(j) \big(\vartheta(i-j) \,A_{j-1,i-1}-A_{j-1,i-1} \,\delta_{i,j}\big)\\[1ex]
    &~~+ \sigma(i) \,\sigma(j) \big(-A_{j+1,i+1} \,\delta_{i,j}-A_{i,i+1} \,\delta_{i,j+1}+\vartheta(i-j)\, A_{j+1,i+1}\big)\,.
\end{aligned} 
\end{equation}
In particular, the entry $(i,j)$ of the projection $(L^2)_{\mathfrak{t}}$ is 
{\small
\begin{equation*}
\begin{aligned} 
    &((L^2)_{\mathfrak{t}})_{ij} = \tfrac{1}{2}\varphi(j) \, \delta_{i,j}\Big( w\big(\!-\!1,\tfrac{j}{2}\big)+w\big(0,\tfrac{j-2}{2}\big) \, w\big(1,\tfrac{j-2}{2}\big)\Big)+ \varphi(i) \, \varphi(j) \Big\{\delta_{i,j+2}\, w\left(0,\tfrac{i-2}{2}\right) \\[1ex]
    &+\tfrac{1}{2}\delta_{i,j-1} \Big( w\left(0,\tfrac{i}{2}\right) \,w\big(\tfrac{i-j+2}{2} ,\tfrac{j}{2}\big)+ w\big(\tfrac{-i+j-2}{2} ,\tfrac{j}{2}\big)\Big)+\delta_{i,j+1} \Big(w\big(\tfrac{i-j-2}{2} ,\tfrac{i}{2}\big)+w\big(0,\tfrac{i-2}{2}\big) \, w\big(\tfrac{-i+j+2}{2} ,\tfrac{i-2}{2}\big)\Big)\\ 
    &+\Big(\vartheta(i-j) -\tfrac{1}{2}\delta_{i,j} \Big)\Big[w\big(\tfrac{-i+j-2}{2} ,\tfrac{j}{2}\big)+ w\big(0,\tfrac{i}{2}\big) \, w\big(\tfrac{i-j+2}{2} ,\tfrac{j}{2}\big)  + \sum_{p=j}^i \sigma(p)\, w\big(\tfrac{-i+p-1}{2} ,\tfrac{p+1}{2}\big) w\big(\tfrac{-j+p+1}{2} ,\tfrac{j}{2}\big) \Big] \Big\} \\
    &+\tfrac{1}{2}\sigma(j)\,\delta_{i,j} \Big( w\big(-1,\tfrac{j+1}{2}\big)+ w\big(0,\tfrac{j+1}{2}\big) w\big(1,\tfrac{j+1}{2}\big)\Big) +\sigma(i) \,\sigma(j) \Big\{\delta_{i,j+2}\,w\big(0,\tfrac{i-1}{2}\big) \\[1ex]
    &+\delta_{i,j+1} \Big(w\big(\tfrac{i-j-2}{2} ,\tfrac{i+1}{2}\big)+w\big(0,\tfrac{j+1}{2}\big) w\big(\tfrac{-i+j+2}{2} ,\tfrac{i+1}{2}\big)\Big) \\[1ex]
    &+\Big(\vartheta(i-j)+\tfrac{1}{2}(\delta_{i,j}-\delta_{i,j+1})\Big)  \Big[w\big(\tfrac{-i+j-2}{2} ,\tfrac{j+1}{2}\big)+w\big(0,\tfrac{j-1}{2}\big) w\big(\tfrac{i-j+2}{2} ,\tfrac{j-1}{2}\big) \\
    &+ \sum_{p=j}^i \varphi(p)\,w\big(\tfrac{i-p+1}{2} ,\tfrac{p}{2}\big)\,  w\big(\tfrac{j-p-1}{2} ,\tfrac{j+1}{2}\big) \Big] \Big\} \,. 
\end{aligned} 
\end{equation*} }

\noindent 
Finally, we calculate the element of the commutator $\left[ (L^2)_{\mathfrak{t}},L \right]_{ij}$ as in~\eqref{pfaff_lattice_in_W},
\begin{equation}
	\left[ (L^2)_{\mathfrak{t}}\,, L \right]_{ij} =\, \varphi(j)\,W_{\varphi}(i,j) + \sigma(j)\,W_{\sigma}(i,j)   + \varphi(i)\,\sigma(j)\,W_{\varphi\sigma}(i,j)+\sigma(i)\,\varphi(j)\,W_{\sigma\varphi}(i,j)\,,
	\end{equation} 
where
{\small 
\begin{align*} 
	W_{\varphi}(i,j) &=  \tfrac{1}{2}\,\delta_{i,j-1} \Big(w(-1,\tfrac{j}{2})+ w(0,\tfrac{j}{2})\,w(1,\tfrac{j}{2})\Big)\,, \\ 
	W_{\sigma}(i,j) &= \tfrac{1}{2}\,\delta_{i,j-1} \Big( w({-1},\tfrac{j-1}{2}) \,w(0,\tfrac{j-1}{2})+ w(0,\tfrac{j-3}{2})\,w(1,\tfrac{j-3}{2})\,w(0,\tfrac{j-1}{2})\Big)\,,  \\
	\begin{split} 
		W_{\sigma\varphi}(i,j) &=\Big\{\delta_{i,j} \Big[-\tfrac{1}{2} \,w\big(0,\tfrac{j-2}{2}\big) \,w\big(\tfrac{i-j+3}{2} ,\tfrac{j-2}{2}\big)-\tfrac{1}{2} \,w\big(\tfrac{1-i+j-3}{2} ,\tfrac{j}{2}\big)\\[1ex]
		&+w\big(0,\tfrac{j}{2}\big) \,w\big(\tfrac{-i+j+1}{2},\tfrac{i+1}{2}\big)-w\big(0,\tfrac{i\!-\!1}{2}\big) \,w\big(\tfrac{-i+j+1}{2} ,\tfrac{i\!-\!1}{2}\big)\Big]-\tfrac{1}{2}\,\delta_{i,j-2} w\big(\tfrac{-i+j-3}{2} ,\tfrac{j}{2}\big)\\[1ex]
		&-\tfrac{1}{2}\,\delta_{i,j\!-\!1} \Big[w\big(0,\tfrac{j-2}{2}\big)\Big(w\big(\!-\!1,\tfrac{j}{2}\big)  \,w\big(1,\tfrac{j-2}{2}\big)+ w\big(\tfrac{i-j+3}{2} ,\tfrac{j-2}{2}\big)\Big)-w\big(0,\tfrac{i+1}{2}\big) \,w\big(\tfrac{i-j+3}{2} ,\tfrac{j}{2}\big)\Big]\\[1ex]
		&+\vartheta(i-j) \Big[\tfrac{1}{2}\,w\big(\tfrac{i-j+1}{2} ,\tfrac{j}{2}\big)\Big( w\big(0,\tfrac{i+1}{2}\big) \,w\big(1,\tfrac{i+1}{2}\big) + w\big(0,\tfrac{j}{2}\big) \,w\big(1,\tfrac{j}{2}\big)-w\big(0,\tfrac{i\!-\!1}{2}\big) \,w\big(1,\tfrac{i\!-\!1}{2}\big)   \\
		&-w\big(0,\tfrac{j-2}{2}\big) \,w\big(1,\tfrac{j-2}{2}\big) \Big) -\sum_{q=j}^i\sum_{p=j}^q \varphi(q)\,\sigma(p)\,w\big( \tfrac{p-j+1}{2},\tfrac{j}{2} \big)\,w\big( \tfrac{i-p+1}{2},\tfrac{q}{2} \big)\,w\big( \tfrac{p-q+1}{2},\tfrac{p+1}{2} \big) \\[-1ex]
		&+\sum_{q=j}^i\sum_{p=q}^i \varphi(p)\,\sigma(q)\,w\big( \tfrac{i-p+1}{2},\tfrac{p}{2} \big)\,w\big( \tfrac{q-j+1}{2},\tfrac{j}{2} \big)\,w\big( \tfrac{q-p-1}{2},\tfrac{q+1}{2} \big) \\[-1ex]
		&+ w\big( 0, \tfrac{j-1}{2} \big) \sum_{p=j}^i \sigma(p)\Big(w\big( \tfrac{i-p+2}{2}, \tfrac{p-1}{2} \big)\,w\big( \tfrac{p-j+1}{2}, \tfrac{j}{2} \big)+w\big( \tfrac{p-i-1}{2}, \tfrac{p+1}{2} \big)\,w\big( \tfrac{p-j+2}{2}, \tfrac{j-1}{2} \big) \Big) \Big]\\
		&+\vartheta(i-j+1) \,w\big(0,\tfrac{j-2}{2}\big) \,w\big(\tfrac{i-j+3}{2} ,\tfrac{j-2}{2}\big) +\Big(\vartheta(i-j+1)-\tfrac{1}{2}\,\delta_{i,j-2} \Big) \Big[ w\big(0,\tfrac{i+1}{2}\big) \,w\big(\tfrac{i-j+3}{2} ,\tfrac{j}{2}\big) \Big]\\
		&+\Big( \vartheta(i-j)+\vartheta(i-j+1)-\tfrac{1}{2}\,\delta_{i,j-1}  \Big) \sum_{p=j}^i \sigma(p)\, w\big( \tfrac{p-i-2}{2},\tfrac{p+1}{2} \big)\,w\big( \tfrac{p-j+1}{2},\tfrac{j}{2} \big)  \\[-1ex]
		&+\Big( \vartheta(i-j+1)-\tfrac{1}{2}\,\delta_{i,j-1} -1  \Big) w\big(0,\tfrac{j-1}{2} \big)\sum_{p=j}^i \sigma(p)\, w\big( \tfrac{p-i-1}{2},\tfrac{p+1}{2} \big)\,w\big( \tfrac{p-j+2}{2},\tfrac{j-1}{2} \big)  \\
		&+\vartheta(i-j-2) \Big(\,w\big(0,\tfrac{i\!-\!1}{2}\big) \,w\big(\tfrac{i-j\!-\!1}{2} ,\tfrac{j}{2}\big)-w\big(0,\tfrac{j}{2}\big) \,w\big(\tfrac{i-j\!-\!1}{2} ,\tfrac{j+2}{2}\big)\Big)\\[-1ex]
		&+\sum_{p=j}^i \sigma(p)\, w\big(0,\tfrac{p}{2} \big)\,w\big(\tfrac{i-p+1}{2},\tfrac{p}{2} \big)\,w\big( \tfrac{p-j+2}{2},\tfrac{j}{2} \big)   \Big\}  \,. 
	\end{split}
\end{align*}
\begin{align*}
	\begin{split} 
		W_{\varphi\sigma}(i,j) &= \Big\{\delta_{i,j+1} \Big[w\big(0,\tfrac{i-2}{2}\big) \,w\big(0,\tfrac{j\!-\!1}{2}\big)-\,w\big(0,\tfrac{i}{2}\big)^2\Big]\\[1ex]
		&+\delta_{i,j} \Big[w\big(0,\tfrac{i}{2}\big)\Big(w\big(\tfrac{i-j\!-\!1}{2},\tfrac{i+2}{2}\big)+\tfrac{1}{2}\, \,w\big(\tfrac{-i+j-3}{2},\tfrac{j+1}{2}\big)-w\big(0,\tfrac{j+1}{2}\big) \,w\big(\tfrac{-i+j+1}{2},\tfrac{i+2}{2}\big)\Big)\\[1ex]
		&+w\big(0,\tfrac{j\!-\!1}{2}\big)\Big(w\big(\tfrac{i-j\!-\!1}{2},\tfrac{i}{2}\big)+\tfrac{1}{2} \,w\big(0,\tfrac{i}{2}\big) \,w\big(\tfrac{i-j+3}{2},\tfrac{j\!-\!1}{2}\big)+\,w\big(0,\tfrac{i-2}{2}\big) \,w\big(\tfrac{-i+j+1}{2},\tfrac{i-2}{2}\big)\Big)\Big]\\[1ex]
		&-\tfrac{1}{2}\,\delta_{i,j\!-\!1} \,w\big(0,\tfrac{i}{2}\big) \Big[  w\big(\!-\!1,\tfrac{j+1}{2}\big) + \,w\big(0,\tfrac{j+1}{2}\big) \,w\big(1,\tfrac{j+1}{2}\big) -w\big(\tfrac{-i+j-3}{2} ,\tfrac{j+1}{2}\big) \Big]\\[1ex]
		&+\tfrac{1}{2} \,\delta_{i,j-2} \,w\big(\tfrac{-i+j-3}{2},\tfrac{j\!-\!1}{2}\big)\,\Big[w\big(0,\tfrac{i}{2}\big) \,w\big(0,\tfrac{j\!-\!1}{2}\big) + \,w\big(0,\tfrac{j\!-\!1}{2}\big) \Big]\\[1ex]
		&+\Big(\vartheta(i-j+1) -\tfrac{1}{2}\,\delta_{i,j-1}\Big) \Big[w\big(0,\tfrac{j\!-\!1}{2}\big) \,w\big(\tfrac{-i+j-3}{2} ,\tfrac{j\!-\!1}{2}\big)-\,w\big(0,\tfrac{i}{2}\big) \,w\big(\tfrac{-i+j-3}{2},\tfrac{j+1}{2}\big)\Big]\\[1ex]
		&-\vartheta(i-j) \Big[\tfrac{1}{2} \,w\big(\tfrac{-i+j\!-\!1}{2} ,\tfrac{j+1}{2}\big)\Big(w\big(0,\tfrac{i}{2}\big) \,w\big(1,\tfrac{i}{2}\big)+w\big(0,\tfrac{j+1}{2}\big) \,w\big(1,\tfrac{j+1}{2}\big)\\
		&-w\big(0,\tfrac{i-2}{2}\big) \,w\big(1,\tfrac{i-2}{2}\big) +w\big(0,\tfrac{j\!-\!1}{2}\big)\,w\big(1,\tfrac{j\!-\!1}{2}\big)\Big)-\sum_{p=j}^i\varphi(p)\,w\big( \tfrac{j-p-1}{2},\tfrac{j+1}{2} \big)\,w\big( \tfrac{p-i-2}{2},\tfrac{p}{2} \big)\\[-1ex]
		&+\sum_{q=j}^i \sum_{p=j}^q \varphi(p)\,\sigma(q)\,w\big( \tfrac{j-p-1}{2} ,\tfrac{j+1}{2} \big)\,w\big( \tfrac{q-i-1}{2} ,\tfrac{q+1}{2} \big)\,w\big( \tfrac{q-p+1}{2} ,\tfrac{p}{2} \big) \\[-1ex]
		&-\sum_{q=j}^i \sum_{p=q}^i \varphi(q)\,\sigma(p)\,w\big( \tfrac{j-q-1}{2} ,\tfrac{j+1}{2} \big)\,w\big( \tfrac{p-i-1}{2} ,\tfrac{p+1}{2} \big)\,w\big( \tfrac{p-q+1}{2} ,\tfrac{q}{2} \big)  \Big]\\[-1ex]
		&+\Big(\vartheta(i-j)-\vartheta(i-j+1)+\tfrac{1}{2}\,\delta_{i,j}+\tfrac{1}{2}\,\delta_{i,j-1} \Big) w\big( 0,\tfrac{i}{2} \big) \sum_{p=j}^i \varphi(p)\,w(\tfrac{i-p+2}{2},\tfrac{p}{2})\,w(\tfrac{j-p-1}{2},\tfrac{j+1}{2})\\[-1ex]
		&+\Big(\vartheta(i-j+1)-\tfrac{1}{2}\,\delta_{i,j-1} -1 \Big) w\big( 0,\tfrac{j-1}{2} \big) \sum_{p=j}^i \sigma(p)\,w(\tfrac{p-j+2}{2},\tfrac{j-1}{2})\,w(\tfrac{p-i-1}{2},\tfrac{p+1}{2})\\
		&+\vartheta(i-j-2) \Big[\,w\big(0,\tfrac{i-2}{2}\big) \,w\big(\tfrac{-i+j+1}{2},\tfrac{j+1}{2}\big)-\,w\big(0,\tfrac{j+1}{2}\big) \,w\big(\tfrac{-i+j+1}{2},\tfrac{j+3}{2}\big)\Big]\\
		&-\sum_{p=j}^i\sigma(p)\,w\big( \tfrac{j-p-2}{2},\tfrac{j+1}{2} \big)\,w\big( \tfrac{p-i-1}{2},\tfrac{p+1}{2} \big)\Big\} \,.
	\end{split}  
\end{align*} }

\section{Initial conditions for the reduced even Pfaff lattice}\label{app:field_variables}
Please note that in this Appendix the elements $w^k_n$ are evaluated at $\mathbf{t} = \mathbf{0}$ and we drop the dependency on time for ease of notation.
Recall the expression for the normalising constant $\nu_n$ for the skew-orthogonal polynomials 
{\begin{equation}\label{eq:hermite_skew_ortho}  
    \nu_n = \frac{\sqrt{\pi}(2n)!}{2^{2n}}.\,,\nonumber
\end{equation} } 

\noindent 
Writing the first equations~\eqref{eq:skew_odd} explicitly, we have
\begin{gather} \label{eq:odd_row_explicit}  
    \begin{aligned} 
        \frac{z\,P_1}{\sqrt{\nu_0}}&= \frac{P_0\,w^{-1}_1}{\sqrt{\nu_0}} + \frac{P_2\,w^{0}_1}{\sqrt{\nu_1}} \\[1ex] 
        \frac{z\,(P_3-P_1)}{\sqrt{\nu_1}}&= \frac{P_0\,w^{-2}_1}{\sqrt{\nu_0}} +\frac{P_2\,w^{-1}_1}{\sqrt{\nu_1}} + \frac{P_4\,w^{0}_2}{\sqrt{\nu_2}} \\[1ex] 
       \frac{z\,(P_5-2\,P_3)}{\sqrt{\nu_2}}&= \frac{P_0\,w^{-3}_1}{\sqrt{\nu_0}} 
       +\frac{P_2\,w^{-2}_2}{\sqrt{\nu_1}} + \frac{P_4\,w^{-1}_3}{\sqrt{\nu_2}} + \frac{P_6\,w^{0}_3}{\sqrt{\nu_3}} \\[1ex] 
       \vdots 
    \end{aligned} 
\end{gather}

\noindent 
Using the three term recurrence relation~\eqref{eq:threepointodd} with $n=1$ and $P_0=1,\,P_{-1}=0$, the first equation gives
\begin{equation*}  
        \frac{1}{\sqrt{\nu_0}}\left({P_{2} + \dfrac{1}{2} \, P_{0}}\right)= \frac{P_0\,w^{-1}_1}{\sqrt{\nu_0}} + \frac{P_2\,w^{0}_1}{\sqrt{\nu_1}}  \,.
\end{equation*} 

\noindent 
We now illustrate the necessary steps to evaluate $w^{-1}_1$ and $w^0_1$ by projecting both sides of the equation above with respect to the same polynomial and exploiting the orthogonality  \eqref{eq:poly_ortho_zero}. Projecting on $P_0$ and $P_2$ we find, respectively,
{\small 
\begin{equation*}  
    \begin{split}
        P_0 \colon \quad \frac{1}{\sqrt{\nu_0}}\left( \left(P_0 \,,\, P_{2} \right)^{(2)}_{\mathbf{0}} + \dfrac{1}{2} \, \left(P_0 \,,\, P_{0} \right)^{(2)}_{\mathbf{0}}\right)&= \frac{\,w^{-1}_1}{\sqrt{\nu_0}}\left(P_0 \,,\, P_{0} \right)^{(2)}_{\mathbf{0}} + \frac{\,w^{0}_1}{\sqrt{\nu_1}}\left(P_0 \,,\, P_{2} \right)^{(2)}_{\mathbf{0}}  \\[1ex]
        \frac{1}{\sqrt{\nu_0}}  \dfrac{1}{2}  &= \frac{\,w^{-1}_1}{\sqrt{\nu_0}}  \quad \implies \quad w^{-1}_1 = \frac{1}{2} \,,  \\[2ex]
        P_2 \colon \quad \frac{1}{\sqrt{\nu_0}}\left( \left(P_2 \,,\, P_{2} \right)^{(2)}_{\mathbf{0}} + \dfrac{1}{2} \, \left(P_2 \,,\, P_{0} \right)^{(2)}_{\mathbf{0}}\right)&= \frac{\,w^{-1}_1}{\sqrt{\nu_0}}\left(P_2 \,,\, P_{0} \right)^{(2)}_{\mathbf{0}} + \frac{\,w^{0}_1}{\sqrt{\nu_1}}\left(P_2 \,,\, P_{2} \right)^{(2)}_{\mathbf{0}}  \\[1ex]
        \frac{1}{\sqrt{\nu_0}}  &=  \frac{\,w^{0}_1}{\sqrt{\nu_1}}  \quad \implies \quad w^{0}_1 = \sqrt{\frac{\nu_1}{\nu_0}} \,.  \\[1ex]
    \end{split}
\end{equation*} }

\noindent 
Given $\nu_0$ and $\nu_1$ from the definition~\eqref{eq:hermite_skew_ortho}, we have
\begin{equation} 
    w^{0}_1 = \sqrt{\frac{\nu_1}{\nu_0}} = \left( \frac{\sqrt{\pi}\, 2!}{2^2}  \, \frac{1}{\sqrt{\pi}} \right)^{1/2} = \frac{1}{\sqrt{2}}\,. 
\end{equation}

\noindent
Iterating the procedure above we obtain the explicit expression for the entries $w^{-k}_n$ with $k>0$ and $w^{0}_n$ in terms of $n$
{\small 
\begin{equation} \label{eq:initial_datum_appendix}
    \begin{split}
        &w^0_n = \sqrt{\frac{\nu_n}{\nu_{n-1}}} = \frac{1}{2} \sqrt{2n(2n-1)} = \sqrt{n\left( n-\frac{1}{2} \right)} \, \\[1ex]
        &w^{-1}_n = \frac{1}{2} \, \\[1ex] 
        &w^{-2}_n = - \frac{1}{2} \sqrt{2n(2n-1)} = - \sqrt{n\left( n-\frac{1}{2} \right)} = - w^0_n\, \\[1ex] 
        &w^{-3}_n = w^{-4}_n = w^{-5}_n = \dots = 0 \,. 
    \end{split}
\end{equation} }

\noindent 
Similarly equations~\eqref{eq:skew_even} read as
{\small 
\begin{equation}
\label{eq:Plista}
    \begin{split}
        \frac{z\,P_0}{\sqrt{\nu_0}} &= \frac{P_1}{\sqrt{\nu_0}}  \\
        \frac{z\,P_2}{\sqrt{\nu_1}} &= \frac{P_3-P_1}{\sqrt{\nu_1}}+\frac{P_1\,w^{1}_1}{\sqrt{\nu_0}}  \\
        \frac{z\,P_4}{\sqrt{\nu_2}} &= \frac{P_5-2\,P_3}{\sqrt{\nu_2}}+\frac{(P_3-P_1)w^1_2}{\sqrt{\nu_1}}+\frac{P_1\,w^{2}_1}{\sqrt{\nu_0}}  \\
        \frac{z\,P_6}{\sqrt{\nu_3}} &= \frac{P_7-3\,P_5}{\sqrt{\nu_3}}+\frac{(P_5-2\,P_3)w^1_3}{\sqrt{\nu_2}}+\frac{(P_3-P_1)w^2_2}{\sqrt{\nu_1}}+\frac{P_1\,w^{3}_1}{\sqrt{\nu_0}}  \\
        \frac{z\,P_8}{\sqrt{\nu_4}} &= \frac{P_9-4\,P_7}{\sqrt{\nu_4}}+\frac{(P_7-3\,P_5)w^1_4}{\sqrt{\nu_3}}+\frac{(P_5-2\,P_3)w^2_3}{\sqrt{\nu_2}}+\frac{(P_3-P_1)w^3_2}{\sqrt{\nu_1}}+\frac{P_1\,w^{4}_1}{\sqrt{\nu_0}}  \\
        \vdots 
    \end{split}
\end{equation} } 

\noindent
As the first equation is a trivial identity between polynomials, we move to the second equation in \eqref{eq:Plista}. Exploiting the three point relation \eqref{eq:threepointeven} with $n=1$ we get
{\small 
\begin{equation*}
        \frac{P_3 +P_1}{\sqrt{\nu_1}} = \frac{P_3-P_1}{\sqrt{\nu_1}}+\frac{P_1\,w^{1}_1}{\sqrt{\nu_0}}  \,.
\end{equation*} }

\noindent
Projecting both sides on $P_1$ we have
{\small 
\begin{equation*}
    \begin{split}
        P_1 \colon& \quad \frac{1}{\sqrt{\nu_1}} = -\frac{1}{\sqrt{\nu_1}}+\frac{w^1_1}{\sqrt{\nu_0}} \quad \implies \quad w^{1}_1 = 2\,\sqrt{\frac{\nu_0}{\nu_1}} \,.
    \end{split}
\end{equation*} }

\noindent 
Similarly, using the three points recurrence relation \eqref{eq:threepointeven} with $n=2$ in the third equation in \eqref{eq:Plista},  we get 
{\small 
\begin{equation*}
        \frac{P_5 +2\,P_3}{\sqrt{\nu_2}} = \frac{P_5-2\,P_3}{\sqrt{\nu_2}} + \frac{(P_3-P_1)w^{1}_2}{\sqrt{\nu_1}}+\frac{P_1\,w^{2}_1}{\sqrt{\nu_0}}  \,,
\end{equation*} }
and projecting on $P_3$ and $P_1$, respectively, we find 
{\small 
\begin{equation*}
    \begin{split}
        P_3 \colon& \quad \frac{2}{\sqrt{\nu_2}} = -\frac{2}{\sqrt{\nu_2}}+\frac{w^1_2}{\sqrt{\nu_1}} \quad \implies \quad w^{1}_2 = 2\cdot 2\,\sqrt{\frac{\nu_1}{\nu_0}} \, \\
        P_1 \colon& \quad 0 = -\frac{w^1_2}{\sqrt{\nu_1}}+\frac{w^2_1}{\sqrt{\nu_0}} \quad \implies \quad w^{2}_1 = \sqrt{\frac{\nu_0}{\nu_1}}\,w^1_2 \,, 
    \end{split}
\end{equation*} }
which imply
{\small
\begin{equation*}
    w^{2}_1 = \sqrt{\frac{\nu_0}{\nu_1}}\,w^1_2 = \sqrt{\frac{\nu_0}{\nu_1}}\,2\,\sqrt{\frac{\nu_1}{\nu_0}} \cdot 2 = 2 \cdot 2 \,. 
\end{equation*} }
As in the previous case, iterating the procedure we obtain the entries $\{w^k_n\}_{n,k\in\mathbb{N}}$ i.e.\ 
{\small 
\begin{equation}\label{eq:w_k_n_generalisation}
    w^k_n = 2\,\frac{(k+n-1)!}{(n-1)!} \, \sqrt{\frac{\nu_{n-1}}{\nu_{k+n-1}}} \,.
\end{equation} }
We also note that the following recursion relation holds
{\small 
\begin{equation} \label{eq:w_k_n_gen_nearest_field}
    w^k_n = n \, \sqrt{\frac{\nu_{n-1}}{\nu_n}} \, w^{k-1}_{n+1}  \,. 
\end{equation} }
By using the explicit form of the the coefficients $\nu_n$ in~\eqref{eq:w_k_n_generalisation} we get 
{\small 
\begin{equation}
    w^k_n = 2^{k+1}\,\frac{(k+n-1)!}{(n-1)!} \, \left(\frac{(2n-2)!}{(2k+2n-2)!}\right)^{\!1/2} \,. 
\end{equation}}
In particular, for $k=1$ we have
{\small
\begin{equation*}
    w^1_n = 2\, \sqrt{n \left( n- \frac{1}{2} \right)^{-1}} \,. 
\end{equation*}} 

Finally, we notice that we can express each entry $w^k_n$ in terms of a product of $w^0_i$. In particular, using the relation~\eqref{eq:w_k_n_gen_nearest_field} and the explicit form of ~$w^0_n$ in~\eqref{eq:initial_datum_appendix}
we have 
{\small
\begin{equation}
\begin{split}
    \prod_{i=n}^{k+n-1} w^0_i \, w^k_n &= 2\, \frac{(k+n-1)!}{(n-1)!}  \quad \implies \quad 
    w^k_n= \left( \prod_{i=n}^{k+n-1} w^0_i  \right)^{\!\!-1} \! 2\, \frac{(k+n-1)!}{(n-1)!}\,. 
\end{split} 
\end{equation} }

\section{The Nijenhuis tensor}\label{app:Nij}
Here we report the non-zero components of the Nijenhuis tensor $\mathcal{N}^i_{jk}$ defined in~\eqref{eq:Nij}, evaluated for the infinite matrix $A(\mathbf{u})$ in Theorem \ref{thm:chain_mixed} in~\eqref{eq:A+-}. 
For $|i|>2$ the non-zero elements are
\begin{align*}
    \mathcal{N}^{\,i}_{0,1}&= \begin{cases} 
    u^0\!\left((i-1)u^{i-1} - (i+1)u^{i+1}\right) \quad &\text{ for }i>2 \\[1.2ex]
    u^0\!\left(i u^{i-1} - (i+2) u^{i+1}\right) \quad &\text{ for }i < -2
    \end{cases}\\[2ex]
            \mathcal{N}^{\,i}_{0,i} &= - 4 u^0\,, ~~\mathcal{N}^{\,i}_{0,i\pm 1} = u^0 u^1,\, ~~\mathcal{N}^{\,i}_{1,i\pm 1} = (u^0)^2,
\end{align*}
\noindent
whereas for $|i|\leq 2$ the non-zero elements are 
\begin{align*}
&\mathcal{N}^{\,2}_{0,1}=u^0 (2 u^1 - 3 u^3)\\[.75ex]
&\mathcal{N}^{\,2}_{0,2}= \mathcal{N}^{\,-1}_{0,-1}= \mathcal{N}^{\,-2}_{0,-2}= - 4 u^0\\[.75ex]
&\mathcal{N}^{\,2}_{0,3} = \mathcal{N}^{\,1}_{0,2} = \mathcal{N}^{\,-1}_{0,-2}= \mathcal{N}^{\,-2}_{0,-3} = u^0 u^1\\[.75ex]
&\mathcal{N}^{\,2}_{1,3}=\mathcal{N}^{\,1}_{1,2}=\mathcal{N}^{\,-1}_{1,-2}=\mathcal{N}^{\,-2}_{1,-3}=(u^0)^2\\[.75ex]
&\mathcal{N}^{\,1}_{0,1}=-2 u^0 (2 + u^2)\\[.75ex]
&\mathcal{N}^{\,-2}_{0,1}=-2 u^0 u^{-3} \\[.75ex]
&
\mathcal{N}^{\,-2}_{-1,1}=-2(u^0)^2\\[.75ex]
&\mathcal{N}^{\,-2}_{0,-1} = 2 u^0 u^1\\[.75ex] 
&\mathcal{N}^{\,-1}_{0,1}=-u^0 u^{-2}.
\end{align*}

\section{\texorpdfstring{Induction step for $w^k_n(\mathbf{t})$}{wk}}\label{app:wk_induction}
We provide details of the induction step to prove the relation~\eqref{eq:wk_red_def} in Theorem \ref{thm:reduction_pfaff} for all $k$. Hence, we assume the relation~\eqref{eq:wk_red_def} holds up to a given $k$ and prove that this implies it holds for $k+1$.

\noindent Let us first observe that the equation for $w^k_n(\mathbf{t})$ in~\eqref{theo:discrete_chain} together with  constraint~\eqref{eq:constr_w0w1}  implies
\begin{equation}\label{eq:wkn_evol} 
    \partial_{t_2} w^k_n = w^{-1}\! \left( c_{n+k}\,w^{k+1}_n - c_{n-1}\,w^{k+1}_{n-1} - c_1 \, w^1_1\, w^k_n + c_n \, w^{k-1}_{n+1}-c_{n+k-1}\,w^{k-1}_n \right).
\end{equation}
Setting $n=1$ and taking into account the boundary condition $w^{k-1}_0(\mathbf{t})\equiv 0$ we have 
\begin{equation} \label{eq:wk1_evol}
    \partial_{t_2} w^k_1 = w^{-1}\! \left( c_{k+1}\,w^{k+1}_1 - c_1 \, w^1_1\, w^k_1 + c_1 \, w^{k-1}_{2}-c_{k}\,w^{k-1}_1 \right) \,.
\end{equation}
According to the inductive hypothesis
$w_2^{k-1}(\mathbf{t})$ is expressed in terms of $w^{k-1}_1$ as follows
\begin{equation}
w_2^{k-1} = \binom{k}{k-1} \frac{c_1\,c_2\, \cdots \, c_{k-1}}{c_2\, c_3\, \cdots \, c_{k}} \,w^{k-1}_1 = k \, \frac{c_1}{c_k} \, w^{k-1}_1 \,. 
\end{equation}
Substituting this expression in equation~\eqref{eq:wk1_evol} we obtain equation~\eqref{eq:wk_evol}. \\

\noindent Let us now consider the left hand side of equation~\eqref{eq:wkn_evol}. Assuming that the relation~\eqref{eq:wk_red_def} holds and using the equation~\eqref{eq:wk1_evol} we get
\begin{equation} \label{eq:lhs_wkn_evol}
\begin{split} 
    \partial_{t_2} w^k_n &= \binom{n+k-1}{k} \, \frac{c_1 \, c_2\, \cdots \, c_k}{c_n \, c_{n+1} \, \cdots \, c_{n+k-1}} \, \partial_{t_2} w^k_1 \\[1ex]
    &= \binom{n+k-1}{k} \, \frac{c_1 \, c_2\, \cdots \, c_k}{c_n \, c_{n+1} \, \cdots \, c_{n+k-1}}\, w^{-1}\! \left( c_{k+1}\,w^{k+1}_1 - c_1 \, w^1_1\, w^k_1 + \frac{k(c_1)^2 - (c_k)^2}{c_k} \,w^{k-1}_1 \right).
\end{split} 
\end{equation}
Similarly, using the relation~\eqref{eq:wk_red_def} into the right hand side of~\eqref{eq:wkn_evol} we obtain
\begin{equation} \label{eq:rhs_wkn_evol}
\begin{split} 
    &w^{-1}\! \left( c_{n+k}\,w^{k+1}_n - c_{n-1}\,w^{k+1}_{n-1} - c_1 \, w^1_1\, w^k_n + c_n \, w^{k-1}_{n+1}-c_{n+k-1}\,w^{k-1}_n \right) = \\[1ex]
    =\,&w^{-1}\! \left\{ c_{n+k}\,w^{k+1}_n - c_{n-1}\,w^{k+1}_{n-1} - c_1 \, w^1_1\, \binom{n+k-1}{k} \, \frac{c_1 \, \cdots \, c_k}{c_n \, \cdots \, c_{n+k-1}} \, w^k_1 \,+ \right. \\[1ex]
     &\left.\hspace{6ex} +  \left(  \, \binom{n+k-1}{k-1} \, \frac{c_1 \, \cdots \, c_{k-1}}{c_{n+1} \, \cdots \, c_{n+k-1}}\,c_n  -\binom{n+k-2}{k-1} \, \frac{c_1 \, \cdots \, c_{k-1}}{c_n \, \cdots \, c_{n+k-2}} \, c_{n+k-1} \right) w^{k-1}_1 \right\}.
\end{split} 
\end{equation}
Equating the expressions in~\eqref{eq:lhs_wkn_evol} and~\eqref{eq:rhs_wkn_evol}, we see that ~$w^{k}_1(\mathbf{t})$ multiplies the same factor in both sides and therefore cancels out. The remaining terms can be rearranged as follows
\begin{equation} \label{eq:red_wkplus1_pre}
\begin{split} 
    &c_{n+k}\,w^{k+1}_n - c_{n-1}\,w^{k+1}_{n-1} =  \binom{n+k-1}{k} \, \frac{c_1 \, \cdots \, c_k}{c_n \, \cdots \, c_{n+k-1}}\,  c_{k+1}\,w^{k+1}_1 + \frac{c_1\, \cdots \, c_{k-1}}{c_{n+1}\, \cdots\, c_{n+k-2}} \times\\[1ex] 
    & \times \left\{ \binom{n+k-1}{k} \frac{k(c_1)^2 - (c_k)^2 }{c_n \, c_{n+k-1}}  +  \binom{n+k-1}{k-1}\, \frac{c_n}{c_{n+k-1}} + \binom{n+k-2}{k-1}\, \frac{c_{n+k-1}}{c_{n}}   \right\} w^{k-1}_1 \,.
\end{split} 
\end{equation}
Multiplying both sides by the product $c_n\, \cdots \, c_{n+k-1}$, we find 
\begin{equation} \label{eq:red_wkplus1}
\begin{split} 
    &c_{n} \, \cdots\, c_{n+k}\,w^{k+1}_n - c_{n-1}\, \cdots \, c_{n+k-1}\,w^{k+1}_{n-1} =  \binom{n+k-1}{k} \, c_1 \, \cdots \, c_{k+1}\,w^{k+1}_1 + c_1\, \cdots \, c_{k-1} \times  \\[1ex] 
    & \times \left\{ \binom{n+k-1}{k} \,(k(c_1)^2 - (c_k)^2 ) +  \binom{n+k-1}{k-1}\, (c_n)^2  + \binom{n+k-2}{k-1}\, (c_{n+k-1})^2   \right\} w^{k-1}_1. 
\end{split} 
\end{equation}
Importantly, the factor multiplying $w^{k-1}_1(\mathbf{t})$ in the second line of~\eqref{eq:red_wkplus1} vanishes, as one can directly verify by using explicit forms of $c_n$ in~\eqref{eq:w0_red_def} and expanding the binomial coefficients:
\begin{equation} \label{eq:red_vanishing}
\begin{split} 
    &  \binom{n+k-1}{k} \,(k(c_1)^2 - (c_k)^2 ) +  \binom{n+k-1}{k-1}\, (c_n)^2  + \binom{n+k-2}{k-1}\, (c_{n+k-1})^2   = \\[1ex]
    & \hspace*{-2ex}= \frac{(n+k-1)!}{k!(n-1)!} (4k-4k^2) - \frac{(n+k-1)!}{(k-1)!\,n!} (4n^2-2n) + \frac{(n+k-2)!}{(k-1)!(n-1)!} (n+k-1) (4n+4k-6) \\[1ex]
    & \hspace*{-2ex}= \frac{(n+k-1)!}{(k-1)!\,(n-1)!} \left( 4-4k- 4n+2 +4n+4k-6\right) = 0\,. 
\end{split} 
\end{equation}
Therefore the expression~\eqref{eq:red_wkplus1_pre} simplifies as follows
\begin{equation}
    c_n\dots c_{n+k}\,w^{k+1}_n-c_{n-1}\dots c_{n+k-1}\,w^{k+1}_{n-1}= \binom{n+k-1}{k}\,c_1 \dots c_{k+1}\,w^{k+1}_1
\end{equation}
Iterating the above recursion relation and using the boundary condition $w^{k+1}_0(\mathbf{t})\equiv 0$, we get
\begin{equation}
\begin{split} 
    &c_{n} \, \cdots\, c_{n+k}\,w^{k+1}_n  = \sum_{j=1}^n \binom{j+k-1}{k} \, c_1 \, \cdots \, c_{k+1}\,w^{k+1}_1 \,,
\end{split} 
\end{equation}
which can be rearranged as
\begin{equation}
\begin{split} 
    &w^{k+1}_n  =  \binom{n+k}{k+1} \, \frac{c_1 \, \cdots \, c_{k+1}}{c_{n} \, \cdots\, c_{n+k}}\,w^{k+1}_1 \,,
\end{split} 
\end{equation}
thus proving that the expression~\eqref{eq:wk_red_def} holds for $k+1$.

\end{appendices}

\bibliography{biblio}
\bibliographystyle{stylesort}

\end{document}